\documentclass[reprint,aps,amsmath,amssymb,twoside,prd,showkeys,superscriptaddress,nofootinbib]{revtex4-2}
\usepackage[letterpaper,top=1.45cm,bottom=1.65cm,left=1.75cm,right=1.75cm]{geometry}
\pdfoutput=1
\usepackage{verbatim}
\usepackage{comment}
\usepackage{graphicx}
\usepackage[T1]{fontenc}
\usepackage{epsfig}
\usepackage{bm}
\usepackage{tensor}
\usepackage{amssymb}
\usepackage{float}
\usepackage{amsmath}
\usepackage{subfigure}
\usepackage{dcolumn}
\usepackage[utf8]{inputenc}
\usepackage{cancel}
\usepackage[colorlinks]{hyperref}
\usepackage[usenames,dvipsnames]{color}
\usepackage{lmodern}
\usepackage[x11names]{xcolor}
\usepackage{array, colortbl, hhline}
\usepackage{xparse}
\DeclareExpandableDocumentCommand\emptycell{O{|c|}m}{\multicolumn{#2}{#1}{}}

\usepackage{fancyhdr}
\fancypagestyle{plain}

\usepackage{placeins}

\hypersetup{
     breaklinks=true,
    pdfstartview={FitH},    
    colorlinks=true,       
    linkcolor=blue,          
    citecolor=red,        
    filecolor=magenta,      
    urlcolor=blue,           
    anchorcolor=green,      
    linktocpage=true
}

\allowdisplaybreaks

\begin{document}
\title{Puncture gauge formulation for Einstein-Gauss-Bonnet gravity and four-derivative scalar-tensor theories in $d+1$ spacetime dimensions}

\author{Llibert Arest\'e Sal\'o}
\email{l.arestesalo@qmul.ac.uk}
\affiliation{School of Mathematical Sciences, Queen Mary University of London, Mile End Road, London, E1 4NS, United Kingdom}

\author{Katy Clough}
\email{k.clough@qmul.ac.uk}
\affiliation{School of Mathematical Sciences, Queen Mary University of London, Mile End Road, London, E1 4NS, United Kingdom}

\author{Pau Figueras}
\email{p.figueras@qmul.ac.uk}
\affiliation{School of Mathematical Sciences, Queen Mary University of London, Mile End Road, London, E1 4NS, United Kingdom}

\begin{abstract}
We develop a modified CCZ4 formulation of the Einstein equations in $d+1$ spacetime dimensions for general relativity plus a Gauss-Bonnet term, as well as for the most general parity-invariant scalar-tensor theory of gravity up to four derivatives. We demonstrate well-posedness for both theories and provide full expressions for their implementation in numerical relativity codes. As a proof of concept, we study the so-called ``stealth-scalarisation'' induced by the spin of the remnant black hole after merger. As in previous studies using alternative gauges we find that the scalarisation occurs too late after merger to impact on the tensor waveform, unless the parameters are finely tuned. Naively increasing the coupling to accelerate the growth of the scalar field risks a breakdown of the effective field theory, and therefore well-posedness, as the evolution is pushed into the strongly coupled regime. Observation of such an effect would therefore rely on the detection of the scalar radiation that is produced during scalarisation. This work provides a basis on which further studies can be undertaken using codes that employ a moving-punctures approach to managing singularities in the numerical domain. It is therefore an important step forward in our ability to analyse modifications of general relativity in gravitational wave observations.
\end{abstract}

\maketitle
\thispagestyle{fancy}

\section{Introduction}
\label{sec:intro}

Gravitational waves from the mergers of compact objects provide an opportunity to study the strong field, highly dynamical regime of general relativity (GR) at higher curvature scales than previous observations. Whilst the curvature scales accessed by current and planned gravitational wave detectors are still well below those where GR is expected to break down due to quantum effects, they nevertheless represent an unexplored part of the parameter space in which deviations could manifest \cite{LISA:2022kgy,Perkins:2020tra,Barausse:2020rsu, Gnocchi:2019jzp,Barack:2018yly,Baker:2014zba}. 
In order to properly test this, we need to understand what deviations could look like in theories beyond GR.

The parameter space of modified theories is highly constrained by a range of astronomical and cosmological observations (see \cite{Baker:2014zba, Koyama:2015vza, Berti:2015itd, Ferreira:2019xrr} for reviews). As discussed in \cite{Baker:2014zba}, there is no unique parameterisation that maps between all different observations and theories, but a well-motivated one is based on the typical length scale of the curvature, for example, as measured by the Kretschmann scalar of the physical system. Using this parameterisation, weaker gravity scenarios like solar system constraints already rule out modifications on larger scales associated with supermassive black holes,\footnote{This assumes that the effects of the modification are not screened at such scales. Another interesting scenario is one in which the modifications act at longer scales (in particular, they can then provide dark energy models), but are screened in high density regions within galaxies (see \cite{Koyama:2015vza} for a review). Numerical studies of such mechanisms are challenging due to the difference in length scales involved, but have produced interesting results in recent years about the effectiveness of the mechanism beyond static, spherical configurations, see \cite{Shibata:2022gec,Lara:2022gof,Bezares:2021dma,Bezares:2021yek}.} 
but the regime of higher curvature (smaller length $\sim$ km) scales as probed by LIGO observations of black holes (BHs) and neutron stars are only just beginning to be constrained \cite{Evstafyeva:2022rve,Perkins:2021mhb,Toubiana:2020vtf,Carson:2019fxr,Yunes:2016jcc,Berti:2015itd}.

Current waveforms are tested for consistency with GR by measuring parameterised deviations to the merger, inspiral and ringdown phases \cite{Maggio:2022hre,Krishnendu:2021fga, LIGOScientific:2021sio,Carson:2019kkh,Cornish:2011ys}. However, it is desirable to obtain predictions for specific models to check whether such parameterised deviations are well-motivated and consistent in alternative theories beyond GR \cite{LISA:2022kgy, Okounkova:2022grv, Johnson-McDaniel:2021yge,Shiralilou:2021mfl,Perkins:2021mhb,Carson:2020ter,Carson:2020cqb}. Such predictions necessitate the use of numerical relativity for the merger section of the signal in near equal mass cases. 
Beyond GR theories also have implications for compact objects such as neutron stars and boson stars, which can undergo scalarisation through mechanisms similar to the BH case (e.g., the recent works \cite{Kuan:2023trn, Ma:2023sok}, see \cite{Doneva:2022ewd} for a review of earlier work).

Lovelock's theorem states that GR is the unique 4-dimensional, local, second derivative theory for a massless spin-2 field \cite{Lanczos:1938sf,Lovelock:1972vz,Lovelock:1971yv}. Therefore modifications to GR require one of these ``pillars'' to be broken. 
From a minimal perspective, one might consider the addition of higher derivatives of the metric to be the most well-motivated. In pure gravity, after considering field redefinitions, the leading correction to GR starts at six or eight derivatives \cite{Endlich:2017tqa}. Such theories of gravity have equations of motion greater than second order and it is not yet understood how to obtain well-posed formulations that capture the physics of interest, despite some recent progress \cite{Cayuso:2017iqc,Allwright:2018rut,Cayuso:2020lca,Franchini:2022ukz,Cayuso:2023aht}. 
The property of well-posedness guarantees that, given some suitable initial data, the solution to the equations of motion exists, is unique and depends continuously on the initial data. It is thus a necessary (but not necessarily sufficient) condition to be able to simulate the theory on a computer and extract waveforms that can then be compared to the predictions of GR.  
Non-trivial modifications in pure gravity that maintain second order equations of motion, and admit well-posed formulations, can be found by going to higher dimensions -- for example, in the $d+1$ dimensional Lovelock theory of GR plus a Gauss-Bonnet term \cite{Lovelock:1971yv}.
Problems with well-posedness also afflict formulations of massive gravity and non-local, Lorentz violating theories such as Einstein-Aether, although some pioneering work is tackling these possibilities \cite{Sarbach:2019yso,deRham:2023ngf}.

\begin{figure*}[t]
    \centering
    \includegraphics[width=0.9\textwidth,trim={3.2cm 7cm 1cm 7cm},clip]{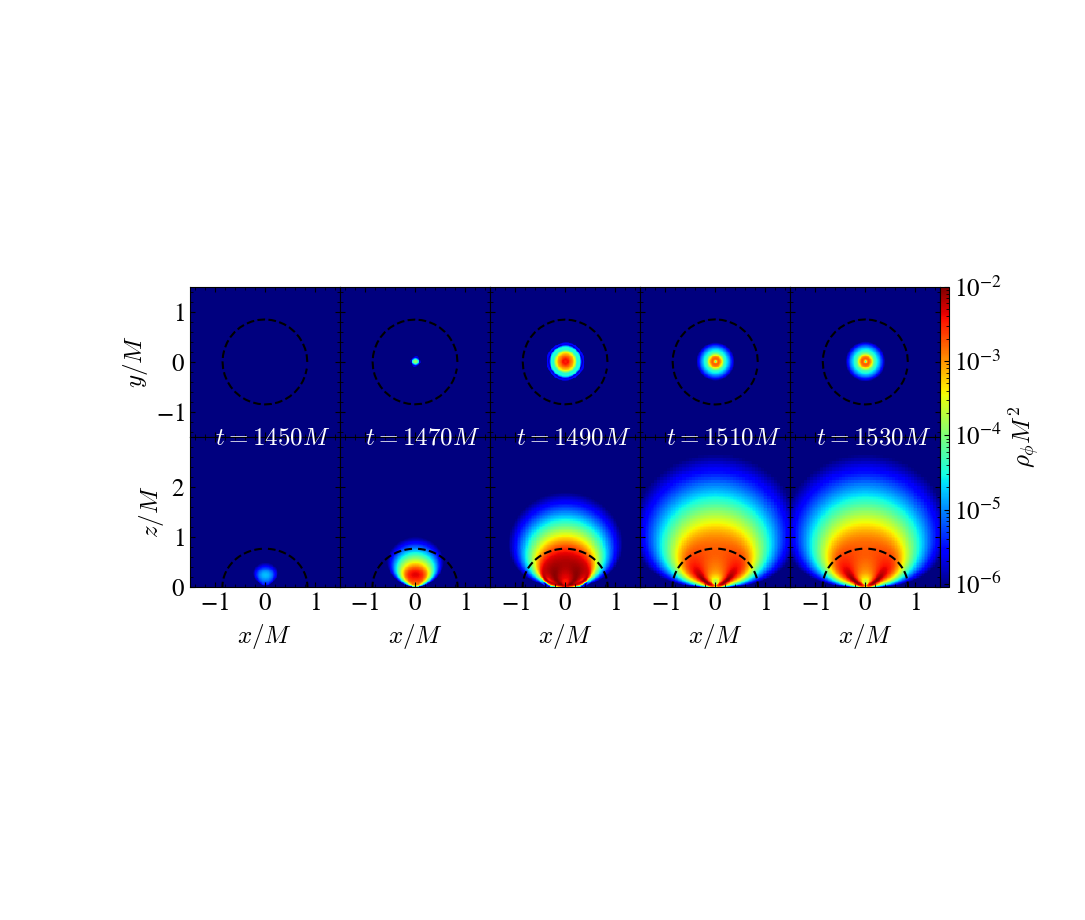}
    \caption{Fully non-linear Stealth Scalarisation: 
    Here we show the time evolution of the scalar cloud after the merger for Einstein-scalar-Gauss-Bonnet theory with exponential quadratic coupling (see eq. \eqref{eq:exp_quad_coupling}) on the rotation plane (upper row) and on a section orthogonal to it (lower row). The colour indicates the contribution of the kinetic and gradient terms to the energy density of the scalar. The dotted black lines denote the location of the Apparent Horizon. We see that the scalar cloud grows by extracting spin from the remnant, and stabilises with a density that is high compared to the curvature scale of the BH.
    }
    \label{fig:cloud}
\end{figure*}


From the perspective of numerical, time-domain studies of GR modifications, one of the simplest modifications is the addition of an additional scalar degree of freedom. 
Some theories can be mapped to GR plus a minimally coupled scalar by a rescaling (going from the Jordan to the Einstein frame), and these clearly admit a well-posed formulation. However, in the absence of an additional mechanism to excite the scalar field, they lack any distinctive features that would distinguish them from GR (since their stationary black hole solutions are those of Kerr). That is, the solutions have ``no hair''.
More general 4-derivative scalar-tensor theories (from the wider class of Horndeski models \cite{Horndeski:1974wa}) may give rise to hairy solutions, but have lacked well-posed formulations until relatively recently. Work has been done using order-reduced methods that  evolve the scalar equation of motion on a fixed GR background \cite{Richards:2023xsr,R:2022tqa,Okounkova:2022grv,Elley:2022ept,Doneva:2022byd,Okounkova:2020rqw,Silva:2020omi,Okounkova:2019zjf,Okounkova:2019dfo,Witek:2018dmd} and hence do not have any well-posedness issues, as long as a certain regularity in the background metric is satisfied \footnote{In particular, the conformal metric is required to be at least ${\mathcal C}^3$ in order to set up an asymptotically regular structure at $\mathcal{I}^+$ \cite{Penrose:1965am}. This is especially relevant to physical situations involving shocks, e.g. in neutron star mergers, where this condition can fail.}. Such simulations can provide an estimate of the scalar dynamics and associated energy losses, but may miss information about the fully non-linear impact on the metric, and potentially suffer from the accumulation of secular errors over long inspirals. Despite their limitations, these studies put initial constraints on the coupling from the merger signal, and have identified many interesting effects such as dynamical descalarisation \cite{Silva:2020omi}, and so-called ``stealth scalarisation'' \cite{Elley:2022ept}, in which the spinning remnant of the merger induces a growth in the scalar field, see Fig. \ref{fig:cloud}.\footnote{These works typically neglect the 4-derivative scalar term, which we see from our work is justified since it is always subdominant to the effect of the Gauss-Bonnet term.}

Progress in simulating the full scalar-tensor theory, including backreaction effects (but limited to the weak-coupling regime), was made possible by Kov\'acs and Reall, who showed that these theories are indeed well-posed in a modified version of the harmonic gauge \cite{Kovacs:2020pns,Kovacs:2020ywu}. Subsequently, studies of some specific scalar-tensor theories within these classes have been probed in their highly dynamical and fully non-linear regimes \cite{East:2020hgw,East:2021bqk,East:2022rqi,Corman:2022xqg}. These studies have probed the limits of the hyperbolicity of the theory in the modified gauge, the stability of the hairy BH solutions, and their imprint on the gravitational wave signal in mergers. They have shown that current post-Newtonian theory is not sufficient to model the dephasing of the gravitational wave signal in the last few orbits \cite{Corman:2022xqg}. However, the work is still in the early stages of development and some questions remain unanswered. In particular, recent work in the spherically symmetric case has shown that there is strong geodesic focusing that is independent of the gauge \cite{R:2022hlf}, but in the general case it is not always clear whether a breakdown in hyperbolicity is due to the gauge or to a problem with the predictivity of the theory.\footnote{Recently \cite{Thaalba:2023fmq} has suggested that by adding extra interactions between the scalar field and the spacetime curvature can ameliorate the loss of hyperbolicity in certain situations.} Instabilities in evolving unequal mass cases to merger using modified Generalised Harmonic Coordinates with excision have also been encountered, and these do not appear to relate to a loss of hyperbolicity but to other numerical issues \cite{Corman:2022xqg}. As we will discuss in this work, our own method requires some tuning of the parameters to achieve stability, and it may require further experimentation before the methods become as well developed and stable as the existing GR formulations.\footnote{An alternative to simulating the full theory that has seen recent success is so-called ``fixing'' of the equations of motion, where the UV behaviour of the equations is explicitly modified, see for example \cite{Cayuso:2023aht,Cayuso:2020lca,Franchini:2022ukz, Lara:2021piy,Bezares:2021yek,Cayuso:2017iqc}.}

Generalised harmonic coordinates (GHC) give a manifest wave-like structure to the equations, but their practical implementation in numerical simulations necessitates excision. The latter, whilst conceptually straightforward, can be difficult to implement in practise. As a consequence,  many groups in the numerical relativity community have opted to use singularity avoiding coordinates such as the BSSN \cite{Nakamura:1987zz,Shibata:1995we,Baumgarte:1998te}, Z4C \cite{Bona:2003fj,Bernuzzi:2009ex} or CCZ4 \cite{Alic:2011gg,Alic:2013xsa} formulations in the puncture gauge \cite{Campanelli:2005dd,Baker:2005vv}, which do not require the excision of the interior of black holes from the computational domain. The extension of the results of \cite{Kovacs:2020pns,Kovacs:2020ywu} to singularity avoiding coordinates would allow such groups to generate waveforms in these models. It would also give an alternative gauge in which to probe questions of hyperbolicity, and may offer advantages for the unequal mass issues found in \cite{Corman:2022xqg} (as we plan to test in future work).

In this paper we extend and generalise the results of our previous Letter \cite{AresteSalo:2022hua}, in which we modified the CCZ4 formulation of the Einstein equations together with the $1+\log$ slicing \cite{Bona:1994dr} and Gamma-driver \cite{Alcubierre:2002kk} gauge conditions and showed that the most general parity-invariant scalar-tensor theory of gravity up to four derivatives (4$\partial$ST) was well-posed, and permitted a stable numerical evolution in singularity avoiding coordinates.
In particular, we extend those results to $d+1$ spacetime dimensions, and give further details on the derivation and numerical implementation of the well-posed formulation. We also treat the case of pure GR with a Gauss-Bonnet term in higher dimensions, for which the same approach is effective.

The article is organised as follows: In Sec. \ref{sec:mCCZ4} we set out the modified CCZ4 formulation for the pure GR case with general matter source terms. This allows us to define the auxiliary metrics required, specify the constraint damping conditions and write down the $d+1$ decomposition and the modified gauge evolution. We then show that this system remains well-posed. In the following sections we extend the analysis in \cite{AresteSalo:2022hua} to two cases of modified theories of gravity:
\begin{itemize}
    \item In Sec. \ref{sec:egb} we detail the case of adding a Gauss-Bonnet term to the usual Einstein-Hilbert Lagrangian in $d+1$ dimensions (with $d>3$).
    \item In Sec. \ref{sec:esgb} we treat the case of an additional scalar degree of freedom in 4$\partial$ST.
\end{itemize}
In both cases, the modifications to the GR case can be accounted for by effective matter source terms -- taking the place of the energy, momentum and stress density terms in the GR case -- but which are not true matter terms but rather specific functions of the curvature quantities. The specific form of these terms is non trivial to write down and not particularly enlightening, and so we give this in the appendices, along with some implementation details regarding the need to invert matrices to obtain certain components in the evolution equations. In the main text we provide an analysis to confirm that the well-posedness of the equations is not spoilt by the additional terms in a suitable weakly coupled regime.
In Sec. \ref{sec:results} we show selected results of simulations in the second case of 4$\partial$ST, demonstrating in particular that different functional forms for the coupling remain well behaved. We conclude in Sec. \ref{sec:discussion}. 

We follow the conventions in Wald's book \cite{Wald:1984rg}. Greek letters $\mu,\nu\,\ldots$ denote spacetime indices and they run from 0 to $d$; Latin letters $i,j,\ldots$ denote indices on the spatial hypersurfaces and they run from 1 to $d$. We set $G=c=1$. 

\section{Modified CCZ4 formulation}
\label{sec:mCCZ4}

In this section we describe in detail the modified CCZ4 formulation that we use and its derivation for $(d+1)$-dimensional Einstein gravity (i.e., GR without modifications) with a general matter source:
\begin{equation}
    I = \tfrac{1}{2\,\kappa}\int d^{d+1}x\sqrt{-g}R+I_{\text{matter}}\,,
    \label{action}
\end{equation}
where $\kappa = 8\pi G$. In order to write the equations in full generality, we include an arbitrary stress tensor in the right hand side (r.h.s.) of the Einstein equations. 

\subsection{Introduction of auxilliary metrics of the modified gauge}

The equations of motion (eoms) that follow from varying \eqref{action} in the modified harmonic gauge introduced by \cite{Kovacs:2020ywu, Kovacs:2020pns} and supplemented by constraint damping terms are given by:
\begin{align} \label{ESFE}
&R^{\mu\nu}-\tfrac{1}{2}\, R \,g^{\mu\nu}+2\,\hat{P}_{\alpha}^{~\beta\mu\nu}\nabla_{\beta}Z^{\alpha}\nonumber
\\
&\hspace{0.7cm}-\kappa_1\big[2n^{(\mu}Z^{\nu)}+\big(\tfrac{d-3}{2+b(x)}+\tfrac{d-1}{2}\kappa_2\big)\,n^{\alpha}Z_{\alpha}\,g^{\mu\nu}\big]\nonumber\\
&= \kappa\,T^{\mu\nu} \,,
\end{align}
where $\hat{P}_{\alpha}^{~\beta\mu\nu}=\delta_{\alpha}^{(\mu}\hat{g}_{\phantom{a}}^{\nu)\beta}-\frac{1}{2}\delta_{\alpha}^{\beta}\hat{g}^{\mu\nu}$,  $Z^{\mu}$ is the vector of constraints, and $\hat{g}^{\mu\nu}$ and $\tilde{g}^{\mu\nu}$ are two auxiliary Lorentzian metrics whose null cones do not intersect with each other and lie outside the light cone of $g^{\mu\nu}$. As shown in \cite{Kovacs:2020ywu}, this can be achieved by setting
\begin{eqnarray}\label{abfunc}
\tilde{g}^{\mu\nu}=g^{\mu\nu}-a(x)\,n^{\mu}n^{\nu}\,, \quad \hat{g}^{\mu\nu}=g^{\mu\nu}-b(x)\,n^{\mu}n^{\nu},
\end{eqnarray}
with $a(x)$ and $b(x)$ being two functions that satisfy $0<a(x)<b(x)$, $0<b(x)<a(x)$ or $-1<a(x)<0<b(x)$, but are otherwise arbitrary, and $n_{\mu}$ is the unit (with respect to $g^{\mu\nu}$) normal to surfaces of constant $x^0$.  In the modified harmonic gauge, the spacetime coordinates $x^\mu$ are fixed by imposing
\begin{equation}
    Z^{\mu}\equiv-\tfrac{1}{2}\big(H^{\mu}+\tilde{g}^{\rho\sigma} \Gamma^{\mu}_{\rho\sigma}\big) =0\,,
\end{equation}
where $H^\mu$ are the source functions, which can be freely chosen. These choices determine the gauge in that formulation, and it amounts to specifying evolution equations for the lapse and shift, which are derived below.

\subsection{$d+1$ decomposition} 
\label{sec:3p1}

We now consider the usual $d+1$ decomposition of spacetime metric:
\begin{equation}
    ds^2=-\alpha^2 dt^2+\gamma_{ij}(dx^i+\beta^i dt)(dx^j+\beta^j dt)\,,
    \label{eq:adm_metric}
\end{equation}
where $\alpha$ and $\beta^i$ are the lapse function and shift vector respectively. In these coordinates, the unit timelike vector normal to the $t\equiv x^0=\text{const}.$ hypersurfaces is given by $n^{\mu}=\frac{1}{\alpha}(\delta_t^{\mu}-\beta^i\delta_i^{\mu})$. The spatial indices are raised and lowered with the physical spatial metric $\gamma_{ij}$.  We also apply the usual conformal decomposition of the evolution variables,
\begin{align}
\chi&=\det(\gamma_{ij})^{-\frac{1}{d}}, \\
\tilde{\gamma}_{ij}&=\chi\,\gamma_{ij}, \\ \tilde{A}_{ij}&=\chi\big(K_{ij}-\tfrac{1}{d}\,\gamma_{ij}K\big)\,,  \\
\hat{\Gamma}^i&=\tilde{\Gamma}^i+2\,\tilde{\gamma}^{ij}Z_j\,  ,
\end{align}
where $\tilde{\Gamma}^i\equiv\tilde\gamma^{kl}\tilde \Gamma^i_{kl}$, $\tilde \Gamma^i_{kl}$ are the Christoffel symbols associated to the conformal spatial metric $\tilde{\gamma}_{ij}$\footnote{The conformal spatial metric $\tilde \gamma_{ij}$ is unrelated to the auxiliary spacetime metric $\tilde g_{\mu\nu}$ defined in \eqref{abfunc}.}, and $K_{ij}=-\frac{1}{2}\mathcal{L}_n\gamma_{ij}$ is the extrinsic curvature of the spatial slices, and $K$ is its trace.

The $d+1$ decomposition of the vector of constraints $Z^\mu$ is given by \cite{Alic:2011gg,Alic:2013xsa}: 
\begin{subequations}\label{constraints_C}
\begin{eqnarray}
&\Theta=Z^0=\tfrac{1}{2}\big[H^{\perp}+K+\frac{1}{\alpha^2}(1+a(x))\partial_{\perp}\alpha\big], \label{eq:def_theta}\\
&\hspace{-0.5cm} Z_i=-\tfrac{1}{2}\big[H_i + \Gamma_i - \frac{1+a(x)}{\alpha}\left(D_i\alpha +\frac{\gamma_{ij}}{\alpha}\partial_{\perp}\beta^j\right)\big], \label{eq:def_Zi} 
\end{eqnarray}
\end{subequations}
where we use the shorthand notation $\partial_{\perp}\equiv \partial_t-\beta^i\partial_i$, $\Gamma^{k}_{ij}$ are the Christoffel symbols of the spatial metric $\gamma_{ij}$ and $\Gamma_i\equiv \gamma_{ij}\gamma^{kl}\Gamma^{j}_{kl}$. Note that only the function $a(x)$ appears in the components of the vector of constraints \eqref{constraints_C}; this will be important in identifying the constraint violating modes when we analyse the hyperbolicity of the evolution equations in the following sections. 

We consider the usual decomposition of the energy momentum tensor of the matter, 
\begin{equation}
\rho=n^{\mu}n^{\nu}T_{\mu\nu}\,,\quad J_i=-n^{\mu}\gamma_i ^{\phantom{i}\nu}T_{\mu\nu}\,,\quad S_{ij}=\gamma_i^{\phantom{i}\mu}\gamma_j^{\phantom{j}\nu}T_{\mu\nu}\,,   
\end{equation}  
In the case of a massless scalar field with stress tensor\footnote{Note that we have considered the coupling of gravity to the scalar field as in the canonical Horndeski Lagrangian, which differs from the normalization used in \cite{East:2020hgw}.}
\begin{equation}
    T^{\phi}_{\mu\nu}=(\nabla_{\mu}\phi)(\nabla_{\nu}\phi) -\tfrac{1}{2}(\nabla\phi)^2g_{\mu\nu}\,,
    \label{eq:scalar_st}
\end{equation}
where $(\nabla\phi)^2=g^{\mu\nu}(\nabla_{\mu}\phi)(\nabla_{\nu}\phi)$, we get
\begin{subequations}
\begin{eqnarray}
   \hspace{-0.5cm}\rho^{\phi}&=&\tfrac{1}{2}\big(K_{\phi}^2+(\partial\phi)^2\big)\,,\\
   \hspace{-0.5cm} J_i^{\phi}&=&K_{\phi}\,\partial_i\phi\,,\\
   \hspace{-0.5cm} S_{ij}^{\phi}&=&(\partial_i\phi)(\partial_j\phi)+\tfrac{1}{2}\,\gamma_{ij}\big(K_{\phi}^2-(\partial\phi)^2\big)\,,
\end{eqnarray}
\end{subequations}
with $(\partial\phi)^2=\gamma^{ij}(\partial_i\phi)(\partial_j\phi)$ and $K_{\phi}=-\tfrac{1}{\alpha}\partial_{\perp}\phi$.

The resulting $d+1$ form of the Einstein field equations coupled to matter is:
\begin{subequations}\label{eqsccz4}
\begin{eqnarray}
\partial_{\perp}\tilde{\gamma}_{ij} &=& -2\alpha\tilde{A}_{ij}+2\tilde\gamma_{k(i}\partial_{j)}\beta^k-\tfrac{2}{d}\tilde{\gamma}_{ij}\partial_k\beta^k, \\
\partial_{\perp}\chi &=& \tfrac{2}{d}\chi\big(\alpha K - \partial_k\beta^k\big), \\
\partial_{\perp}K&=&-D^iD_i\alpha +\alpha\left[R+2\,D_iZ^i +K(K-2\Theta)\right] \nonumber\\
&&-d\,\kappa_1(1+\kappa_2)\,\alpha\,\Theta+\tfrac{\kappa\,\alpha}{d-1}\big[S-d\,\rho\big]\nonumber\\
&&\textstyle-\frac{d\,\alpha\,b(x)}{2(d-1)(1+b(x))}\Big[R-\tilde{A}_{ij}\tilde{A}^{ij}+\frac{d-1}{d}K^2 \nonumber\\
&&\textstyle\hspace{1.5cm}-(d-1)\kappa_1(2+\kappa_2)\,\Theta -2\,\kappa\,\rho\Big]\,,\\
\partial_{\perp}\Theta &=&{\textstyle\frac{\alpha}{2}}\big[R-\tilde A_{ij}\,\tilde A^{ij}+{\textstyle\frac{d-1}{d}}\,K^2 
+2\,D^iZ_i-2\,\Theta\,K\big]\nonumber\\
&&-Z_i\,D^i\alpha-\tfrac{\kappa_1}{2}\big(d+1+(d-1)\kappa_2\big)\,\alpha\,\Theta-\kappa\,\alpha\,\rho \nonumber\\
&&-\textstyle\frac{b(x)}{1+b(x)}\Big\{ \tfrac{\alpha}{2}\big( R-\tilde A_{ij}\,\tilde A^{ij}+{\textstyle\frac{d-1}{d}}\,K^2 \big) \nonumber\\
&&\hspace{1.75cm} -\tfrac{\kappa_1}{2}\big[\tfrac{d-3}{2+b}+d+1+(d-1)\kappa_2\big]\,\alpha\,\Theta\nonumber\\
&&\hspace{1.75cm} 
- \kappa\,\alpha\,\rho
\Big\}\,,\\
\partial_{\perp}\tilde{A}_{ij}&=&~\alpha[\tilde{A}_{ij}(K-2\Theta)-2\,\tilde{A}_{ik}\tilde{A}^k_{~j}]+2\tilde{A}_{k(i}\partial_{j)}\beta^k\nonumber\\&&+\chi\left[\alpha\left(R_{ij} + 2D_{(i}Z_{j)}-\kappa\,S_{ij}\right)-D_iD_j\alpha\right]^{\text{TF}} \nonumber\\&&- \tfrac{2}{d}(\partial_k\beta^k)\tilde{A}_{ij}\,,\\
\partial_{\perp}\hat\Gamma^i
&=&2\,\alpha\big(\tilde\Gamma^i_{\phantom i kl}\tilde A^{kl}-{\textstyle\frac{d-1}{d}}\tilde\gamma^{ij}\partial_jK-{\textstyle\frac{d}{2\,\chi}}\,\tilde A^{ij}\partial_j\chi\big) \nonumber\\&&
-2\,\tilde A^{ij}\partial_j\alpha-\hat\Gamma^j\partial_j\beta^i + {\textstyle\frac{2}{d}}\,\hat\Gamma^i\partial_j\beta^j \nonumber\\&&+ \tfrac{d-2}{d}\,\tilde\gamma^{ik}\partial_k\partial_j\beta^j + \tilde\gamma^{jk}\partial_j\partial_k\beta^i \nonumber\\
&&+2\,\alpha\,\tilde\gamma^{ij}\big(\partial_j\Theta- {\textstyle\frac{1}{\alpha}}\,\Theta\,\partial_j\alpha - \tfrac{2}{d}\,K\,Z_j\big)\nonumber\\
&&-2\,\kappa_1\,\alpha\,\tilde\gamma^{ij}Z_j\,-2\,\kappa\,\alpha\,\tilde\gamma^{ij}J_j\nonumber\\
&&-\textstyle\frac{2\alpha\,b(x)}{1+b(x)}\Big[
\tilde D_j\tilde A^{ij}-\big(\tfrac{d-1}{d}\big)\tilde\gamma^{ij}\partial_jK-\tfrac{d}{2\,\chi}\tilde A^{ij}\partial_j\chi \nonumber\\
&&\hspace{1.2cm}+\tilde\gamma^{ij}\big(\partial_j\Theta-\tfrac{1}{d}\,K\,Z_j\big) - \tilde A^{ij}Z_j \nonumber\\
&&\hspace{1.2cm}-\kappa_1\,\tilde\gamma^{ij}\,Z_j-\kappa\,\tilde\gamma^{ij}J_j
\Big]\,.
\end{eqnarray}
\end{subequations}
Setting $b(x)=0$ and $d=3$ in \eqref{eqsccz4} we recover the equations in \cite{brown}. 

In the case of a massless scalar field coupled to GR, we would also have the corresponding equation of motion for the scalar field
\begin{equation}
    \Box\phi=0\,.
\end{equation} 
This equation can be written as two  first order (in time) equations for the scalar field $\phi$ and its curvature $K_{\phi}$; in the  $d+1$ decomposition of the spacetime metric \eqref{eq:adm_metric}, they are given by
\begin{subequations}
    \begin{eqnarray}
        \partial_{\perp}\phi &=& -\alpha\, K_\phi, \\
        \partial_{\perp}K_{\phi} &=& \alpha(-D^iD_i\phi + KK_{\phi}) - (D^i\phi) D_i\alpha\,.
    \end{eqnarray}
    \label{eq:scalar_eqs}
\end{subequations}

\subsection{Gauge evolution equations} \label{sec:gauge}

Recall that the choice of source functions $H_i$ and $H^{\perp}$, which is completely free, results in the evolution equations for the gauge variables. For instance, in the standard GR case, i.e., $a(x)=b(x)=0$, by choosing
\begin{subequations}\label{sources_choice}
\begin{eqnarray}
H^{\perp}&=&(2\Theta-K)\big(1-\tfrac{2}{\alpha} \big)\,, \\ 
H_i &=& \tfrac{D_i\alpha}{\alpha}+\tfrac{d-2}{2}\partial_i\chi+\hat{\Gamma}_i\big(\tfrac{d}{2(d-1)\alpha^2}-\chi \big)\,,
\end{eqnarray}
\end{subequations}
the conditions $\Theta=0$ and $Z_i=0$ in \eqref{eq:def_theta}--\eqref{eq:def_Zi} lead to the usual  $1+\log$ slicing and the (integrated) Gamma-driver evolution equations for the lapse and the shift respectively,\footnote{The integrated Gamma-driver equation \eqref{eq:gamma_driver} contains an integration constant that we did not include. If the initial data is not conformally flat, one has to take this constant into account to obtain smooth coordinates throughout the evolution. See for instance \cite{Figueras:2015hkb,Figueras:2017zwa} for examples where this is important.}
\begin{subequations}\label{gauge}
\begin{eqnarray}
\partial_{\perp}\alpha  &=& - 2\alpha(K-2\Theta), \\
\partial_t\beta^i &=&\beta^j\partial_j \beta^i +\tfrac{d}{2(d-1)}\hat{\Gamma}^i\,,\label{eq:gamma_driver}
\end{eqnarray}
\end{subequations}
where the factor $\frac{d}{2(d-1)}$  in \eqref{eq:gamma_driver} comes from imposing that the shift propagates at the speed of light in the asymptotic region.\footnote{This factor is a gauge choice and in some instances, e.g., higher dimensions, other choices may be more convenient for numerical stability.} However, this gauge is not adequate for our purposes since it does not have any dependency on the function $a(x)$ and hence it does not take advantage of the corresponding auxiliary metric that we have introduced. As we shall show in Section \ref{sec:hyp}, the function $a(x)$ plays a crucial role in the proof of strong hyperbolicity in the modified theories of gravity that we consider. A suitably modified version of the $1+\log$ slicing and Gamma-driver equations \eqref{gauge} can be found by the choice of source functions in \eqref{sources_choice} and setting to zero the constraints \eqref{constraints_C} with $a(x)\neq 0$. 
The resulting modified gauge evolution equations become
\begin{subequations}\label{mgauge}
\begin{eqnarray}
&\partial_{\perp}\alpha = -\frac{2\alpha}{1+a(x)}(K-2\Theta), \\
&\partial_{\perp}\beta^i = \frac{d}{2(d-1)} \frac{\hat{\Gamma}^i}{1+a(x)}-\frac{a(x)\,\alpha}{1+a(x)}\,D^i\alpha\,.
\end{eqnarray} 
\end{subequations}

Equations \eqref{eqsccz4}, \eqref{mgauge} together with \eqref{eq:scalar_eqs} provide the closed system of evolution equations whose hyperbolicity we will analyse in Section \ref{sec:hyp}.

\subsection{Constraint damping} \label{sec:damping}

We have included constraint damping terms \cite{Gundlach:2005eh,Pretorius:2004jg,Alic:2011gg,Alic:2013xsa} in the Einstein equation \eqref{ESFE}. One can recover the usual CCZ4 equations \cite{Alic:2011gg,Alic:2013xsa} by considering the trace-reversed version of \eqref{ESFE} and setting $\phi=a(x)=b(x)=0$. By analysing the propagation of the constraint violating modes around Minkowski space as in \cite{Gundlach:2005eh}, we obtain the bounds
\begin{equation}
    \kappa_1>0\,,\quad  \kappa_2>-\tfrac{2}{2+b(x)}\,,
\end{equation} 
that guarantee that constraint violating modes are exponentially suppressed (around Minkowski space). In this section we give the details of the calculation of these bounds so the impatient reader can skip this subsection. 

Note that the $b(x)$ term appearing in equation \eqref{ESFE} has been manually inserted so that the bound on $\kappa_2$ does not depend on $d$.  Taking the divergence of \eqref{ESFE}, one gets
\begin{eqnarray}
\Box Z_{\mu}+R_{\mu\nu}Z^{\nu}-\kappa_1\nabla^{\nu}\left(2n_{(\mu}Z_{\nu)}+\hat{\kappa}_2g_{\mu\nu}n^{\rho}Z_{\rho}\right)\nonumber\\=\nabla^{\nu}\left[b(x)\left(2n_{\beta}n_{(\mu}\delta^{\alpha}_{\nu)}\nabla^{\beta}Z_{\alpha}-n_{\mu}n_{\nu}\nabla^{\rho}Z_{\rho}\right)\right]\,,
\end{eqnarray}
where we have defined $\hat{\kappa}_2=\tfrac{d-3}{2}+\tfrac{d-1}{2}\kappa_2$ as a shorthand notation. Now, linearizing around a background solution $g_{\mu\nu}^{(0)}$ such that $R_{\mu\nu}^{(0)}=0$ and $Z_{\mu}^{(0)}=0$ and going to a frame where $n^{\mu}=(1,0,...,0)$ without loss of generality, one gets
\begin{subequations}
\begin{eqnarray}
\big(\Box-b(x)\partial_t^2\big) Z_0 -\kappa_1\big[(2+\hat{\kappa}_2)\partial_tZ_0-\partial^iZ_i \big]&=& 0\,, \hspace{0.6cm} \\
\big(\Box-b(x)\partial_t^2\big) Z_i -\kappa_1\big(\partial_tZ_i+\hat{\kappa}_2\partial_iZ_0 \big)&=& 0\,.
\end{eqnarray}
\end{subequations}

Then, using a plane-wave ansatz $Z_{\mu}=e^{st+i\,k_ix^i}\hat{Z}_{\mu}$, we are led to the following eigenvalue problem
\begin{eqnarray}\label{mat_damping}
\begin{pmatrix}
\xi-\kappa_1(1+\hat{\kappa}_2)s && i\,\kappa_1\,k && 0 \\
-i\,\kappa_1\hat{\kappa}_2\,k && -\xi && 0 \\
0 && 0 && -\xi
\end{pmatrix}
\begin{pmatrix}
\hat{Z}_0 \\ \hat{Z}_n \\ \hat{Z}_A
\end{pmatrix}
=0\,,
\end{eqnarray}
where $\xi=-s^2(1+b(x))-k^2-\kappa_1s$, $\hat{Z}_n$ is the component of $\hat{Z}_i$ in the direction of $k_i$ and $\hat{Z}_A$ are the components orthogonal to $k_i$.

The eigenvalues for $\hat{Z}_A$ are given by
\begin{eqnarray}
s=\tfrac{-\kappa_1}{2(1+b(x))}\pm\sqrt{\big(\tfrac{\kappa_1}{2(1+b(x))}\big)^2-\tfrac{k^2}{1+b(x)}}\,,
\end{eqnarray}
while the corresponding eigenvalues for $\hat{Z}_0$ and  $\hat{Z}_n$ are more complicated and can be found by setting to zero the determinant of the upper-left block of the matrix in \eqref{mat_damping}, which yields the following quartic polynomial equation,
\begin{eqnarray}
    \left((1+b(x))s^2+k^2\right)^2 +\kappa_1^2\left(-k^2\hat{\kappa}_2+s^2(2+\hat{\kappa}_2)\right) \nonumber\\+ \kappa_1\,s \left((1+b(x))s^2+k^2\right)(3+\hat{\kappa}_2)=0\,.
\end{eqnarray}
For the special case $\hat{\kappa}_2=0$ they take the simple form
\begin{equation}
s=-\tfrac{\kappa_1}{1+b(x)}\pm\sqrt{\big(\tfrac{\kappa_1}{1+b(x)}\big)^2-\tfrac{k^2}{1+b(x)}}\,.
\label{eq:eigen_ZA}
\end{equation}
In this case one has that for large wavenumbers, $k\gg \kappa_1$,
\begin{equation}
s\approx -\tfrac{\kappa_1}{1+b(x)}\pm \tfrac{i\,k}{\sqrt{1+b(x)}}\,,\,  s\approx -\tfrac{\kappa_1}{2(1+b(x))}\pm\tfrac{i\,k}{\sqrt{1+b(x)}}\,,
\end{equation}
while for small wavenumbers, $k\ll \kappa_1$, we get
\begin{eqnarray}
s\approx -\tfrac{\kappa_1}{1+b(x)}\,,~ -\tfrac{k^2}{\kappa_1}\,,~ -\tfrac{2\kappa_1}{1+b(x)}\,,~ -\tfrac{k^2}{2\kappa_1}\,,
\end{eqnarray}
Clearly from \eqref{eq:eigen_ZA}, the real part is always negative and hence these modes are always damped.  We have verified that the eigenvalues for $\hat{Z}_0$ and $\hat{Z}_n$ are undamped for  $\hat{\kappa}_2<-\tfrac{2}{2+b(x)}$; for $\hat{\kappa}_2=-\tfrac{2}{2+b(x)}$, they also have a simple form, namely
\begin{eqnarray}
s&=&\pm\tfrac{i\,k}{\sqrt{1+b(x)}}, \\
s&=&-\tfrac{(4+3b(x))\kappa_1}{2(1+b(x))(2+b(x))}\nonumber\\&&\pm\sqrt{\big(\tfrac{b(x)\kappa_1}{2(2+b(x))(1+b(x)))}\big)^2-\tfrac{k^2}{1+b(x)}}\,,
\end{eqnarray}
and hence they are undamped for all values of $k_i$. Therefore we conclude that damping occurs for $\kappa_1>0$ and $\hat{\kappa}_2>-\frac{2}{2+b(x)}$, which implies that $\kappa_2>-\frac{2}{2+b(x)}$.

\subsection{Hyperbolicity analysis}\label{sec:hyp}

In this section we prove the strong hyperbolicity of the $d+1$ Einstein-scalar-field equations in the same way as in \cite{brown}. We start by writing the principal part of the equations. For this purpose we need to introduce an orthonormal $d$-bein (triad in $d=3$) consisting of a unit covector $\xi_i$, such that $\xi_i\gamma^{ij}\xi_j=1$, and unit vectors $e_A^i$ with $A=1,...,d-1$ such that $\xi_ie_A^i=0$ and $e_A^i\gamma_{ij}e_B^j=\delta_{AB}$. Then, keeping the highest derivative terms in the equations \eqref{eqsccz4},  \eqref{eq:scalar_eqs} and \eqref{mgauge}, and replacing $\partial_{\mu}\to i\xi_{\mu}\equiv i(\xi_0,\xi_i)$,\footnote{Note that this $i$ factor differs from the conventions in \cite{brown}; here we follow the conventions of \cite{Kovacs:2020pns,Kovacs:2020ywu}.} we obtain the system
\begin{eqnarray}
i\xi_0U={\mathbb M}(\xi_k)U, 
\label{eq:principle_part}
\end{eqnarray}
where $U$ is a $2(3d+2)$-dimensional vector accounting for the principal part of the CCZ4 variables plus the scalar field $\phi$ and its curvature $K_{\phi}$, where we have taken into account the constraints $\det(\tilde{\gamma}_{ij})=1$ and $\text{Tr}(\tilde{A}_{ij})=0$. Explicitly, the principal part \eqref{eq:principle_part} for the Einstein-scalar-field system in our gauge is given by
\begin{subequations}\label{shypccz4}
\begin{eqnarray}
i\check{\xi}_0\hat{\tilde{\gamma}}_{ij} &=& 2i\tilde{\gamma}_{k(i}\xi_{j)}\hat{\beta}^k-2\alpha\hat{\tilde{A}}_{ij}-\tfrac{2i}{d}\tilde{\gamma}_{ij}\xi_k\hat{\beta}^k\,, \\
i\check{\xi}_0\hat{\chi} &=&\tfrac{2}{d}\chi\big(\alpha\hat{K}-i\xi_k\hat{\beta}^k \big)\,, \\
i\check{\xi}_0\hat{\phi} &=&-\alpha \hat{K}_{\phi}\,, \\
i\check{\xi}_0\hat{K} &=& \hat{\alpha} + i\alpha\chi \xi_i\hat{\hat{\Gamma}}^i-\alpha\big(\tfrac{d-1}{\chi}\hat{\chi}-\tfrac{\tilde{\gamma}^{jk}\hat{\tilde{\gamma}}_{jk}}{2}\big)\nonumber\\&&+\tfrac{b(x)d\alpha}{2\chi(d-1)(1+b(x))}\big(\xi^l\xi^k\hat{\tilde{\gamma}}_{kl}-\tilde{\gamma}^{jk}\hat{\tilde{\gamma}}_{jk}\nonumber\\&&\hspace{2.7cm}+(d-1)\hat{\chi} \big) \,, \\
i\check{\xi}_0\hat{K}_{\phi} &=& \alpha\hat{\phi}\,, \\
i\check{\xi}_0\hat{\Theta} &=& -\tfrac{\alpha}{2}\big(\tfrac{d-1}{\chi}\hat{\chi}-\tfrac{\tilde{\gamma}^{ij}\hat{\tilde{\gamma}}_{ij}}{2} \big)+ \tfrac{i\alpha\chi \xi_i\hat{\hat{\Gamma}}^i}{2} \nonumber\\&& +\tfrac{b(x)\alpha}{2\chi(1+b(x))}\big(\xi^l\xi^k\hat{\tilde{\gamma}}_{kl}-\chi\tilde{\gamma}^{jk}\hat{\tilde{\gamma}}_{jk}\nonumber\\&&\hspace{2cm}+(d-1)\hat{\chi} \big)\,, \\
i\check{\xi}_0\hat{\tilde{A}}_{ij} &=& \big(\xi_i\xi_j - \tfrac{1}{d}\tfrac{\tilde{\gamma}_{ij}}{\chi}\big)\big(\chi\hat{\alpha} - \tfrac{(d-2)\alpha}{2}\hat{\chi}\big) \nonumber\\&&+i\alpha\chi\big(\tilde{\gamma}_{k(i}\xi_{j)}\hat{\hat{\Gamma}}^k-\tfrac{\tilde{\gamma}_{ij}\xi_k\hat{\hat{\Gamma}}^k}{d} \big)\nonumber\\&&+\tfrac{1}{2}\alpha \big(\hat{\tilde{\gamma}}_{ij}-\tfrac{1}{d}\tilde{\gamma}_{ij}\tilde{\gamma}^{kl}\hat{\tilde{\gamma}}_{kl} \big)\,, \\
i\check{\xi}_0\hat{\hat{\Gamma}}^i &=& \tfrac{2i\alpha\xi^i}{\chi}\big(\hat{\Theta}-\tfrac{d-1}{d}\hat{K}\big)- \tfrac{1}{\chi}\big(\hat{\beta}^i + \tfrac{d-2}{d}\xi^i\xi_l\hat{\beta}^l \big)\nonumber\\&&-\tfrac{2i\alpha b(x)}{1+b(x)}\Big(\tfrac{\xi^i}{\chi}\big(\hat{\Theta}-\tfrac{d-1}{d}\hat{K}\big)
+\xi_j\hat{\tilde{A}}^{ij} \Big)\,,\\
i\check{\xi}_0\hat{\alpha} &=& -\tfrac{2\alpha}{1+a(x)} (\hat{K}-2\hat{\Theta})\,, \\
i\check{\xi}_0\hat{\beta}^i &=& \tfrac{d}{2(d-1)}\hat{\hat{\Gamma}}^i+\tfrac{a(x)}{1+a(x)}\big(\tfrac{d}{2(d-1)}\hat{\hat{\Gamma}}^i-i\alpha\xi^i\hat{\alpha}\big)\,, 
\end{eqnarray}
\end{subequations}
where $\check{\xi}_0=\xi_0-\beta^i\xi_i$ and the hat $\hat\,$ denotes the background values of the corresponding variables.

In the following,  we use the notation $\perp$ to denote the contraction of any tensor $T^i$ with the normal covector $\xi_i$, e.g., $T_\perp=T^i\xi_i$; therefore,  $\hat{\gamma}_{\perp\perp}=\hat{\gamma}_{ij}\xi^i\xi^j$ and so on. Similarly, upper case Latin indices are defined by contractions with the components of the $d$-bein; for instance, $\hat{\gamma}_{AB}=\hat{\gamma}_{ij}e_A^ie_B^j$ and analogously for the other variables. Having introduced the notation, we can now decompose the principal part of the equations \eqref{shypccz4} into a scalar, vector and traceless tensor blocks depending on how they transform with respect to transformations of the $d$-bein vectors $e_A^i$.  The tensor block  is given by
\begin{subequations}\label{hyp_tensor}
\begin{eqnarray}
i\check{\xi}_0\hat{\tilde{\gamma}}_{AB}^{\text{TF}} &=& -2\alpha\hat{\tilde{A}}_{AB}^{\text{TF}}, \\ i\check{\xi}_0\hat{\tilde{A}}_{AB}^{\text{TF}} &=& \tfrac{\alpha}{2}\hat{\tilde{\gamma}}_{AB}^{\text{TF}}\, ,
\end{eqnarray}
\end{subequations}
with eigenvalues $\check{\xi}_0=\pm\alpha$. Note that this block is unchanged with respect to the GR case in standard puncture gauge.%

The vector block is
\begin{subequations}\label{hyp_vector}
\begin{eqnarray}
i\check{\xi}_0\hat{\tilde{\gamma}}_{\perp A}&=&i\chi\hat{\beta}_A-2\alpha\hat{\tilde{A}}_{\perp A}\,, \\ i\check{\xi}_0\hat{\tilde{A}}_{\perp A} &=& \tfrac{\alpha}{2}\hat{\tilde{\gamma}}_{\perp A} + \tfrac{i\alpha\chi^2}{2}\hat{\hat{\Gamma}}_A\,, \\ i\check{\xi}_0\hat{\beta}_A &=& \tfrac{d}{2(d-1)(1+a(x))}\hat{\hat{\Gamma}}_A\,, \\ i\check{\xi}_0\hat{\hat{\Gamma}}_A &=& -\tfrac{1}{\chi}\hat{\beta}_A -\tfrac{2ib(x)\alpha}{\chi^2(1+b(x))}\hat{\tilde{A}}_ {\perp A}\,,
\end{eqnarray}
\end{subequations}
with eigenvalues $\check{\xi}_0=\pm\tfrac{\alpha}{\sqrt{1+b(x)}}$ and $~\pm\sqrt{\tfrac{d}{2(d-1)\chi(1+a(x))}}$. These eigenvalues are degenerate for $\alpha^2\chi=\tfrac{d}{2(d-1)}\tfrac{1+b(x)}{1+a(x)}$,  which can be avoided if we choose $b(x)>\tfrac{d-2}{d}+\frac{2(d-1)a(x)}{d}$ given the ranges of $\alpha$ and $\chi$. This degeneracy reduces to the one already present in the standard CCZ4 formulation of GR when $a(x)=b(x)=0$, which does not cause problems in practical applications. The same appears to happen in our new formulation. Therefore, in practice we can replace this constraint by $b(x)\neq\tfrac{d-2}{d}+\frac{2(d-1)a(x)}{d}$ so as to avoid the degeneracy at spatial infinity.

Finally, the scalar block is
\begin{subequations}\label{hyp_scalar}
\begin{eqnarray}
i\check{\xi}_0\hat{\tilde{\gamma}}_{\perp\perp} &=& \tfrac{2i(d-1)}{d}\chi\hat{\beta}^{\perp} - 2\alpha\hat{\tilde{A}}_{\perp\perp}\,,\\
i\check{\xi}_0\hat{\chi} &=& \tfrac{2}{d}\chi(\alpha\hat{K} - i\hat{\beta}^{\perp} )\,, \\
i\check{\xi}_0\hat{\phi} &=& -\alpha\hat{K}_{\phi}\,,\\
i\check{\xi}_0\hat{K} &=& \hat{\alpha}+i\alpha\chi\hat{\hat{\Gamma}}^{\perp}+\tfrac{\alpha}{2\chi}\big(\hat{\tilde{\gamma}}_{\perp\perp}+\hat{\tilde{\gamma}}_{AB}\delta^{AB}\big)-\tfrac{\alpha}{d-1}\tfrac{\hat{\chi}}{\chi}\nonumber\\&&-\tfrac{b(x)d\alpha}{2\chi(1+b(x))}\big(\tfrac{1}{d-1}\hat{\tilde{\gamma}}_{AB}\delta^{AB}-\hat{\chi}\big)\,, \\
i\check{\xi}_0\hat{K}_{\phi} &=& \alpha\hat{\phi}, \\
i\check{\xi}_0\hat{\Theta} &=& \tfrac{i}{2}\alpha\chi\hat{\hat{\Gamma}}^{\perp} +\tfrac{\alpha}{4\chi}\big(\hat{\tilde{\gamma}}_{\perp\perp}+\hat{\tilde{\gamma}}_{AB}\delta^{AB}\big)- \tfrac{(d-1)\alpha}{2}\tfrac{\hat{\chi}}{\chi}\nonumber\\&&-\tfrac{\alpha b(x)}{2(1+b(x))\chi}\big(\hat{\tilde{\gamma}}_{AB}\delta^{AB}-(d-1)\hat{\chi}\big)\,,\\
i\check{\xi}_0\hat{\tilde{A}}_{\perp\perp}&=&\tfrac{d-1}{d}\chi\hat{\alpha} - \tfrac{(d-1)(d-2)\alpha}{2d}\hat{\chi}+i\tfrac{(d-1)\alpha}{d}\chi^2\hat{\hat{\Gamma}}^{\perp}\nonumber\\&&  - \tfrac{\alpha}{2d}\big(\hat{\tilde{\gamma}}_{AB}\delta^{AB}-(d-1)\hat{\tilde{\gamma}}_{\perp\perp})\big)\,,\\
i\check{\xi}_0\hat{\alpha} &=& -\tfrac{2\alpha}{1+a(x)} (\hat{K}-2\hat{\Theta})\,,\\
i\check{\xi}_0\hat{\beta}^{\perp}&=&\tfrac{d}{2(d-1)(1+a(x))}\hat{\hat{\Gamma}}^{\perp}-\tfrac{ia(x)}{1+a(x)}\alpha \hat{\alpha}\,,\\
i\check{\xi}_0\hat{\hat{\Gamma}}^{\perp} &=& \tfrac{2i\alpha}{\chi}\left(\hat{\Theta}-\tfrac{d-1}{d}\hat{K}\right)-\tfrac{2(d-1)}{d}\tfrac{1}{\chi}\hat{\beta}^{\perp}\nonumber\\&& -\tfrac{2i\alpha b(x)}{\chi(1+b(x))}\big(\hat{\Theta}-\tfrac{d-1}{d}\hat{K}+\tfrac{1}{\chi}\hat{\tilde{A}}_{\perp\perp}\big)\,,
\end{eqnarray}
\end{subequations}
with eigenvalues $\check{\xi}_0 = \pm\tfrac{1}{\sqrt{\chi(1+a(x))}}\,,~\pm\sqrt{\tfrac{2\alpha}{1+a(x)}}\,,~\pm\alpha$ and $\pm\tfrac{\alpha}{\sqrt{1+b(x)}}$, with the last pair of multiplicity $2$. The eigenvalues degenerate for $\alpha=\frac{1}{2\chi}$, $\alpha^2=\frac{1+b(x)}{\chi(1+a(x))}$ and $\alpha=\frac{2(1+b(x))}{1+a(x)}$,  which do not spoil the hyperbolicity of the system as long as $a(x)\neq b(x)$, so that again we avoid degeneracy at spatial infinity.\footnote{We note that while for $\alpha^2=\frac{1}{\chi(1+a(x))}$ some of the eigenvalues coincide, the corresponding eigenvectors remain distinct and hence there is no degeneracy in this case.}

To summarise, the constraints on the functions $a(x)$ and $b(x)$ that guarantee the hyperbolicity of the system are 
\begin{equation}
\begin{aligned}
&0<b(x)\neq\tfrac{d-2}{d}+\tfrac{2(d-1)a(x)}{d} \text{ and}\,\Bigg\{\hspace{-0.5cm}\begin{array}{cc}-1<a(x)<0 \\ \text{or} \\ \hspace{0.7cm}0 < a(x) < b(x)\end{array}\\
&\text{or}\\
&a(x)\neq 1+2b(x) \text{ and } 0 < b(x) < a(x)\,.
\end{aligned}
\end{equation}

Following \cite{Kovacs:2020pns,Kovacs:2020ywu}, we can classify the eigenvalues into three types:\footnote{Note that the plus and minus signs of $\check{\xi}_0$ correspond to the ongoing and outgoing modes.}
\begin{itemize}
    \item Physical eigenvalues: $\check{\xi}_0=\pm\alpha$ with multiplicity $d$, consisting of the $2(d-1)$  polarisations of the gravitational field plus the additional $2$ polarizations from the scalar field.
    \item ``Gauge-condition violating'' eigenvalues: $\check{\xi}_0=\pm\sqrt{\tfrac{2\alpha}{1+a(x)}}\,,~\pm\tfrac{1}{\sqrt{\chi(1+a(x))}}$ and $\pm\sqrt{\tfrac{d}{2(d-1)\chi(1+a(x))}}$, with the last pair of multiplicity $d-1$.
    \item ``Pure-gauge'' eigenvalues: $\check{\xi}_0=\pm\tfrac{\alpha}{\sqrt{1+b(x)}}$ with multiplicity $d+1$.
\end{itemize}
Their corresponding eigenvectors have been explicitly written in Appendix \ref{app:eigenvectors}. Clearly the eigenvalues are real (recall that in all cases $a(x)>-1$ and $b(x)>0$), they smoothly depend on $\xi_i$ and so do corresponding eigenvectors. Hence, we conclude that $\mathbb M$ is diagonalisable. Moreover, the propagation of the constraints (see Appendix \ref{app:constraints}) is strongly hyperbolic, showing that if they are satisfied at the initial time they will continue to be satisfied at future times. Therefore, we have proved that the system is strongly hyperbolic and, thus, well-posed. In the following two sections we extend this well-posedness result to certain modified theories of gravity.

\section{Einstein-Gauss-Bonnet theory of gravity}
\label{sec:egb}

In this section we consider Einstein-Gauss-Bonnet gravity, which is a Lovelock theory, in a $(d+1)$-dimensional spacetime $({\mathcal M}, g_{\mu\nu})$, with $d>3$. The action for this theory is given by \cite{torii}
\begin{eqnarray}
S=\tfrac{1}{2\kappa}\int d^{d+1}x\sqrt{-g}(R+\lambda^{\text{GB}}{\mathcal L}_{\text{GB}}),
\label{eq:action_Lovelock}
\end{eqnarray}
where $\lambda^{\text{GB}}$ is the coupling constant of the theory and
\begin{equation}
    {\mathcal L}_{\text{GB}}=R^2-4R_{\mu\nu}R^{\mu\nu}+R_{\mu\nu\rho\sigma}R^{\mu\nu\rho\sigma}\,,
    \label{eq:Gauss_Bonet_term}
\end{equation}
is the Gauss-Bonnet term. We view \eqref{eq:action_Lovelock}  as a low energy EFT of gravity, with the Gauss-Bonet term being the first correction in an otherwise infinite series; therefore, we will demand that it has to be suitably small compared to the Einstein-Hilbert term. This holds in the weakly coupled regime, which is defined by
\begin{equation}
    L\gg \sqrt{|\lambda^{\text{GB}}}|\,, \label{eq:weak_coupling_lovelock}
\end{equation}
where $L$ is any characteristic length scale of the system associated to the spacetime curvature. 

The eoms that follow from varying \eqref{eq:action_Lovelock} supplemented with our modified harmonic gauge fixing terms as well as constraint damping terms are given by
\begin{align}
&R^{\mu\nu}-\tfrac{1}{2}\, R \,g^{\mu\nu}+2\hat{P}_{\alpha}^{~\beta\mu\nu}\nabla_{\beta}Z^{\alpha}\nonumber
\\
&\hspace{0.7cm}-\kappa_1\big[2n^{(\mu}Z^{\nu)}+\big(\tfrac{d-3}{2+b(x)}+\tfrac{d-1}{2}\kappa_2\big)\,n^{\alpha}Z_{\alpha}\,g^{\mu\nu}\big]\nonumber\\
&= \lambda^{\text{GB}}{\mathcal H}_{\mu\nu}\,,
\label{eq:eoms_lovelock}
\end{align}
where
\begin{equation}
    \begin{aligned}
    {\mathcal H}_{\mu\nu}=-2\Big(&R\,R_ {\mu\nu}-2\,R_{\mu\alpha}\,R^{\alpha}_{\phantom{\alpha}\nu}-2R^{\alpha\beta}\,R_ {\mu\alpha\nu\beta}\\
    &+R_{\mu}^{\phantom{\mu}\alpha\beta\gamma}\,R_{\nu\alpha\beta\gamma}\Big)+\tfrac{1}{2}g_{\mu\nu}{\mathcal L}_{\text{GB}}\,.
    \label{eq:def_H_GB}
    \end{aligned}
\end{equation} 
In practice, ${\mathcal H}_{\mu\nu}$ can be thought of as an effective stress-energy tensor and we treat it as such in the $d+1$ decomposition. The explicit form of the contributions of $\mathcal{H}_{\mu\nu}$ to the stress-energy tensor in $d+1$ form can be found in Appendix \ref{app:eomegb}, along with implementation details of the evolution equations.

For $d=3$,  the symmetry properties of the curvature tensor imply that the equations of motion \eqref{eq:eoms_lovelock} reduce to the standard Einstein equations in vacuum \cite{Kovacs:2020ywu}, so this theory is only different from GR for $d>3$. In this section we will 
explicitly prove well-posedness of \eqref{eq:eoms_lovelock} in our formulation for $d=4$ and in the weakly coupled regime \eqref{eq:weak_coupling_lovelock}. We expect the formulation to remain well-posed for other values of $d>3$.

To show that \eqref{eq:action_Lovelock} is well-posed in our mCCZ4 formulation, we need to find the principal part of the full theory, \eqref{eqsccz4}--\eqref{mgauge}, which can be written as
\begin{align}
    \mathbb{M}=\mathbb{M}_0+\delta\mathbb{M}\,,
    \label{eq:M_full}
\end{align}
where ${\mathbb M}_0$ is the principal part of the Einstein theory, already computed in \eqref{hyp_tensor}--\eqref{hyp_scalar} (without the contributions of the scalar field) and $\delta\mathbb{M}=\lambda^{\text{GB}}\mathbb{M}^{\text{GB}}$ are the contributions from the higher derivative terms, which are small compared to $\mathbb M_0$ in the weakly coupled regime. The explicit form of ${\mathbb M}^{\text{GB}}$ can be found in the Mathematica notebook \cite{supp} which is provided as supplementary material.

Therefore, to prove that the full theory is well-posed in an open neighbourhood around the Einstein theory, we can proceed by explicitly computing the eigenvalues and eigenvectors of \eqref{eq:M_full} perturbatively and showing that $\mathbb M$ has real eigenvalues and is diagonalisable.

Consider one of the eigenvalues\footnote{Here we suppress the subscript $0$ on $\xi_0$ to simplify the notation.} of the unperturbed principal part $\mathbb{M}_0$, namely $\xi$ with multiplicity $N^\xi$; let the associated right and left eigenvectors be $\{{\bf v^{\xi}_{\text{\tiny R},i}}\}_{i=1}^{N^{\xi}}$ and $\{{\bf v^{\xi}_{\text{\tiny L},i}}\}_{i=1}^{N^{\xi}}$ respectively. The perturbed eigenvalues $\left\{\xi+\delta\zeta^{\xi}_i\right\}_{i=1}^{N^{\xi}}$ and eigenvectors $\left\{{\bf \alpha^{\xi}_i}\cdot{\bf v_{\text{\tiny R}}^{\xi}} + \delta{\bf w^{\xi}_i} \right\}_{i=1}^{N^{\xi}}$ can be obtained by solving the eigenvalue problem \cite{hinch},
\begin{align}
{\mathcal T}^{\xi}{\bf\alpha^{\xi}_i} &= i\delta\zeta^{\xi}_i{\bf\alpha^{\xi}_i} \,, \label{eq:system_hinch1}\\ 
\left(\mathbb{M}_0-i\xi{\mathbb I} \right)\delta{\bf w^{\xi}_i} &=\left(i\delta\zeta^{\xi}_i{\mathbb I}-\delta\mathbb{M} \right)({\bf \alpha^{\xi}_i}\cdot{\bf v_{\text{\tiny R}}^{\xi}})\,, \label{eq:system_hinch2}
\end{align}
where ${\mathcal T}^{\xi}_{ij}= \frac{{\bf v_{\text{\tiny L},i}^{\xi\dagger}}\delta\mathbb{M}\,{\bf v_{\text{\tiny R},j}^{\xi}}}{{\bf v_{\text{\tiny L},i}^{\xi\dagger}}{\bf v_{\text{\tiny R},i}^{\xi}}}$. Note that \eqref{eq:system_hinch1} ensures that the r.h.s. of \eqref{eq:system_hinch2} has no components parallel to $\xi$. Therefore,  the matrix  ${\mathbb M}_0-i\xi{\mathbb I}$ on the l.h.s. of \eqref{eq:system_hinch2} is invertible \cite{hinch}.

To prove well-posedness we need to verify that the matrices $\left\{{\mathcal T}^{\xi}\right\}_{\xi\in\text{Spec}({\mathbb M}_0)}$ are diagonalisable and that the perturbed eigenvectors depend smoothly on $\xi_k$. From the projection matrices ${\mathcal T}$ corresponding to each type of eigenvalues (see the attached Mathematica notebook \cite{supp}), one can see that the only non-trivial contributions  occur for the physical eigenvalues. In this case, the explicit form of the projection matrix is 
\begin{eqnarray}
{\mathcal T}^{\pm\alpha}\hspace{-0.1cm}=\hspace{-0.1cm}\pm\hspace{-0.1cm}\begin{pmatrix}
2P_{00} & -2P_{01} & -2P_{02} & -2P_{12} & -2P_{12} \\
-2P_{01} & 2P_{11} & -2P_{12} & 0 & 2P_{02} \\
-2P_{02} & -2P_{12} & 2P_{22} & 2P_{01} & 0 \\
-2P_{12} & 0 & 2P_{01} & 2P_{11} &  \text{\footnotesize $\begin{matrix} P_{11}+P_{22}\\-P_{00} \end{matrix}$}\\
-2P_{12} & 2P_{02} & 0 & \text{\footnotesize $\begin{matrix} P_{11}+P_{22} \\ -P_{00} \end{matrix}$} & 2P_{22}
\end{pmatrix},\hspace{0.2cm}
\end{eqnarray}
with
\begin{equation}
\begin{aligned}
    P_{AB}=\lambda^{\text{GB}}\,e_A^ie_B^j(&{\mathcal L}_nK_{ij}+\frac{1}{\alpha}D_iD_j\alpha+K_{ik}K^k_{~j}\\
    &-2\xi^kN_{ikj}+\xi^k\xi^lM_{ikjl})\,,
\end{aligned}
\end{equation}
and ${\mathcal L}_n$ denotes the Lie derivative along $n^{\mu}$. Finding explicit expressions of the first order corrections to the physical eigenvalues is not necessary since we know that they exist and that they are real given that ${\mathcal T}^{\pm\alpha}$ are real and  symmetric. Therefore, this fact together with the smoothness of all the coefficients in ${\mathbb M}^{\text{GB}}$ ensures the well-posedness of the weakly coupled EGB theory in the $4+1$ modified CCZ4 formulation that we have developed.

\section{Four-derivative scalar tensor theory}
\label{sec:esgb}

The next modified theory of gravity that we consider is the most general parity-invariant scalar-tensor theory of gravity up to four derivatives (4$\partial$ST), whose action is \cite{Weinberg:2008hq}
\begin{equation}
\label{eq:action}
    \begin{aligned}
    I = \frac{1}{16\pi}\int d^4x\sqrt{-g}&\big[-V(\phi)+R+X\\
    &+g_2(\phi)X^2+\lambda(\phi)\mathcal{L}_\text{GB}\big]\,,
    \end{aligned}
\end{equation}
where $X\equiv-\frac{1}{2}(\nabla_\mu\phi)(\nabla^\mu\phi)$, $V(\phi)$ is the scalar potential, $g_2(\phi)$ and $\lambda(\phi)$ are smooth functions of the scalar field $\phi$ (but not of its derivatives), and $\mathcal{L}_\text{GB}$ is the Gauss-Bonet term \eqref{eq:Gauss_Bonet_term}. 

The form of the coupling $\lambda(\phi)$ determines the presence of scalar hair. Previous works have divided the classes of coupling functions into the two following cases \cite{Elley:2022ept,R:2022hlf}:
\begin{itemize}
    \item Type I: $\lambda(\phi) \sim \phi + O(\phi^2)$. In this case the scalar field is always sourced by the presence of curvature and so the Kerr family of black holes are not stationary solutions of the theory. Since all the stationary black hole spacetimes in the theory must have hair, this case is strongly constrained by observations of astrophysical BHs. This case includes the so called shift-symmetric and dilatonic couplings.
    \item Type II: $\lambda(\phi) \sim \phi^2 + O(\phi^3)$. In this case Kerr black holes can be stationary solutions of the theory, but in certain regions of the parameter space there can also exist black holes with non-trivial scalar configurations, i.e., hairy black holes. This means that astrophysical black holes may be on either the hairy or non-hairy branches, which makes them more difficult to constrain.
\end{itemize}

The eoms derived from \eqref{eq:action} in the modified harmonic gauge, analogously to \eqref{ESFE}, yield
\begin{align}
&\textstyle R^{\mu\nu}-\frac{1}{2}\, R \,g^{\mu\nu}+2\hat{P}_{\alpha}^{~\beta\mu\nu}\nabla_{\beta}Z^{\alpha} \nonumber\\
&\hspace{0.7cm}-\kappa_1\textstyle\big[2n^{(\mu}Z^{\nu)}+\kappa_2\,n^{\alpha}Z_{\alpha}\,g^{\mu\nu}\big] \nonumber \\
&\hspace{0.35cm}= T^{\phi\,\mu\nu} +{\mathcal H}^{\mu\nu} 
+T^{X\,\mu\nu} -\tfrac{1}{2}V(\phi)g^{\mu\nu}\,, \label{eq:eom_metric} \\
&[1+2g_2(\phi)X]\,\Box \phi -V'(\phi)-3X^2g_2'(\phi) \, \nonumber\\
&\hspace{0.35cm}- 2g_2(\phi)(\nabla^{\mu}\phi)(\nabla^{\nu}\phi)\nabla_{\mu}\nabla_{\nu}\phi = -\lambda'(\phi){\mathcal L}_{\text{GB}} \,, \label{eq:eom_sf}
\end{align}
where
\begin{align}
    T^X_{\mu\nu}=&~\textstyle g_2(\phi)X(\nabla_{\mu}\phi)(\nabla_{\nu}\phi)+\tfrac{1}{2}g_2(\phi)X^2g_{\mu\nu}\,, \\
    {\mathcal H}_{\mu\nu}= &\textstyle -4\big[2R^{\rho}_ {~(\mu}{\mathcal C}_{\nu)\rho}-{\mathcal C}(R_{\mu\nu}-\frac{1}{2}R\,g_{\mu\nu})-\frac{1}{2}R\,{\mathcal C}_{\mu\nu}\,\nonumber\\&+{\mathcal C}^{\alpha\beta}\left(R_{\mu\alpha\nu\beta}-g_{\mu\nu}R_{\alpha\beta}\right)\big]\, ,
\end{align}
with 
\begin{align}\label{Cmunu}
    {\mathcal C}_{\mu\nu}\equiv\lambda'(\phi)\nabla_{\mu}\nabla_{\nu}\phi+\lambda''(\phi)(\nabla_{\mu}\phi)(\nabla_{\nu}\phi)\,,
\end{align}
and ${\mathcal C}\equiv g^{\mu\nu}\mathcal{C}_{\mu\nu}$.

For the remainder of this paper we consider for simplicity a $4\partial$ST theory with no potential for the scalar field and with the coupling functions being $\lambda(\phi)=\tfrac{\lambda^{\text{GB}}}{4}f(\phi)$ and $g_2(\phi)=g_2$, where $f(\phi)$ is an arbitrary function (which is either linear, quadratic or exponential in our simulations) and $\lambda^{\text{GB}}$ and $g_2$ are constants that we assume to satisfy the weak coupling conditions, namely 
\begin{equation}
    L\gg \sqrt{|\lambda'(\phi)}|\,,\sqrt{|g_2|}\,,
\end{equation}
where $L$ accounts here as well for any characteristic length scale of the system associated to the spacetime curvature and the gradients of the scalar field.

As in the previous case the modifications give rise to effective stress-energy contributions in the $d+1$ decomposition. The explicit form of these contributions in $d+1$ form can be found in Appendix \ref{app:eomesgb}, along with implementation details regarding the evolution equations.

In order to show well-posedness for the $4\partial$ST in our modified CCZ4 gauge, we proceed with the same perturbation analysis done for the previous case. Here we write again the principal part of the theory as,
\begin{eqnarray}
    \mathbb{M}=\mathbb{M}_0+\delta\mathbb{M}\,,
\end{eqnarray}
where in this case ${\mathbb M}_0$ is the principal part of the Einstein-scalar-field theory (here also including the scalar field part) and $\delta\mathbb{M}=\lambda^{\text{GB}}\mathbb{M}^{\text{GB}}+g_2\mathbb{M}^{X}$ are the contributions from the higher derivative terms, which are small compared to $\mathbb M_0$ in the weakly coupled regime.
${\mathbb M}^{\text{GB}}$ is also written down in the Mathematica Notebook attached \cite{supp} and, as for ${\mathbb M}^{X}$, its only contribution comes from
\begin{align}
{\mathbb M}^X\hat{K}_{\phi}=2\big[&K_{\phi}^2-\xi^i\xi^j(D_i\phi)(D_j\phi)\big]\hat{\phi}\nonumber\\&+2iK_{\phi}(\xi^iD_i\phi)\,\hat{K}_{\phi}\,,
\end{align}

Here again the only non trivial contributions to the eigenvalues occur for the physical eigenvalues. Setting $\epsilon=\pm1$, we have that the corresponding projection matrices are
\begin{eqnarray}
{\mathcal T}^{\epsilon\alpha}=\begin{pmatrix}2\sigma_{\epsilon}&\tfrac{2\epsilon}{\chi}\psi_{01}&-\tfrac{\epsilon}{\chi}\big(\psi_{00}-\psi_{11}\big)\\
\tfrac{\epsilon\chi}{2}\psi_{01}&2\eta_{\epsilon}&0\\
-\tfrac{\epsilon\chi}{4}\big(\psi_{00}-\psi_{11}\big)&0&2\eta_{\epsilon}\end{pmatrix},
\end{eqnarray}
where
\begin{align}
     \eta_{\epsilon}=&\textstyle\left[2\xi_i\gamma^i_{\mu}n_{\nu}-\epsilon\left(n_{\mu}n_{\nu} + \xi_i\xi_j\gamma^i_{\mu}\gamma^j_{\nu} \right)\right]{\mathcal C}^{\mu\nu}\,,\\
     \sigma_{\epsilon}=&~\frac{g_2}{2}\left[\xi_i(D^i\phi) K_{\phi}+\epsilon\left(K_{\phi}^2-\xi_i\xi_j(D^i\phi)(D^j\phi) \right) \right] \,,\\
     \psi_{AB}=&~\lambda^{\text{GB}}e_A^ie_B^j\Big[\textstyle{{\mathcal L}_nK_{ij}+\frac{1}{\alpha}D_iD_j\alpha} \\
     &\textstyle{+R_{ij}+KK_{ij}-K_i^{~k}K_{jk}
     +2\xi_k\big(D^kK_{ij}-D_{(i}K_{j)}^{~k} \big)}\Big]\nonumber\,.
\end{align} 
Apart from proving that ${\mathcal T}^{\pm\alpha}$ diagonalises, we can explicity compute the six physical eigenvalues of the $4\partial$ST theory in mCCZ4 perturbatively up to first order; the two corresponding to the purely gravitational sector\footnote{The corresponding eigenvectors are null with respect to the effective metric $C^{\mu\nu}=g^{\mu\nu}-4{\mathcal C}^{\mu\nu}$ as described in \cite{Reall:2021voz}.} are given by
 \begin{align}
 \label{eq:eigen_phys1}
 \xi_0=&~\alpha\left(\epsilon+2\eta_{\epsilon}\right)\,,
 \end{align} 
 and the four corresponding to the mixed gravitational-scalar field polarisations are 
 \begin{align}
 \label{eq:eigen_phys2}
 \xi_0=&~\alpha\bigg(\epsilon+\eta_{\epsilon}+\sigma_{\epsilon}\\
 &\hspace{0.6cm}\textstyle{\pm\sqrt{\left(\eta_{\epsilon}-\sigma_{\epsilon}\right)^2+\psi_{12}^2+\left(\frac{\psi_{11}-\psi_{22}}{2}\right)^2}\bigg)}\,,\nonumber
 \end{align}
where for simplicity we have shifted $\xi_0-\beta^k\xi_k\to \xi_0$. Furthermore, it is straightforward to see that the smoothness conditions are satisfied. Hence, this proves well-posedness in the weakly coupled $4\partial$ST theory in the $3+1$ formalism for the modified CCZ4 formulation that we have developed.

\section{Results of simulations for 4$\partial$ST}
\label{sec:results}

In this section we extend the results shown in \cite{AresteSalo:2022hua} for the $4\partial$ST theory, which we have implemented (using the equations in Appendix \ref{app:eomesgb}) as an extension to \texttt{GRChombo} \cite{Clough:2015sqa,Andrade:2021rbd}. The implementation and study of the Gregory-Laflamme instability of black strings in the higher-dimensional Einstein-Gauss-Bonnet theory is under way and will be presented elsewhere.

\begin{figure*}[t]
    \centering
    \includegraphics[width=7.5cm]{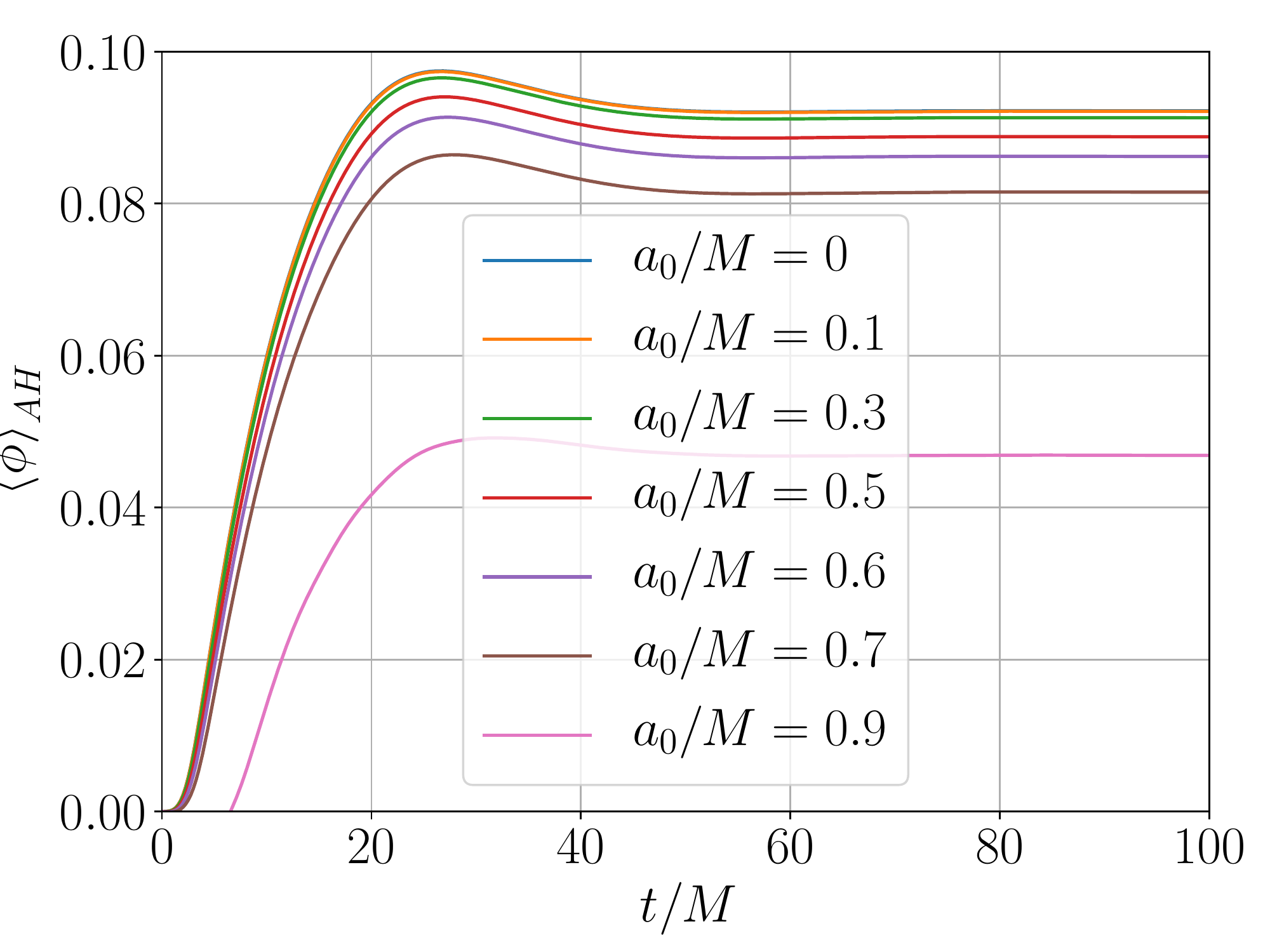}
    \includegraphics[width=7.5cm]{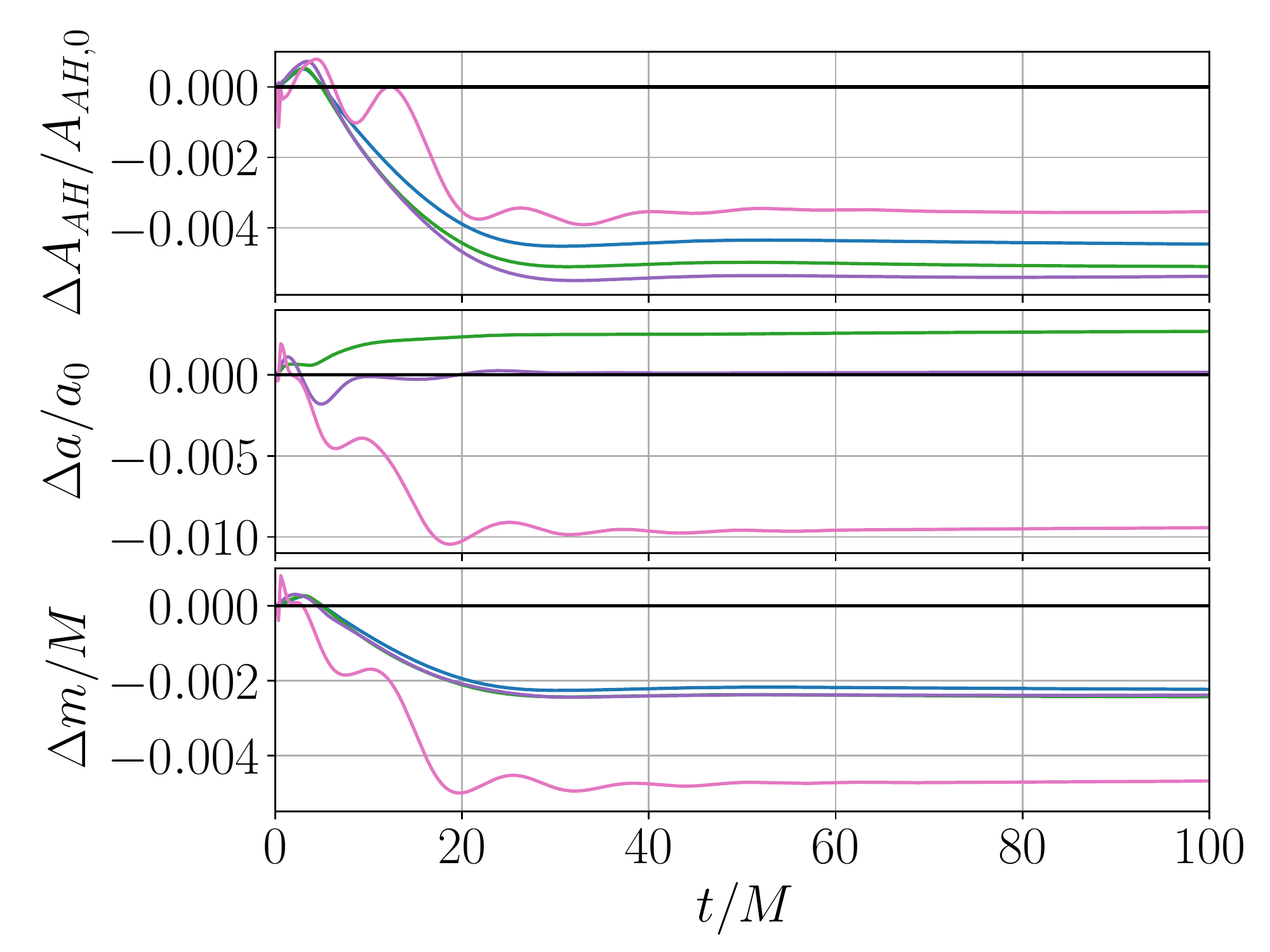}
    \caption{Simulations of single BHs with spin in (Type I) shift-symmetric Einstein-scalar-Gauss-Bonnet theory for $\lambda^{\text{GB}}/M^2=0.2$. From an initial zero value of the scalar field, the curvature sources a growth of the scalar hair to the stationary state, with higher spins sourcing smaller average field values as expected. The energy for the scalar hair is extracted from the BH, which results in a decrease in its AH area (which is permitted in these modified theories of gravity) and mass. In cases with spin, angular momentum may also be extracted by the scalar field.
    \emph{Left}: Average  
    value of the scalar field at the AH for different values of the initial dimensionless spin $a_0/M$. \emph{Right}: Change in the AH area, spin and mass relative to their initial values.
    }
    \label{fig:kerr}
\end{figure*}

\subsection{Technical details}

\subsubsection{Gauge parameters}
We have chosen the functions appearing in our modified CCZ4 gauge to be spatial constants, with $a(x)=0.2$ and $b(x)=0.4$ in all our simulations.  This choice gives reasonable results but our initial investigations suggest that tuning these values or choosing metric dependent functions may give better constraint conservation.\footnote{Making $a(x)$ and $b(x)$ functions of the evolution variable $\chi$ so that they interpolate between zero in the asymptotic region and the values quoted above near black holes does not seem to make any significant difference in the numerical stability of our simulations.} This will be investigated further in future work.

\subsubsection{Excision of the EsGB terms}

As in \cite{AresteSalo:2022hua,Figueras:2020dzx,Figueras:2021abd} we smoothly switch off some of the higher derivative terms in the eoms well inside the Apparent Horizon (AH) by replacing $\lambda^{\text{GB}}\to\tfrac{\lambda^{\text{GB}}}{1+e^{-100(\chi-\chi_0)}}$ with $\chi_0=0.15$ for spinless Black Holes (BHs). The specific value of $\chi_0$ needs to be adjusted (smaller) for higher spin, with the value chosen to be within the AH in our chosen coordinates (see Figure C1 in \cite{Radia:2021smk} for the values in typical puncture gauge simulations). In Binary Black Hole (BBH) merger simulations we have found that it helps to change the value of $\chi_0$ after the merger, since the final remnant has a dimensionless spin of the order of $a/M\gtrsim 0.7$. Provided that the excision happens well within the AH, this should not change the physical behaviour outside it.

\subsubsection{Constraint damping parameters}

We have also noted that the values of the damping constraint coefficients $\kappa_1$ and $\kappa_2$ play an essential role for keeping the violation of the Hamiltonian and Momentum constraints of the system under control and in particular the best values appear to depend on the final spin of the stationary BH solution that the system evolves to. Therefore, we also increase the values after the remnant is formed -- more details are given in the following subsections. We use the usual rescaling   $\kappa_1\to\kappa_1/\alpha$ that allows for stable evolutions of BHs as in \cite{Alic:2013xsa}.

\subsubsection{Numerical set-up}

For the runs with single BHs we use a computational domain of $L = 256M$ with the BH situated at the centre of the grid, and $N=128$ grid points on the coarsest level. We use $6$ levels of refinement, which results in a finest resolution of $dx_{finest} = M/16$ on the finest grid, giving $\sim 35$ grid points across the BH horizon in the quasi-isotropic Kerr coordinates \cite{Liu:2009al} that we use to set the initial conditions for the metric.  These coordinates are a generalisation of the wormhole-like isotropic Schwarzschild coordinates, and similarly evolve into a trumpet-like solution for the (modified) puncture gauge within the first $\sim 10M$ of the simulation. At this point the BH horizon is located at $r\sim 0.98M$ in the zero spin case, which is similar to the GR puncture gauge value \cite{Radia:2021smk}.

For the BBH mergers we have chosen $L=512M$, with $N=128$ grid points on the coarsest level. We use $9$ levels of refinement, which results in a resolution of $dx_{finest} = M/64$ on the finest grid, which gives roughly 60 points across the horizon of each BH prior to their merger. We anticipate that higher resolutions would be required for detailed waveform templates, but for this study we are mainly interested in the overall phenomenology. For both type of simulations we use $6^\text{th}$ order finite differences to discretise the spatial derivatives and a standard RK4 time integrator to step forward in time.  We have checked convergence for these parameters, as shown in \cite{AresteSalo:2022hua}.

We study two cases for the BBH mergers:
\begin{itemize}
    \item Case 1: The BHs have equal masses $m_{(1)} = m_{(2)} = 0.49M$, initial separation $11M$ and initial velocities $v_{(i)} = (0, \pm0.09, 0)$. These initial conditions were tuned to have roughly circular initial orbits in GR such that the two black holes merge in approximately ten orbits. For this case we superpose the solutions for two boosted black holes as described in \cite{Baumgarte:2010ndz,Bowen:1980yu}, using a perturbative solution for the conformal factor that is accurate up to order $(P^i P_i)^2$.
    \item Case 2: An equal mass binary where the component BHs each have non-zero initial (dimensionless) spin of $a_0/M=0.4$ aligned with the orbital axis. In this case we use a standalone version of the TwoPunctures code \cite{Ansorg:2004ds} to generate Bowen-York initial data \cite{Bowen:1980yu} with a separation of $11M$, initial velocities $v_{(i)}=(0,\pm 0.08, 0)$, equal masses of $m_1=m_2=0.31$ (so that the total ADM mass is approximately $1$) and angular momentum $J_{(i)}/m_{(i)}^2=(0,0,0.4)$. In this case the orbits are only roughly circular and we have around eight orbits prior to merger in the GR case.
\end{itemize}

Note that in both cases we use GR initial data, which remains a solution of the constraint equations only in the case in which the additional scalar degree of freedom is zero. In some cases below we add a small perturbation in the field to source its growth after the merger. In these cases, where that the constraints are initially violated, the violations are small and quickly damped away by the damping terms in the eoms. A generalization of the initial data solver of \cite{Aurrekoetxea:2022mpw} to the $4\partial\text{ST}$ theory will be presented elsewhere \cite{Brady:2023dgu}.

\subsection{Type I Coupling -- shift-symmetric EsGB}

We start by considering the simplest case of scalarisation in the $4\partial$ST theory by adding a linear coupling $f(\phi)=\phi$, which is often referred to as shift-symmetric Einstein-scalar-Gauss-Bonnet (EsGB) theory (usually in the absence of the $g_2$ term, although this term also respects the shift-symmetry). 

As discussed above, due to the curvature sourcing the scalar field, BH solutions in this theory differ from the Kerr solution in that they possess a non-trivial scalar configuration, that is, they have scalar hair.

As an initial test, we set the initial conditions to be the single Kerr BH with mass parameter $M=1$ as described above, and set the scalar field to zero initially. The values of the constraint damping coefficients have been chosen to be $\kappa_1=0.35-1.7$ (higher values for this parameter are found to be required for higher spins in order to stabilise the final state) and $\kappa_2=-0.5$. 

Figure \ref{fig:kerr} \footnote{In this and subsequent figures, we present the average value of certain quantities across the Apparent Horizon (AH). We denote 
\begin{eqnarray}
    \langle\psi\rangle_{\text{AH}}=\tfrac{\int\tfrac{1}{\chi}\,\psi\,r^2(\theta,\phi)\sin\theta d\theta d\phi}{\int\tfrac{1}{\chi}\,r^2(\theta,\phi)\sin\theta drd\theta d\phi}\,,
\end{eqnarray}
where $r,\theta,\phi$ account for the spherical coordinates, $r=r(\theta,\phi)$ is the Apparent Horizon and $1/\chi$ is a factor coming from the determinant of the induced metric on a $2$-dimensional surface, which we find is a good approximation to the exact value since the determinant of the (Cartesian) conformal metric in our formulation is $1$.} shows that a stationary hairy BH solution is obtained for all the values of the dimensionless spin parameter $a_0/M$ after an initial transient period of growth of the scalar hair. These final stationary states are consistent with the results in \cite{East:2020hgw}.

\begin{figure}[t]
    \centering
    \includegraphics[width=8cm]{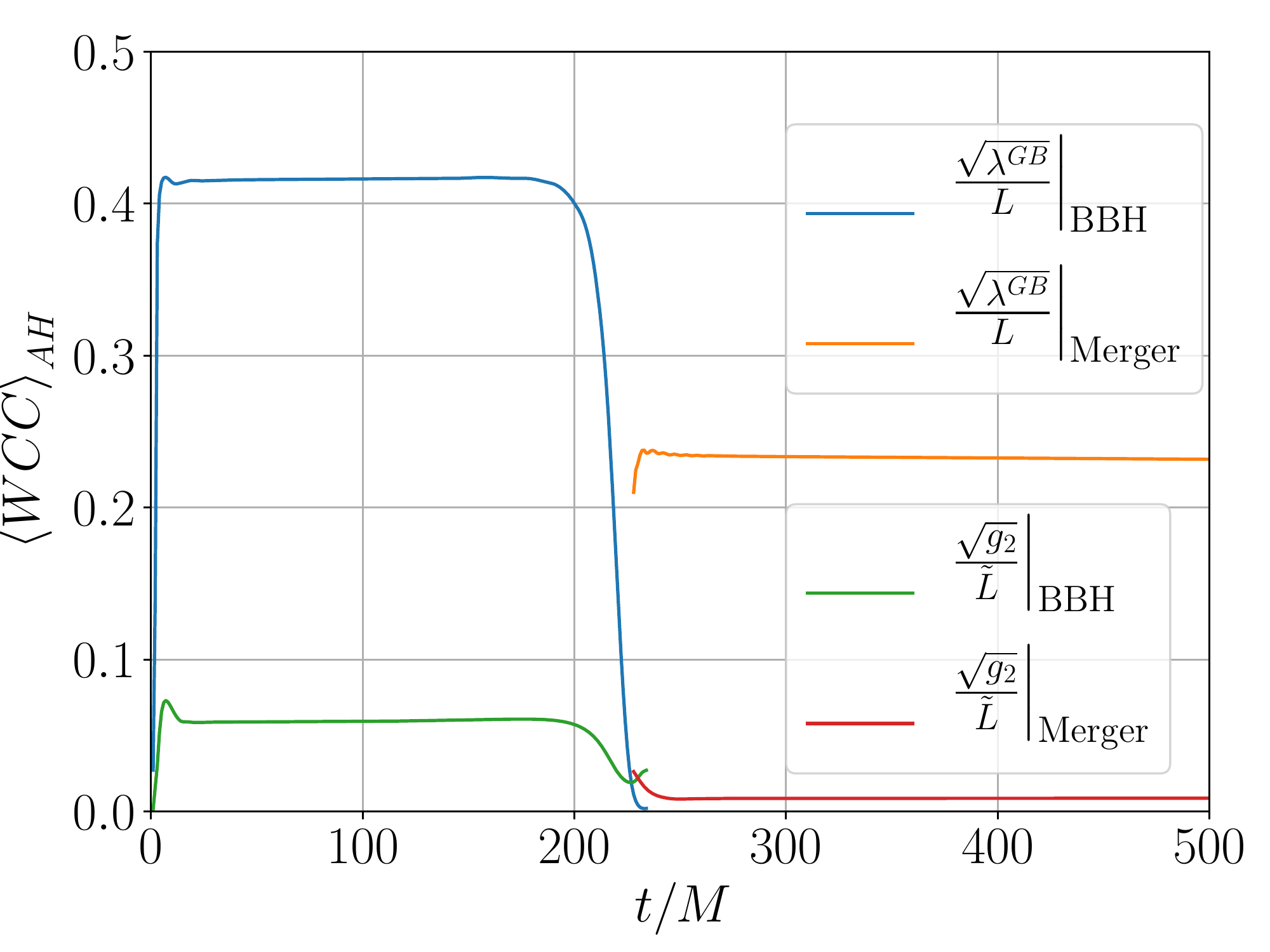}
    \caption{Here we show that the weak coupling regime holds throughout the BBH merger simulation in shift-symmetric $4\partial$ST theory with $\lambda^{\text{GB}}=0.05M^2$ and $g_2=M^2$, namely the one with the highest coupling constants considered in \cite{AresteSalo:2022hua}. 
    We depict the evolution of the WCCs in Equation \eqref{eq:wfc} for both the BBHs throughout the inspiral and the merger during the ringdown, seeing that they are not violated for these values of the couplings. As expected, the highest values are before the merger, due to the smaller curvature scales of the initial black holes compared to the final remnant.}
    \label{fig:wfc}
\end{figure}

Next, we extend the results for BBH Mergers in shift-symmetric $4\partial$ST in \cite{AresteSalo:2022hua} by studying whether the weak coupling condition (WCC) still holds for the case with the highest values of the couplings, namely $\lambda^{\text{GB}}=0.05M^2$ and $g_2=M^2$.
We use equal mass, non spinning  BHs (Case 1 described above), finding that the number of orbits reduces to three for the chosen values of the couplings. The constraint damping coefficients are set to $\kappa_1=0.35/M$ and $\kappa_2=-0.1$.
We observe in Figure \ref{fig:wfc} that the WCC
\begin{eqnarray}\label{eq:wfc}
    \sqrt{\lambda^{\text{GB}}}/L\ll1\,, \qquad \sqrt{g_2}/\tilde{L}\ll1\,,
\end{eqnarray}
still holds (even though it is close to the limit).
Here the relevant length scales that represent the curvature quantities of the metric and scalar sectors are
\begin{eqnarray}
    &L^{-1}=\max\{|R_{ij}|^{1/2},|\nabla_{\mu}\phi|,|\nabla_{\mu}\nabla_{\nu}\phi|^{1/2},|{\mathcal L}_{\text{GB}}|^{1/4}\} \,,\nonumber\\
    &\tilde{L}^{-1}=\max\{|K_{\phi}|,|D_i\phi D^i\phi|^{1/2}\} \,.
\end{eqnarray}
As expected, the highest values of the weak coupling conditions occur right before the merger, given that the curvature scales are larger near the initial BHs in comparison to the final remnant. Therefore if the WCC is not breached during the inspiral, and it appears to be safe during merger and ringdown phases. Note that the WCC is not a well-defined mathematical condition, but is however a heuristic condition that helps us identify that we are in the regime of validity of the theory where the eigenvalues of the principal symbol do not differ significantly from GR.

For this same coupling function, we also test our ability to stably evolve equal mass BBH cases with non zero initial component spins (Case 2 above).
We used the following values of the constraint damping coefficients, $\kappa_1=1.4/M$ and $\kappa_2=-0.1$, which we changed to $\kappa_1=1.7/M$ and $\kappa_2=0$ after merger. We also decrease the value of $\chi_0$ from $\chi_0=0.15$ to $\chi_0=0.05$ after the merger.

\begin{figure}
    \centering
    \includegraphics[width=8cm]{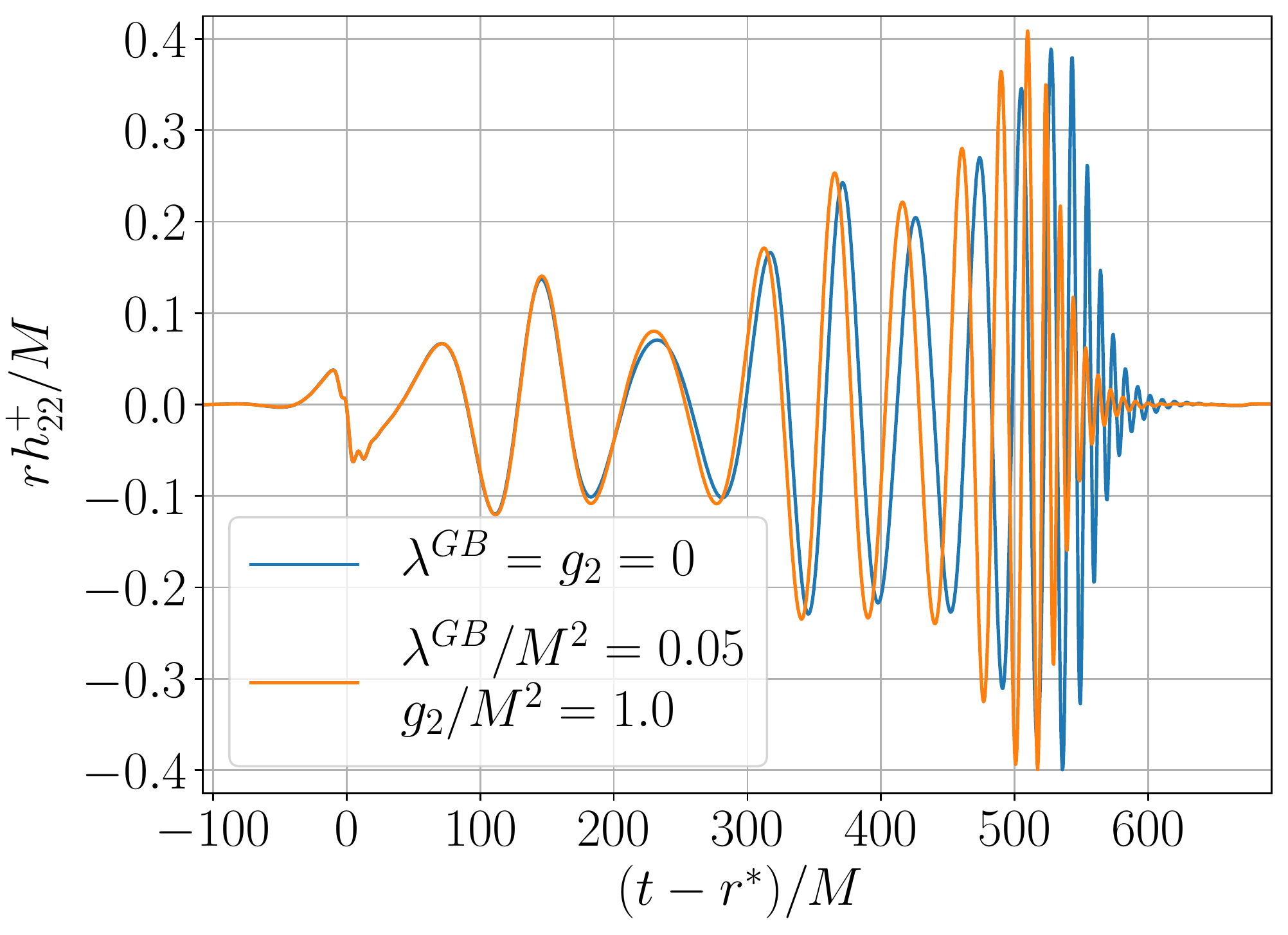}
    \caption{Initial spinning Binary Black Hole mergers (with spins initially aligned along the orbital momentum, $a_0^{\pm}/M=0.4$) in GR (blue) and shift-symmetric $4\partial$ST (orange), for the following values of the coupling constants, $\lambda^{\text{GB}}=0.05M^2$ and $g_2=M^2$. We show the $(2,2)$ mode of the gravitational strain in retarded time, $u=t-r^*$ (where $r^*$ is the tortoise coordinate), observing that the additional extraction and radiation of energy via the scalar channel induces the merger to happen sooner 
    compared to GR.}
    \label{fig:spinbbh}
\end{figure}

The result is shown in Figure \ref{fig:spinbbh}, where we compare the $(2,2)$ mode of the gravitational strain with GR by extracting the gravitational waves at $r=100M$.\footnote{For the accuracy purposes of this article, we only considered the extraction at this radius but in a number of cases we checked that this result is essentially the same as the one obtained by extrapolating to null infinity.}  We find that for the chosen parameters the final spin reduces from $\sim0.85$ in GR to $\sim0.84$ in shift-symmetric $4\partial$ST theory, as expected from the extraction of spin caused by the non-trivial scalar field.

\subsection{Type II Coupling -- tachyonic growth and stealth scalarisation}

At this point we turn to the second class of coupling function, Type II, in which the coupling results in a (spatially dependent) mass term. These admit both scalarised and non-scalarised BH solutions. 

In this case we study the binary case directly, and we choose the coupling parameters so that the scalar hair is generated as a result of the merger. We use Case 1 of the BBH configurations described above throughout this section.

The simplest case of a Type II coupling is a quadratic coupling, namely $f(\phi)=\phi^2$. As studied in \cite{Silva:2020omi,Elley:2022ept}, this coupling function can have a tachyonic instability which leads to a spin-induced scalarisation or descalarisation. We study the case in which the remnant scalarises after the merger due to its spin for a high enough negative value of the coupling $\lambda^{\text{GB}}$. 

We used as constraint damping coefficients $\kappa_1=0.35/M$ and $\kappa_2=-0.1$ initially, but after merger changed them to $\kappa_1=1.7/M$ and $\kappa_2=0$, together with reducing the initial value of $\chi_0=0.15$ to $\chi_0=0.05$.
We also needed to add an initial  perturbation in the scalar field to seed the instability, for which we choose the (arbitrary) form $\phi(r)=10^{-3}(1+0.01r^2e^{-r^2})$.

\begin{figure}
    \centering
    \includegraphics[width=8cm]{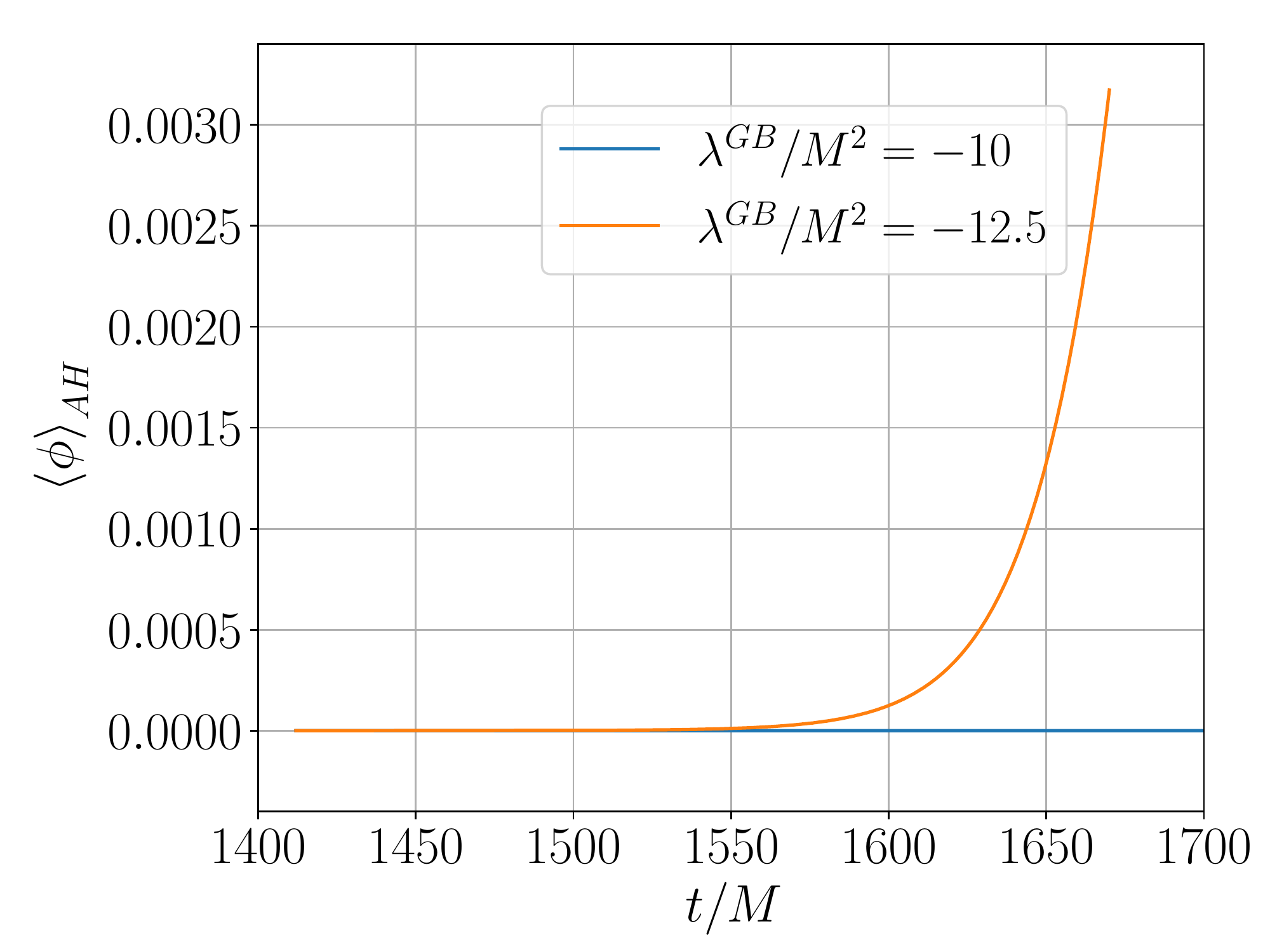}
    \caption{Evolution of the average value of the scalar field across the AH, after the merger occurred in a Binary Black Hole simulation for Einstein-scalar-Gauss-Bonnet theory with quadratic coupling for two different values of the coupling $\lambda^{\text{GB}}$. We see that for a critical coupling value of around $\lambda^{\text{GB}} = -10 M^2$ the remnant (with $a/M\sim 0.7$) scalarises and the value of the scalar field grows exponentially. Eventually the simulation breaks down, since the weak field condition (and hence well-posedness) does not hold anymore.
    }
    \label{fig:phiquad}
\end{figure}

Given that the initial BHs have zero spin, there is initially no scalarisation for this sign in the coupling and the scalar field dissipates. Only after the merger does the scalar field have a non-trivial evolution. In Figure \ref{fig:phiquad} we show the two possible behaviours -- exponential growth or zero growth, and find that the critical value of $\lambda^{\text{GB}}$ for which the transition occurs happens at around $10 M^2$. For the values of the coupling that induces exponential growth we observe that the weak coupling condition is eventually violated and, thus, at some point along the evolution the theory ceases to be well-posed, which results in the breakdown of the simulation (see also \cite{Doneva:2023oww} for further results).

A more phenomenologically interesting class of Type II coupling functions was proposed in \cite{Doneva:2022byd,Doneva:2022yqu}, with the form
\begin{equation}
    f(\phi)=\omega^{-1}(1-e^{-\omega\phi^2})\,, \label{eq:exp_quad_coupling}
\end{equation} 
which we refer as exponential quadratic. This type of coupling has the same initial behaviour as the quadratic one, but the tachyonic instability is saturated by the non-linearities at larger amplitudes, meaning that one can follow the growth of the scalar hair and settling of the solution into a steady hairy BH state after the merger while the theory remains weakly coupled throughout the evolution. This is the case referred to as ``stealth scalarisation'' in previous works \cite{East:2021bqk,Elley:2022ept}.

\begin{figure*}
    \centering
    \includegraphics[width=16cm]{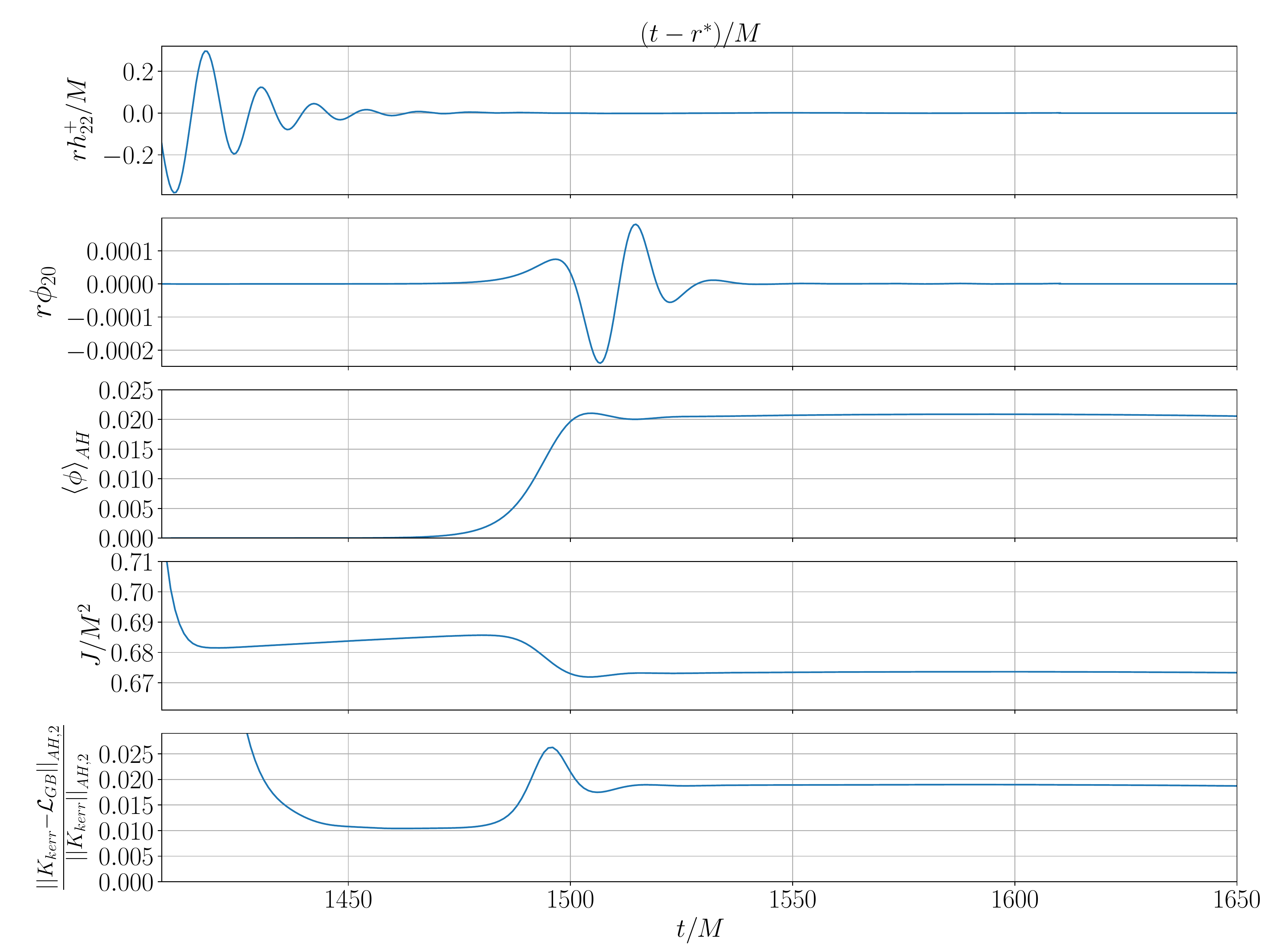}
    \caption{Here we summarise the key results from the post merger phase of spin-induced scalarisation in the EsGB theory with exponential quadratic coupling. 
    We see that the spin of the remnant following merger generates a tachyonic mass, by which the scalar field acquires a non-trivial configuration. This happens late in the ringdown of the tensor modes. It is accompanied by a burst of radiation in the scalar mode $(2,0)$, which coincides with the extraction of spin from the merger and the highest deviation of the Gauss-Bonnet curvature with respect to the Kretschmann scalar of a Kerr Black Hole (note that the initial deviation in this quantity is due to the merger state being far from Kerr).
    \emph{From top to bottom}: $(2,2)$ mode of the gravitational strain, $(0,2)$ scalar mode in retarded time, average value of the scalar field at the AH, evolution of the spin and $L^2$ norm of the Gauss-Bonnet curvature relative to the Kerr Kretschmann scalar.
    }
    \label{fig:exp_phi}
\end{figure*}

Here we used again the same set-up as in the quadratic coupling case with $\omega=200$ and $\lambda^{\text{GB}}/M^2=-20$ \footnote{The motivation for this large value of $\omega$ is that it leads to $\omega\phi^2\sim1$ at the apparent horizon when the black hole has scalarised, which is where we expect the theory to start to break down. A further study of the impact of different values of this parameter has been carried out in \cite{Doneva:2023oww}.} . The results are depicted in Figures \ref{fig:cloud} and \ref{fig:exp_phi}. 

Figure \ref{fig:exp_phi} shows that the single BH that results from the merger scalarises after the ringdown of the tensor modes, which coincides with the a burst of radiation in the scalar mode $(2,0)$. At this point we observe the largest deviation of the Gauss-Bonnet curvature scalar with respect to the Kretschmann scalar of a Kerr BH (with the same angular momentum and mass as measured from the quasilocal quantites at the AH). This scalarisation process extracts spin from the remnant  BH, which decreases its intrinsic spin before settling into an equilibrium state. The end result is a stable hairy BH, but an observation of the effect would rely on the scalar mode being detectable as a secondary signal, since the tensor modes are emitted during a period in which the theory cannot be distinguished from GR and thus are unaffected -- at least to the precision to which we are able to measure the quasinormal modes (QNMs) here. This is consistent with the behaviour observed for the scalarisation of isolated Kerr BHs in \cite{East:2021bqk}.

\begin{figure}
    \centering
    \includegraphics[width=8cm,trim={0.9cm 3.5cm 0.2cm 3cm},clip]{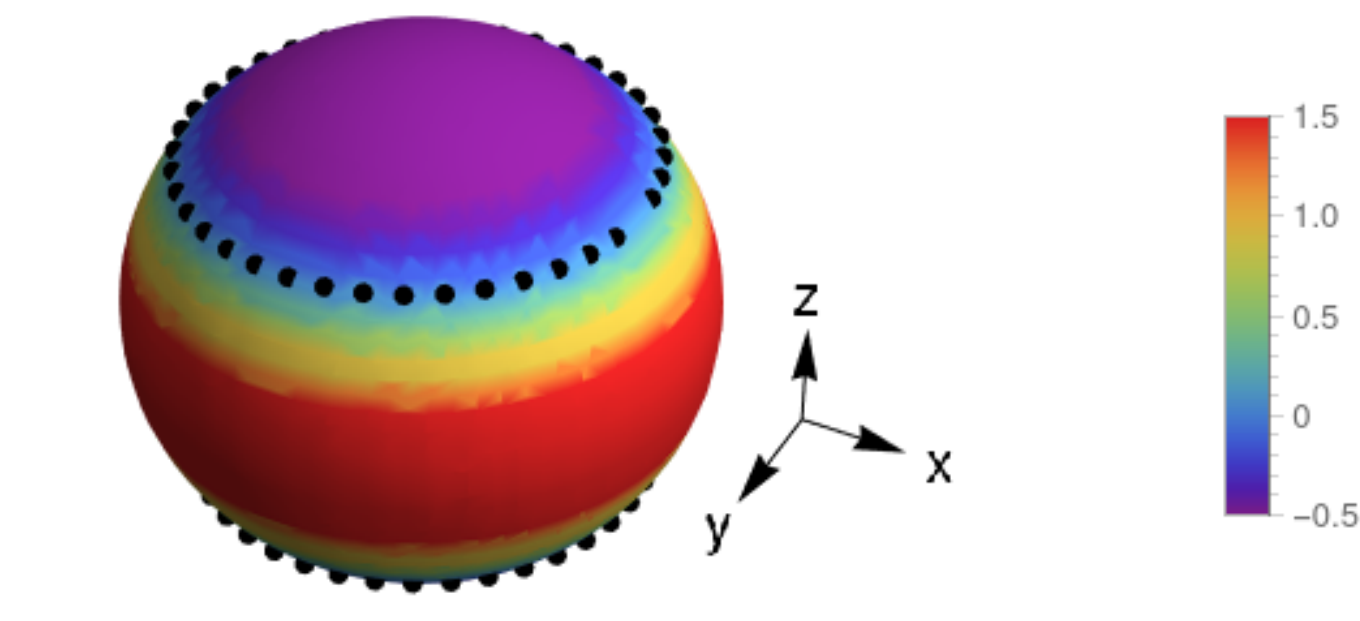}
    \caption{The effective mass of the scalar field is proportional to the Gauss-Bonnet curvature ${\mathcal L}_{\text{GB}}$ (see \cite{Elley:2022ept} for a discussion). Hence, a change of sign (which occurs for high spins) gives rise to the spin-induced scalarisation in the Type II couplings. Here we show the value of the Gauss-Bonnet curvature around the AH of the final BH of our BBH merger simulation in the exponential quadratic EsGB theory at $t=1530M$ (when the value of the scalar field has already settled down). The dotted points denote the region where ${\mathcal L}_{\text{GB}}=0$. We observe that the negative regions coincide with those where the scalar field has a non-trivial contribution as from Figure \ref{fig:cloud}.
    }
    \label{fig:sphere}
\end{figure}

\begin{figure*}
    \centering
    \includegraphics[width=0.9\textwidth,trim={3.2cm 9.5cm 1cm 9.5cm},clip]{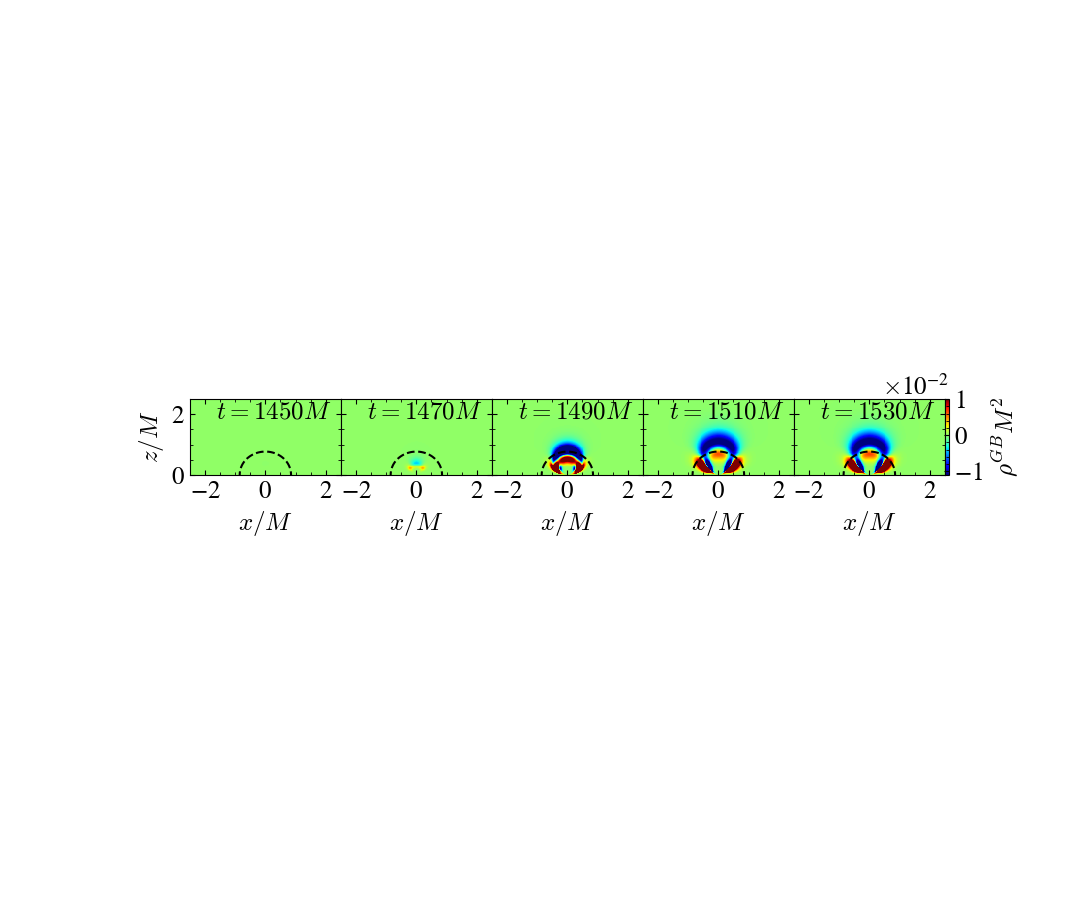}
    \caption{Spatial configuration of the Gauss-Bonnet contribution to the effective energy density $\rho^{\text{GB}}$ throughout the ringdown of an equal mass BBH merger in the EsGB theory with an exponential quadratic coupling. The profile is shown in a section orthogonal to the rotation plane. We observe that in some regions around the AH $\rho^{\text{GB}}$  becomes negative, which explains the violation of the Null Curvature Condition (NCC) in this theory of gravity.}
    \label{fig:rhoGB}
\end{figure*}

In Figure \ref{fig:cloud} we see that the scalar field is localised around the poles of the AH, which is consistent with  the Gauss-Bonnet curvature acting as the source term for the scalar, as depicted in Figure \ref{fig:sphere}. We also show in Figure \ref{fig:rhoGB} the contribution of the Gauss-Bonnet term to the energy density, namely $\rho^{\text{GB}}$, from which we can see that it gives rise to a negative contribution to the total energy density in some regions around the AH. This permits a violation of the Null Curvature Condition (NCC) in this modified theory of gravity \cite{R:2022hlf}.

We note that for the chosen coupling function, the absolute value of the overall coupling constant $\lambda^{\text{GB}}$ could be increased beyond the value that we have used in order to increase the speed of growth of the scalar hair. This would push the field growth closer to the ringdown, potentially having an impact on the emission of tensor modes in this phase. However, in order to avoid breaking the hyperbolicity of the equations and weak coupling conditions during the evolution, the value of $\omega$ in the coupling function \eqref{eq:exp_quad_coupling} must also be increased in proportion to $\lambda^\text{GB}$ (i.e., keeping $\lambda^\text{GB}/(M^2\omega^{1/2})$ constant). As a result, the final maximum scalar field value will be smaller, and whilst the $\rho^{\text{GB}}$ values at maximum should remain the same, as in Fig. \ref{fig:rhoGB}, the usual kinetic contribution of the field to the energy density, as shown in Fig. \ref{fig:cloud}, will be reduced.  
Due to this trade off, there should exist optimum values of $\omega$ and $\lambda^{\text{GB}}$ that maximise the overlap of the growth of  the scalar hair and the ringdown of the BH, thus resulting in the largest modification of the tensor QNMs. We leave a full analysis of this to future work.

\section{Discussion}
\label{sec:discussion}
In this article we have developed a modified CCZ4 formulation of the Einstein equations in $d+1$ spacetime dimensions for GR plus a Gauss-Bonnet term, as well as for the most general parity-invariant scalar-tensor theory of gravity up to four derivatives ($4\partial \text{ST}$). We used a modified version of the CCZ4 formulation of the Einstein equations based on \cite{Kovacs:2020pns,Kovacs:2020ywu}, together with a modification of the puncture gauge extensively used in numerical relativity. We demonstrated well-posedness for both theories and provided full expressions for their implementation in numerical relativity. 

In the $4\partial \text{ST}$ theory, we studied both Type I and Type II couplings, including the so-called ``stealth-scalarisation'' effect where the scalar cloud arises due to the spin of the remnant after merger. As in previous studies using alternative gauges we found that the scalarisation generically occurs too late after merger to impact on the tensor waveform. Too large values of the coupling -- to accelerate the growth of the scalar hair -- result in a breakdown of the theory as it is pushed into the strongly coupled regime in which well-posedness is no longer assured. However, we point out that this can be compensated in our chosen coupling function by tuning the values of the higher order interactions. Without such tuning, observation will rely on detection of the scalar GWs that we show accompany the scalarisation post merger.

Since the formalism is still in its infancy, it is likely that the methods -- in particular the choices for the functions $a(x)$ and $b(x)$ and the damping parameters -- can be optimised further. We found that we needed to be careful in tracking the constraint violations and tuning the parameters in order to get sensible results, especially in cases of BHs with high intrinsic spins. It will be interesting to test whether the puncture gauge provides greater robustness in studies of unequal mergers, which have been found to be challenging in the modified GHC gauge (see \cite{Corman:2022xqg}). 

This work provides a basis on which further studies can be undertaken using codes that employ a moving-punctures approach to managing singularities in the numerical domain. In particular, it seems likely that one can extend our well-posedness results in singularity avoiding coordinates to the general Horndeski theory \cite{Kovacs:2020pns,Kovacs:2020ywu}. 
This article, and our previous work \cite{AresteSalo:2022hua} (see also \cite{Cayuso:2023aht}), allow puncture based codes to compute theoretical gravitational waveforms in certain alternative theories of gravity of interest and compare them to observations, which is of strong interest to the GW community.
With the full range of second order theories opened up for numerical study, a key question that should be answered is where best to focus the research effort, given the large parameter space, and the high computational cost of the simulations.

\subsection*{Acknowledgements}
We would like to thank Aron Kov\'acs and Harvey Reall for numerous discussions about well-posedness. We also want to thank Tiago Fran\c{c}a and Josu Aurrekoetxea for their support in some of the technical aspects of the paper and Ulrich Sperhake and Miren Radia for help with the initial GR binary parameters. We also thank Miguel Bezares, Daniela Doneva, Klaas De Kinder, Tjonnie Li and Shunhui Yao for helpful conversations and correspondence related to our previous paper.
We thank the entire \texttt{GRChombo}\footnote{\texttt{www.grchombo.org}} collaboration for their support and code development work. PF would like to thank the Enrico Fermi Institute and the Department of Physics of the University of Chicago for hospitality during the final stages of this work. Some of this work was presented at the ``Kings and Queens of Gravity'' workshop in London in 2023 and at the ``Journe\'es Relativistes de Tours 2023'' at the University of Tours (France); we would like to thank the organizers for inviting us and the participants for stimulating discussions. PF and KC are supported by an STFC Research Grant ST/X000931/1 (Astronomy at Queen Mary 2023-2026). PF is supported by a Royal Society University Research Fellowship  No. URF\textbackslash R\textbackslash 201026, and No. RF\textbackslash ERE\textbackslash 210291. KC is supported by an STFC Ernest Rutherford fellowship, project reference ST/V003240/1. LAS is supported by a QMUL Ph.D. scholarship. The simulations presented used the ARCHER2 UK National Supercomputing Service\footnote{\texttt{https://www.archer2.ac.uk}} under the EPSRC HPC project no. E775, the CSD3 cluster in Cambridge under Projects No. DP128. The Cambridge Service for Data Driven Discovery (CSD3), partially operated by the University of Cambridge Research Computing on behalf of the STFC DiRAC HPC Facility. The DiRAC component of CSD3 is funded by BEIS capital via STFC capital Grants No. ST/P002307/1 and No. ST/ R002452/1 and STFC operations Grant No. ST/R00689X/1. DiRAC is part of the National e-Infrastructure.\footnote{\texttt{www.dirac.ac.uk}} The authors gratefully acknowledge the Gauss Centre for Supercomputing e.V.\footnote{\texttt{www.gauss-centre.eu}} for providing computing time on the GCS Supercomputer SuperMUC-NG at Leibniz Supercomputing Centre.\footnote{\texttt{www.lrz.de}} Calculations were performed using the Sulis Tier 2 HPC platform hosted by the Scientific Computing Research Technology Platform at the University of Warwick. Sulis is funded by EPSRC Grant EP/T022108/1 and the HPC Midlands+ consortium. This research also
utilised Queen Mary’s Apocrita HPC facility, supported
by QMUL Research-IT \cite{apocrita}. For some computations we have also used the Young Tier 2 HPC cluster at UCL; we are grateful to the UK Materials
and Molecular Modelling Hub for computational resources, which is partially funded by
EPSRC (EP/P020194/1 and EP/T022213/1). For the purpose of Open Access, the author has applied a
CC BY public copyright licence to any Author Accepted
Manuscript version arising from this submission.

\appendix

\section{Eigenvectors of the Einstein-scalar-field principal part}
\label{app:eigenvectors}

In this appendix we display the expression of the eigenvectors of the Einstein-scalar-field principal part in the modified CCZ4 gauge in $d+1$ spacetime dimensions in Tables \ref{tab:phys}, \ref{tab:viol} and \ref{tab:pure}, corresponding respectively to the physical, ``gauge-condition violating'' and ``pure-gauge'' categories:
\begin{table}[H]
    \centering
    \begin{tabular}{c|c|c|c c}
        $\hat{\tilde{\gamma}}_{ij}$ &  $\hat{\tilde{A}}_{ij}$ & $\hat{\phi}$ & $\hat{K}_{\phi}$  \\
        \cline{1-4}
        0 & 0 & $\mp1$ & 1 \\
        $\mp2e_A^ie_B^j$ & $e_A^ie_B^j$ & 0 & 0  \\
        \vspace{-0.3cm}
        & & & & $\forall A\neq B$ \\
            \vspace{-0.2cm}
        $\pm2(e_A^ie_A^j-e_B^ie_B^j)$ & $-e_A^ie_A^j+e_B^ie_B^j$ & 0 & 0 
    \end{tabular}
    \caption{Physical eigenvectors.}
    \label{tab:phys}
    \end{table}
    \begin{table}[H]
    \centering
    \begin{tabular}{c|c|c|c}
        $\hat{\tilde{\gamma}}_{ij}$ & $\hat{\chi}$ & $\hat{\beta}^i$ & $\hat{\alpha}$  \\
        $\hat{\tilde{A}}_{ij}$ & $\hat{K}$ & $\hat{\hat{\Gamma}}^i$  \\
        \hline
        $-\frac{\chi^2}{d-1}e_A^ie_B^i\delta^{AB}$ & $-\frac{\chi^2}{d-1}$ &  $\pm\frac{d\sqrt{\chi}}{2(d-1)\sqrt{1+a(x)}}\xi_i$ & 0 \\
        0 & 0 & $\xi_i$\\
        \hline
        $\chi^2e_A^i\xi^j$ & 0 & $\pm\sqrt{\frac{d\chi}{2(d-1)(1+a(x))}}e^A_i$ & 0\\
        0 & 0 & $e^A_i$ \\
        \hline
         $-\frac{\chi^2}{d-1}e_A^ie_B^j\delta^{AB}$ & $-\frac{\chi^2}{d-1}$ & $\pm\frac{d\sqrt{2}\left(1+a(x)(1-2\alpha\chi)\right)}{4(d-1)\sqrt{\alpha(1+a(x))}}\xi_i$ & $\frac{d(-1+2\alpha\chi)}{2(d-1)\alpha}$ \\
         $\mp\sigma e_A^ie_B^j\delta^{AB}$  & $\pm \tfrac{d\sigma}{\chi}$  & $\xi_i$ 
    \end{tabular}
    \caption{``Gauge-condition violating'' eigenvectors, where $\sigma=\tfrac{\chi\sqrt{2(1+a(x))}(1-2\alpha\chi)}{4(d-1)\alpha^{3/2}}$.}
    \label{tab:viol}
    \end{table}
    \begin{table*}[ht]
    \centering
    \begin{tabular}{c|c|c}
        $\hat{\tilde{\gamma}}_{ij}$ & $\hat{\chi}$ & $\hat{\beta}^i$ \\ 
        $\hat{\tilde{A}}_{ij}$ &$\hat{K}$ & $\hat{\hat{\Gamma}}^i$ \\
        $\hat{\Theta}$ & $\hat{\alpha}$ \\
        \hline
       $\frac{\chi(1+b(x))}{b(x)}\left(-\frac{d}{2(d-1)(1+a(x))\alpha^2}+\chi \right)e_A^i\xi_j$ & 0 & $\pm\frac{d\sqrt{1+b(x)}}{2(d-1)\alpha(1+a(x))}e^A_i$ \\ $\pm\frac{\chi\sqrt{1+b(x)}}{2b(x)\alpha^2}\left(\frac{d}{2(d-1)}\frac{1+b(x)}{1+a(x)}-\alpha^2\chi\right)e_A^i\xi^j$ & 0 & $e^A_i$ \\
       0 & 0 \\
       \hline
        $-\frac{\chi^2(1+b(x))}{d-2+2(d-1)b(x)}\left(\frac{d\left(2-\alpha(1+a(x))\right)}{4\alpha^2\chi a(x)b(x)}+d-1\right)e_A^ie_B^j\delta^{AB}$ & $\chi\frac{1+b(x)}{2a(x)\alpha^2}\frac{4\left(d-(d-2)a(x)\alpha^2\chi\right)-\frac{d^2}{d-1}\alpha\frac{1+a(x)}{1+b(x)}}{(d-2)(d-2+2(d-1)b(x))}$ & $\frac{d}{2(d-1)a(x)\alpha}\xi_i$  \\
         $\pm\chi^2\frac{\sqrt{1+b(x)}}{d-2+2(d-1)b(x)}\left(\frac{d\left(2-\alpha(1+a(x))\right)}{4(d-1)\alpha^2\chi a(x)b(x)}+1 \right)e_A^ie_B^j\delta^{AB}$  &  $\pm\frac{d\sqrt{1+b(x)}}{4a(x)\alpha^2}\frac{4(d-(d-2)a(x)\alpha^2\chi)-\frac{d^2}{d-1}\alpha\frac{1+a(x)}{1+b(x)}}{(d-2)(d-2+2(d-1)b(x))}$ & $\xi_i$ \\
       $\pm\frac{d\sqrt{1+b(x)}}{4a(x)\alpha^2}\frac{2(d-(d-2)a(x)\alpha^2\chi)-\alpha(2-(d-2)b(x))\frac{1+a(x)}{1+b(x)}}{(d-2)(d-2+2(d-1)b(x))}$ & 0 \\
        \hline
        $\pm\frac{\chi\sqrt{1+b(x)}}{a(x)b(x)\alpha}\frac{2(1+a(x)(1+b(x)))-\alpha(1+a(x))^2}{d-2+2(d-1)b(x)}e_A^ie_B^j\delta^{AB}$ & $\mp\frac{\sqrt{1+b(x)}\chi}{a(x)\alpha}\frac{2(2(d-1)+da(x))-d\alpha\frac{(1+a(x))^2}{1+b(x)}}{(d-2)(d-2+2(d-1)b(x))}$ & $\xi_i$\\ $-\chi\frac{1+a(x)\left(1+2\frac{d-1}{d}b(x)(1+b(x))\right)-\frac{\alpha}{2}(1+a(x))^2}{a(x)b(x)(d-2+2(d-1)b(x))\alpha}e_A^ie_B^j\delta^{AB}$ & $\frac{4(d-1)\left(a(x)(1+b(x))-\frac{d}{d-2}(1+a(x))\right)+d^2\alpha\frac{(1+a(x)^2}{1+b(x)}}{2\alpha a(x)(d-2+2(d-1)b(x))}$ & 0 \\
        $\frac{d-1}{2a(x)\alpha}\frac{\frac{(1+a(x))^2}{1+b(x)}(2-(d-2)b(x))\alpha-2\left(d+a(x)(2-(d-2)b(x))\right)}{(d-2)(d-2+2(d-1)b(x))}$ & $\mp\frac{1+a(x)}{\alpha\sqrt{1+b(x)}}$
    \end{tabular}
    \caption{``Pure-gauge'' eigenvectors.}
    \label{tab:pure}
\end{table*}

\section{Propagation of the constraints}\label{app:constraints}
Below we consider the propagation of the constraints in the modified CCZ4 formulation in our gauge in $d+1$ spacetime dimensions. Let the Hamiltonian and momentum constraints be denoted by  ${\mathcal H}$ and ${\mathcal M}_i$ respectively. Then, we find that the constraints obey the following evolution equations: 
\begin{subequations}\label{eq_constraints}
\begin{eqnarray}
\partial_{\perp}{\mathcal H}&=&\left(\tfrac{2+b(x)}{1+b(x)}\right)\alpha\,K\,{\mathcal H}-\tfrac{2}{\alpha}D^i\big(\alpha^2{\mathcal M}_i\big)\nonumber\\
&&+4\alpha\big(K\gamma^{ij}-K^{ij}\big)\big(D_iZ_j-\Theta\,K_{ij}\big)\nonumber\\
&&-\tfrac{2(d-1)}{1+b(x)}\kappa_1\alpha\big[1+\tfrac{\kappa_2}{2}(2+b(x))\big]K\Theta\,,\\
\partial_{\perp}{\mathcal M}_i&=&\alpha\,K\,{\mathcal M}_i-\tfrac{1}{2\alpha}D_i(\alpha^2{\mathcal H})+\tfrac{b(x)}{2(1+b(x))}D_i(\alpha{\mathcal H})\nonumber\\
&&-2D^j\big[\alpha\big(D^kZ_k\gamma_{ij}-D_{(i}Z_{j)}+\Theta(K_{ij}-K\gamma_{ij})\big)\big]\nonumber\\
&&+\tfrac{d-1}{1+b(x)}\kappa_1\big[1+\tfrac{\kappa_2}{2}(2+b(x))\big]D_i(\alpha\Theta)\,,\\
\partial_{\perp}\Theta &=& \tfrac{\alpha}{2(1+b(x))}{\mathcal H}+\alpha(D_iZ^i-K\Theta)-
Z^iD_i\alpha\nonumber\\&&-\tfrac{\alpha\kappa_1}{1+b(x)}\big(\tfrac{d+1+2b(x)}{2+b(x)}+\tfrac{d-1}{2}\kappa_2\big)\Theta\,, \\
\partial_{\perp}Z_i &=& -\Theta D_i\alpha+\tfrac{\alpha}{1+b(x)}\big(D_i\Theta+{\mathcal M}_i\nonumber\\&&\hspace{1.3cm}-Z_jK_i^{~j}(2+b(x))-\kappa_1Z_i\big)\,.
\end{eqnarray}
\end{subequations}

We consider the principal part of \eqref{eq_constraints} and decompose it into its scalar and vector sectors respectively, as in Section \ref{sec:hyp}. The scalar sector is given by
\begin{subequations}
  \begin{eqnarray}
    \check{\xi}_0\hat{\mathcal H}&=&-2\alpha\,\hat{\mathcal M}_{\perp}\,,\\
    \check{\xi}_0\hat{\mathcal M}_{\perp}&=&-\tfrac{\alpha}{2(1+b(x))}\hat{\mathcal H}\,,\\
    \check{\xi}_0\hat{Z}_{\perp}&=&\tfrac{\alpha}{1+b(x)}(\hat{\mathcal M}_{\perp}+\hat{\Theta})\,,\\
    \check{\xi}_0\Theta&=&\tfrac{\alpha}{2(1+b(x))}\hat{\mathcal H}+\alpha\hat{Z}_{\perp}\,.
\end{eqnarray}  
\end{subequations}
The respective eigenvalues are $\xi_0=\beta^{\perp}\pm\tfrac{\alpha}{\sqrt{1+b(x)}}$, each of them with multiplicity $2$ but there is no degeneracy in the corresponding eigenvectors. The vector sector of the principal part is given by
\begin{subequations}
    \begin{eqnarray}
        \check{\xi}_0\hat{\mathcal M}_A&=&\alpha\hat{Z}_A\,,\\
        \check{\xi}_0\hat{Z}_A&=&\tfrac{\alpha}{1+b(x)}{\mathcal M}_A\,,
    \end{eqnarray}
\end{subequations}
with eigenvalues $\xi_0=\beta^{\perp}\pm\tfrac{\alpha}{\sqrt{1+b(x)}}$. Therefore, the system is strongly hyperbolic and, thus, it follows that if the constraints are satisfied initially then they continue to hold throughout the evolution.

\section{Equations of motion of the Einstein-Gauss-Bonnet Theory in modified CCZ4}
\label{app:eomegb}

The tensor \eqref{eq:def_H_GB} that appears on the r.h.s. of the equations of motion that result from varying the action \eqref{eq:action_Lovelock} with respect to the metric plays the role of an effective stress-energy tensor. Therefore, its $d+1$ decomposition gives\footnote{The signs have been chosen so that the quantities in the $d+1$ decomposition of $\mathcal{H}_{\mu\nu}$ enter the equations of motion with the same signs as the analogous quantities for a standard stress-energy tensor.}
\begin{subequations}
    \begin{eqnarray}
        \kappa\,\rho =&n^\mu\,n^\nu\,\mathcal{H}_{\mu\nu}\,,\\
        \kappa\,J_i =&-n^{\mu}\gamma_i^{\phantom{i}\nu}{\mathcal H}_{\mu\nu}\,,\\
        \kappa\,S_{ij} = &\gamma_i^{\phantom{i}\mu}\gamma_j^{\phantom{j}\nu}\mathcal{H}_{\mu\nu}\,,
    \end{eqnarray}
\end{subequations}
where 
\begin{subequations}\label{egbcomp}
\begin{eqnarray}
\kappa\,\rho&=&-\tfrac{\lambda^{\text{GB}}}{2}\big(M^2-4M_{ij}M^{ij}+M_{ijkl}M^{ijkl} \big), \\
\kappa\,J_i&=&-2\lambda^{\text{GB}}\big(MN_i-2M_i^{~j}N_j+2M^{jk}N_{ijk} \nonumber\\ && \hspace{1cm}-M_i^{~ljk}N_{jkl} \big)\,, \\
\kappa\,S_{ij}&=& 2\lambda^{\text{GB}}\Big[4M^k_{(i}F_{j)k}+2M_{i~j}^{~k~l}F_{kl}-MF_{ij}-2M_{ij}F\nonumber
        \\ &&\hspace{-1.2cm} + 2N_iN_j-4N^kN_{k(ij)} - N_{kli}N^{kl}_{~~j}-2N_{ikl}N_j^{~kl}\nonumber
        \\&&\hspace{-1.2cm}+ MM_{ij}-2\big(M_{ik}M^k_{~j}+M^{kl}M_{ikjl}\big) + M_{iklm}M_j^{~klm}\nonumber
        \\&&\hspace{-1.2cm}+\gamma_{ij}\Big(MF-2M^{kl}F_{kl}+N_{klm}N^{klm}-2N_kN^k\nonumber
        \\&&\hspace{-0.6cm}-\tfrac{1}{4}\big(M^2-4M_{kl}M^{kl}+M_{klmn}M^{klmn}\big) \Big)\Big]\,,
\end{eqnarray}
\end{subequations}
with
\begin{subequations}
\begin{eqnarray}
M_{ijkl}&=&R_{ijkl}+K_{ik}K_{jl}-K_{il}K_{jk}\,, \\
N_{ijk}&=&D_iK_{jk}-D_{j}K_{ik}\,, \\
F_{ij}&=&{\mathcal L}_nK_{ij}+\frac{D_iD_j\alpha}{\alpha}+K_{ik}K_j^{~k}\,,
\end{eqnarray}
\end{subequations}
where ${\mathcal L}_n$ denotes the Lie derivative along $n^{\mu}$, and $M_{ij}=\gamma^{kl}M_{ikjl}$, $M=\gamma^{ij}M_{ij}$ and $N_i=\gamma^{jk}N_{jik}$, as they are also defined in Eq. \eqref{MNeq}.

The $d+1$ equations are obtained by inserting the above quantities in Eqs. \eqref{eqsccz4} except for $\tilde{A}_{ij}$ and $K$, whose evolution equations are given by the following coupled system,
\begin{eqnarray}\label{mat_egb}
\begin{pmatrix} X^{kl}_{ij} & Y_{ij} \\ X^{kl}_K & Y_K \end{pmatrix}
\begin{pmatrix} \partial_t \tilde{A}_{kl} \\ \partial_tK  \end{pmatrix}=
\begin{pmatrix} Z_{ij}^{\tilde{A}} \\ Z^K  \end{pmatrix}\,,
\end{eqnarray}
where the elements of the matrix are
\begin{subequations}
\begin{eqnarray}
X_{ij}^{kl}&=&\gamma_i^{~k}\gamma_j^{~l}+2\lambda^{\text{GB}}\big[M\gamma_i^{~k}\gamma_j^{~l}+\tfrac{6}{d}\gamma_{ij}M^{kl} \nonumber\\&&\hspace{1.5cm}-2\big(M_{i~j}^{~k~l}+2M_{(i}^{~k}\gamma_{j)}^{~l} \big) \big]\,, \\
X_K^{kl}&=&-\tfrac{4(d-3)}{(d-1)\chi}\lambda^{\text{GB}}M^{kl}\,, \\
Y_{ij}&=&\tfrac{4(d-3)}{d}\lambda^{\text{GB}}\chi\big(M_{ij}-\tfrac{1}{d}\gamma_{ij}M\big)\,, \\
Y_K &=& 1+\tfrac{2(d-2)(d-3)}{d(d-1)}\lambda^{\text{GB}}M\,,
\end{eqnarray}
\end{subequations}
whereas the r.h.s. terms are
\begin{subequations}
\begin{eqnarray}
Z_{ij}^{\tilde{A}}&=&{\mathcal L}_{\beta}\tilde{A}_{ij}-2\alpha\tilde{A}_{il}\tilde{A}^l_{~j}\nonumber\\&&+ \chi\big[\alpha\big(R_{ij} + 2D_{(i}Z_{j)} -\kappa\,\bar{S}_{ij}\big)-D_iD_j\alpha\big]^{\text{TF}}  \nonumber\\&&+\alpha\tilde{A}_{ij}(K-2\Theta) - \tfrac{2}{d}\partial_k\beta^k\tilde{A}_{ij}\,, \\
Z^K&=&\beta^i\partial_iK-D^iD_i\alpha +\alpha\left[R+2\,D_iZ^i +K(K-2\Theta)\right] \nonumber\\
&&-d\,\kappa_1(1+\kappa_2)\,\alpha\,\Theta+\tfrac{\kappa\,\alpha}{d-1}\big(\bar{S}-d\rho\big)\nonumber\\
&&\textstyle-\frac{d\,\alpha\,b(x)}{2(d-1)(1+b(x))}\Big[R-\tilde{A}_{ij}\tilde{A}^{ij}+\frac{d-1}{d}K^2 \nonumber\\
&&\textstyle\hspace{0.6cm}-(d-1)\kappa_1(2+\kappa_2)\,\Theta -2\,\kappa\,\rho\Big]\,,
\end{eqnarray}
\end{subequations}
with $\bar{S}_{ij}$ and $\bar{S}$, which are obtained by subtracting the time derivatives of $\tilde{A}_{ij}$ and $K$ from $S_{ij}$ (which are in turn computed from Eq. \eqref{mat_egb}), given by
\begin{subequations}
\begin{eqnarray}
\kappa\,\bar{S}_{ij}^{\text{TF}}&=&\lambda^{\text{GB}}\big\{\tfrac{8}{\chi}M_{(i}^{~k}\hat{\mathcal O}_{j)k}+\tfrac{4}{\chi}M_{i~j}^{~k~l}\hat{\mathcal O}_{kl}-4\hat{\mathcal O} M_{ij}\nonumber\\
&&\hspace{0.6cm}-\tfrac{2}{\chi}M\hat{\mathcal O}_{ij}+\tfrac{12}{d}\tfrac{\tilde{\gamma}_{ij}}{\chi}\big(\tfrac{M\hat{\mathcal O}}{2}-\tfrac{M^{kl}\hat{\mathcal O}_{kl}}{\chi}\nonumber\\
&&\hspace{0.6cm}+N_kN^k-\tfrac{N_{klm}N^{klm}}{2} \big)\big\}-\tfrac{4\,\kappa}{d}\tfrac{\tilde{\gamma}_{ij}}{\chi}\rho\nonumber\\
&&\hspace{-0.5cm}-2\lambda^{\text{GB}}\big[M_{iklm}M_ j^{~klm}-2(M_{ik}M^k_{~j}+M^{kl}M_{ikjl})\nonumber\\&&\hspace{0.7cm}+MM_{ij}+2\big(N_iN_j-2N^kN_{k(ij)} \nonumber\\&&\hspace{1.8cm}-\tfrac{1}{2}N_{kli}N^{kl}_{~~j}-N_{ikl}N_j^{~kl}\big)\big] \,, \\
\kappa\,\bar{S}&=& 2\lambda^{\text{GB}}(d-3)\big(M\hat{\mathcal O} - \tfrac{2}{\chi}M^{ij}\hat{\mathcal O}_{ij}+2N_iN^i\nonumber\\&&\hspace{0.8cm}-N_{ijk}N^{ijk} \big)+4\,\kappa\,\rho\,,
\end{eqnarray}
\end{subequations}
where $\hat{\mathcal O}_{ij}=\tfrac{1}{\alpha}{\mathcal L}_{\beta}\tilde{A}_{ij}-\tilde{A}_{ik}\tilde{A}_j^{~k}+\frac{2}{d}K\tilde{A}_{ij}+\tilde{\gamma}_{ij}\left(\frac{K}{d}\right)^2-\tfrac{\chi}{\alpha} D_iD_j\alpha-\frac{2}{d}\tilde{A}_{ij}\tfrac{\partial_k\beta^k}{\alpha}+\frac{\tilde{\gamma}_{ij}}{d}\tfrac{1}{\alpha}{\mathcal L}_{\beta}K$ and $\hat{\mathcal O}=\tfrac{1}{\alpha}{\mathcal L}_{\beta}K+\tilde{A}_{ij}\tilde{A}^{ij}+\frac{K^2}{d}-\tfrac{1}{\alpha}D_kD^k\alpha$\,.

\section{Equations of motion of the 4$\partial$ST in modified CCZ4}
\label{app:eomesgb}
In this appendix we write down the equations of motion of the theory, eqs. \eqref{eq:eom_metric}--\eqref{eq:eom_sf}, in the 3+1 form as we have implemented in our code.

We start writing the $3+1$ decomposition of $T_{\mu\nu}^X$ appearing in Eq. \eqref{eq:eom_metric},
\begin{subequations}
\begin{eqnarray}
    \rho^X&=&\tfrac{1}{8}\big(K_{\phi}^2-(\partial\phi)^2\big)\big(3K_{\phi}^2+(\partial\phi)^2\big)\,,\\
    J_i^X&=&\tfrac{1}{2}K_{\phi}\partial_i\phi\big(K_{\phi}^2-(\partial\phi)^2\big)\,,\\
    S_{ij}^X&=&\tfrac{1}{2}\big(K_{\phi}^2-(\partial\phi)^2\big)\big[(D_i\phi) D_j\phi\nonumber\\
    &&\hspace{0.5cm}+\tfrac{1}{4}\gamma_{ij}\big(K_{\phi}^2-(\partial\phi)^2\big)\big]\,,
\end{eqnarray}
\end{subequations}
and from ${\mathcal H}_{\mu\nu}$ as well,
\begin{subequations}\label{edgbcomp}
\begin{eqnarray}
\rho^{\text{GB}}&=&\tfrac{\Omega M}{2} - M_{kl}\Omega^{kl}, \, \\
J^{\text{GB}}_i&=&\tfrac{\Omega_iM}{2}-M_{ij}\Omega^j - 2\big(\Omega^j_ {~[i}N_{j]}-\Omega^{jk}D_{[i}K_{j]k}\big),\, \\
S^{\text{GB}}_{ij}&=&2\gamma^k_{~(i}\Omega_{j)}^{\text{TF},l}\big({\mathcal L}_nA_{kl}+\tfrac{1}{\alpha}(D_kD_l\alpha)^{\text{TF}}+A_{km}A^m_{~l}\big)\nonumber\\
&&\hspace{-0.3cm}-\Omega_{ij}^{\text{TF}}\big({\mathcal L}_nK+\tfrac{1}{\alpha}D^kD_k\alpha-3A_{kl}A^{kl}-\tfrac{K^2}{3}\big) \nonumber\\
&&\hspace{-0.3cm}-\tfrac{\Omega}{3}\big({\mathcal L}_nA_{ij}+\tfrac{1}{\alpha}(D_iD_j\alpha)^{\text{TF}}+A_{im}A^m_{~j} \big)\nonumber\\
&&\hspace{-0.3cm}-\Omega_{nn}M_{ij}+N_{(i}\Omega_{j)}-2\epsilon_{(i}^{~kl}B_{j)k}\Omega_l\nonumber\\
&&\hspace{-0.3cm}+\gamma_{ij}\big[\rho^{\text{GB}}-N^k\Omega_k+\tfrac{M}{6}\big(\Omega_{nn}+\tfrac{\Omega}{3}\big)-\tfrac{1}{3}\Omega^{\text{TF},kl}M_{kl}\nonumber\\
&&\hspace{0.3cm}-\Omega^{\text{TF},kl}\big({\mathcal L}_nA_{kl}+\tfrac{1}{\alpha}(D_kD_l\alpha)^{\text{TF}}+A_{km}A^m_{~l}\big)\nonumber\\
&&\hspace{0.3cm}+\tfrac{2\Omega}{9}\big({\mathcal L}_nK+\tfrac{D^kD_k\alpha}{\alpha}-\tfrac{3}{2}A_{kl}A^{kl}-\tfrac{K^2}{3}\big)\big]\,,
\end{eqnarray}
\end{subequations}
with
\begin{subequations}
\begin{eqnarray}\label{MNeq}
\hspace{-0.5cm}M_{ij}&=&R_{ij}+\tfrac{1}{\chi}\left(\tfrac{2}{9}\tilde{\gamma}_{ij}K^2+\tfrac{1}{3}K\tilde{A}_{ij}-\tilde{A}_{ik}\tilde{A}_j^{~k} \right)\,, \\
\hspace{-0.5cm}N_i&=&\tilde{D}_j\tilde{A}_i^{~j}-\tfrac{3}{2\chi}\tilde{A}_i^{~j}\partial_j\chi-\tfrac{2}{3}\partial_iK\,, \\
\hspace{-0.5cm}B_{ij}&=&\epsilon_{(i}^{~kl}D_kA_{j)l}\,,\\
\hspace{-0.5cm}\Omega_i&=&f'\big(\partial_iK_{\phi}-\tilde{A}^j_{~i}\partial_j\phi-\tfrac{K}{3}\partial_i\phi \big)+f''K_{\phi}\partial_i\phi\,,\\
\hspace{-0.5cm}\Omega_{ij}&=&f'\left(D_iD_j\phi-K_{\phi}K_{ij}\right)+f''(\partial_i\phi) \partial_j\phi\,,\\
\hspace{-0.5cm}\Omega_{nn}&=&f''K_{\phi}^2-\tfrac{f'}{\alpha}D^k\alpha D_k\phi-\tfrac{f'}{\alpha}\partial_{\perp}K_{\phi}\,,
\end{eqnarray}
\end{subequations}
where $N_i$ is the GR momentum constraint, $B_{ij}$ is the magnetic part of the Weyl tensor and $\Omega_i$, $\Omega_{ij}$ and $\Omega_{nn}$ come from the $3+1$ decomposition of ${\mathcal C}_{\mu\nu}$ in Eq. \eqref{Cmunu}. In addition, we have 
\begin{subequations}
\begin{align}
    M^{\text{TF}}_{ij} &\equiv M_{ij}-\tfrac{1}{3}\gamma_{ij}M\,,\\
    \Omega^{\text{TF}}_{ij} &\equiv \Omega_{ij}-\tfrac{1}{3}\gamma_{ij}\Omega\,,
\end{align}
\end{subequations}
where $M=\gamma^{kl}M_{kl}$ is the GR Hamiltonian constraint and $\Omega=\gamma^{kl}\Omega_{kl}$. 

So, using that 
\begin{subequations}
\begin{eqnarray}
\kappa\,\rho&=&\tfrac{1}{2}\rho^{\phi}+g_2\rho^X+\lambda^{\text{GB}}\rho^{\text{GB}}\,,\\ \kappa\,J_i&=&\tfrac{1}{2}J_i^{\phi}+g_2J_i^X+\lambda^{\text{GB}}J_i^{\text{GB}}\,,\\
\kappa\,\bar{S}_{ij}&=&\tfrac{1}{2}S_{ij}^{\phi}+g_2S_{ij}^X+\lambda^{\text{GB}}\bar{S}_{ij}^{\text{GB}}\,,
\end{eqnarray}
\end{subequations}
where the bar denotes again that the terms depending on the time derivatives of ${\tilde{A}}_{ij}$ and $K$ are substracted since they are taken into account in the matrix on the l.h.s. in Eq. \eqref{mat_esgb}, we can obtain the equations of motion in the $3+1$ form by replacing those quantities in Eqs. \eqref{eqsccz4} and \eqref{eq:scalar_eqs} with $d=3$, except for $K$, $\tilde{A}_{ij}$ and $K_{\phi}$, which satisfy the following system of coupled partial differential equations:
\begin{eqnarray}\label{mat_esgb}
\begin{pmatrix} X^{kl}_{ij} & Y_{ij} & 0 \\ X^{kl}_K & Y_K & 0 \\ X^{kl}_{K_{\phi}} & Y_{K_{\phi}} & I \end{pmatrix}
\begin{pmatrix} \partial_t \tilde{A}_{kl} \\ \partial_tK \\ \partial_tK_{\phi} \end{pmatrix}=
\begin{pmatrix} Z_{ij}^{\tilde{A}} \\ Z^K \\ Z^{K_{\phi}} \end{pmatrix},
\end{eqnarray}
where the elements of the matrix are defined as follows,
\begin{subequations}
\begin{eqnarray}
X_{ij}^{kl}&=& \gamma_i^k\gamma_j^l\big(1-\tfrac{\lambda^{\text{GB}}}{3}\Omega \big)+2\lambda^{\text{GB}}\big(\gamma_{(i}^k\Omega_{j)}^{\text{TF},l}\nonumber\\&& -\tfrac{\gamma_{ij}}{3}\Omega^{\text{TF},kl}-\tfrac{\lambda^{\text{GB}}}{\Sigma} f'^2M^{\text{TF}}_{ij}M^{\text{TF},kl}\big)\,,\\
X_K^{kl}&=&\tfrac{\lambda^{\text{GB}}}{2\chi}\big(\Omega^{\text{TF},kl}-\tfrac{\lambda^{\text{GB}}}{\Sigma}f'^2M\,M^{\text{TF},kl}\big)\,, \\
X_{K_{\phi}}^{kl}&=&\tfrac{\lambda^{\text{GB}}}{2\chi}f'M^{\text{TF},kl}\,, \\
Y_{ij}&=&\tfrac{\lambda^{\text{GB}}}{3}\chi\big(-\Omega^{\text{TF}}_{ij}+\tfrac{\lambda^{\text{GB}}}{\Sigma}f'^2M\,M^{\text{TF}}_{ij} \big)\,, \\
Y_K&=&1+\tfrac{\lambda^{\text{GB}}}{3}\big(-\Omega+\tfrac{\lambda^{\text{GB}}}{4\Sigma}f'^2M^2 \big)\,,\\
Y_{K_{\phi}}&=&-\tfrac{\lambda^{\text{GB}}}{12}f'M\,, \\
I&=&\Sigma\,,
\end{eqnarray}
\end{subequations}
where $\Sigma=1+g_2(3K_{\phi}^2-(\partial\phi)^2)$, while the terms of the r.h.s. are
\begin{subequations}
\begin{eqnarray}
Z_{ij}^{\tilde{A}}&=&\chi\big[-D_iD_j\alpha+\alpha\left(R_{ij} + 2D_{(i}Z_{j)} -\kappa\,\bar{S}_{ij}\right) \big]^{\text{TF}}\nonumber\\
&&+\alpha\left[\tilde{A}_{ij}(K-2\Theta)-2\tilde{A}_{il}\tilde{A}^l_{~j}\right]\nonumber\\
&&+\beta^k\partial_k\tilde{A}_{ij}+2\,\tilde A_{k(i}\partial_{j)}\beta^k-\tfrac{2}{3}\tilde{A}_{ij}(\partial_k\beta^k)\,,\\
Z^K&=&\beta^i\partial_iK-D^iD_i\alpha +\alpha\left[R+2\,D_iZ^i +K(K-2\Theta)\right] \nonumber\\
&&-3\,\kappa_1(1+\kappa_2)\,\alpha\,\Theta+\tfrac{\kappa\,\alpha}{2}(\bar{S}-3\rho)\nonumber\\
&&\textstyle-\frac{3\,\alpha\,b(x)}{4(1+b(x))}\Big[R-\tilde{A}_{ij}\tilde{A}^{ij}+\frac{2}{3}K^2 \nonumber\\
&&\textstyle\hspace{1.5cm}-2\kappa_1(2+\kappa_2)\,\Theta -2\,\kappa\,\rho\Big]\,,\\
Z^{K_{\phi}}&=& \Sigma\big[\beta^i\partial_i K_\phi+\alpha\big(-D^iD_i\phi+K\,K_\phi\big)-(D^i\phi)D_i\alpha\big]\nonumber\\&&+\alpha\,g_2\,Z^{g_2}-\tfrac{\lambda^{\text{GB}}}{4}\,\alpha\,f'\,\bar{\mathcal L}_{\text{GB}}\,,
\end{eqnarray}
\end{subequations}
where
\begin{eqnarray}
    Z^{g_2}&=&2K_{\phi}^2(D^iD_i\phi-KK_{\phi})+2 D^i\phi \left[(D^j\phi) D_iD_j\phi\right.\nonumber
    \\&&\left.-K_{\phi}\big(2D_iK_{\phi}-D^j\phi\tfrac{1}{\chi}\tilde{A}_{ij}-\tfrac{1}{3}KD_i\phi\big)\right]\,,
\end{eqnarray}
with $\bar{\mathcal L}_{\text{GB}}$ also denoting that we are substracting the terms with time derivatives, which are take into account in the elements of the matrix above. Finally the expression of these remaining quantities yields
\begin{subequations}
\begin{eqnarray}
\bar{S}_{ij}^{\text{GB,TF}}&=&-\tfrac{1}{3}\left(\Omega^{TF}_{ij}-\tfrac{\lambda^{\text{GB}}}{\Sigma}f'^2MM^{\text{TF}}_{ij}\right)\nonumber\\
&&\hspace{0.5cm}\times\big[-\tfrac{1}{\alpha}\beta^i\partial_iK
+\tfrac{1}{\alpha}D_iD^i\alpha-\tilde{A}_{kl}\tilde{A}^{kl}-\tfrac{K^2}{3} \big]\nonumber\\
&&-M_{ij}^{TF}\left[\Omega+f''(K_{\phi}^2-(\partial\phi)^2)\right.\nonumber\\
&&\hspace{1.2cm}\left.-\tfrac{g_2}{\Sigma}f'Z^{g_2}-\tfrac{\lambda^{\text{GB}}}{\Sigma}f'^2H \right]\nonumber\\
&&-\tfrac{1}{3}\,\Omega\left[\tfrac{1}{\alpha}D_iD_j\alpha+\tfrac{1}{\chi}\big(\tilde{A}_{im}\tilde{A}^m_{~j}-\hat{\Theta}_{ij}\big) \right]^{\text{TF}} \nonumber\\
&&+2\,\Omega_{(i}^{\text{TF},k}\left[\tfrac{1}{\alpha}D_{j)}D_k\alpha+\tfrac{1}{\chi}\big(\tilde{A}_{j)m}\tilde{A}_k^{~m}-\hat{\Theta}_{j)k}\big) \right] \nonumber\\
&&-\tfrac{2}{3}\,\Omega_{ij}^{\text{TF}}\left(\tfrac{1}{\alpha}D_kD^k\alpha-\tilde{A}_{kl}\tilde{A}^{kl} \right)+\left[N_{(i}\Omega_{j)}\right]^{\text{TF}}\nonumber\\
&&-2\,\left(\tfrac{1}{3}\,\gamma_{ij}\,\Omega^{\text{TF},kl}+\tfrac{\lambda^{\text{GB}}}{\Sigma}f'^2M_{ij}^{\text{TF}}M^{\text{TF},kl} \right)\nonumber\\
&&\hspace{0.5cm}\times\left[\tfrac{1}{\alpha}D_kD_l\alpha
+\tfrac{1}{\chi}\big(\tilde{A}_{km}\tilde{A}^m_{~l}-\hat{\Theta}_{kl}\big) \right]\nonumber\\
&&-2\left(D_kA_{ij}-D_{(i}A_{j)k}\right)\Omega^k\nonumber\\
&&-\gamma_ {ij}\,(D^kA_{kl})\,\Omega^l+\Omega_{(i}D^kA_{j)k} \,,\\
\bar{S}^{\text{GB}}&=&\tfrac{2}{3}\left(\Omega-\tfrac{\lambda^{\text{GB}}}{4\Sigma}f'^2M^2\right)\nonumber\\
&&\hspace{0.5cm}\times\left[-\tfrac{1}{\alpha}\beta^i\partial_iK+\tfrac{1}{\alpha}D_iD^i\alpha-\tilde{A}_{ij}\tilde{A}^{ij}-\tfrac{K^2}{3} \right]\nonumber\\
&&+2\, M\left(\tfrac{1}{4}\,f''(K_{\phi}^2-(\partial\phi)^2)-\tfrac{g_2}{4\Sigma}f'Z^{g_2}\right.\nonumber\\
&&\hspace{1.1cm}\left.-\tfrac{\lambda^{\text{GB}}}{4\Sigma}f'^2H+\tfrac{1}{3}\,\Omega\right)\nonumber\\
&&-2\Omega^iN_i-\Omega^{\text{TF},ij}\,M^{\text{TF}}_{ij} -\rho^{\text{GB}}
\nonumber\\
&&+\big(\Omega^{\text{TF},kl}-\tfrac{\lambda^{\text{GB}}}{\Sigma}f'^2MM^{\text{TF},kl}\big)\nonumber\\
&&\hspace{0.5cm}\times\left(\tfrac{1}{\alpha}D_kD_l\alpha+\tfrac{1}{\chi}\tilde{A}_{km}\tilde{A}_{~l}^m -\tfrac{\hat{\Theta}_{kl}}{\chi} \right)\,,  \\
\bar{\mathcal L}_{\text{GB}}&=&-\tfrac{4}{3}\,M\left[-\tfrac{1}{\alpha}\beta^i\partial_iK + \tfrac{1}{\alpha}D_iD^i\alpha-\tilde{A}_{ij}\tilde{A}^{ij}-\tfrac{K^2}{3} \right] \nonumber\\
&&+8\,M^{\text{TF},kl}\left[\tfrac{1}{\alpha}D_kD_l\alpha+\tfrac{1}{\chi}\left(\tilde{A}_{kj}\tilde{A}^j_{~l}-\hat{\Theta}_{kl}\right) \right]\nonumber\\
&&-4\, H\, ,
\end{eqnarray}
\end{subequations}
where we have used $\hat{\Theta}_{kl}=\tfrac{1}{\alpha}{\mathcal L}_{\beta}\tilde{A}_{kl}+\tfrac{2}{3}\left(K-\tfrac{1}{\alpha}\partial_i\beta^i \right)\tilde{A}_{kl}$ with ${\mathcal L}_{\beta}\tilde{A}_{ij}=\beta^k\partial_k \tilde A_{ij}+2\tilde A_{k(i}\partial_{j)}\beta^k$ and 
\begin{equation}
    \begin{aligned}
    H=~2\,B_{ij}B^{ij}+N_i&N^i 
    =
    -\tfrac{4}{3}D_iK\big(N^i+\tfrac{D^iK}{3}\big) \\
    &+2\,D_iA_{jk}\big(D^iA^{jk}-D^jA^{ik} \big)\,.
    \end{aligned}
\end{equation}


\bibliographystyle{apsrev4-2}
\bibliography{ref}

\begin{thebibliography}{99}%
\makeatletter
\providecommand \@ifxundefined [1]{%
 \@ifx{#1\undefined}
}%
\providecommand \@ifnum [1]{%
 \ifnum #1\expandafter \@firstoftwo
 \else \expandafter \@secondoftwo
 \fi
}%
\providecommand \@ifx [1]{%
 \ifx #1\expandafter \@firstoftwo
 \else \expandafter \@secondoftwo
 \fi
}%
\providecommand \natexlab [1]{#1}%
\providecommand \enquote  [1]{``#1''}%
\providecommand \bibnamefont  [1]{#1}%
\providecommand \bibfnamefont [1]{#1}%
\providecommand \citenamefont [1]{#1}%
\providecommand \href@noop [0]{\@secondoftwo}%
\providecommand \href [0]{\begingroup \@sanitize@url \@href}%
\providecommand \@href[1]{\@@startlink{#1}\@@href}%
\providecommand \@@href[1]{\endgroup#1\@@endlink}%
\providecommand \@sanitize@url [0]{\catcode `\\12\catcode `\$12\catcode
  `\&12\catcode `\#12\catcode `\^12\catcode `\_12\catcode `\%12\relax}%
\providecommand \@@startlink[1]{}%
\providecommand \@@endlink[0]{}%
\providecommand \url  [0]{\begingroup\@sanitize@url \@url }%
\providecommand \@url [1]{\endgroup\@href {#1}{\urlprefix }}%
\providecommand \urlprefix  [0]{URL }%
\providecommand \Eprint [0]{\href }%
\providecommand \doibase [0]{https://doi.org/}%
\providecommand \selectlanguage [0]{\@gobble}%
\providecommand \bibinfo  [0]{\@secondoftwo}%
\providecommand \bibfield  [0]{\@secondoftwo}%
\providecommand \translation [1]{[#1]}%
\providecommand \BibitemOpen [0]{}%
\providecommand \bibitemStop [0]{}%
\providecommand \bibitemNoStop [0]{.\EOS\space}%
\providecommand \EOS [0]{\spacefactor3000\relax}%
\providecommand \BibitemShut  [1]{\csname bibitem#1\endcsname}%
\let\auto@bib@innerbib\@empty
\bibitem [{\citenamefont {Arun}\ \emph {et~al.}(2022)\citenamefont {Arun} \emph
  {et~al.}}]{LISA:2022kgy}%
  \BibitemOpen
  \bibfield  {author} {\bibinfo {author} {\bibfnamefont {K.~G.}\ \bibnamefont
  {Arun}} \emph {et~al.} (\bibinfo {collaboration} {LISA}),\ }\href
  {https://doi.org/10.1007/s41114-022-00036-9} {\bibfield  {journal} {\bibinfo
  {journal} {Living Rev. Rel.}\ }\textbf {\bibinfo {volume} {25}},\ \bibinfo
  {pages} {4} (\bibinfo {year} {2022})},\ \Eprint
  {https://arxiv.org/abs/2205.01597} {arXiv:2205.01597 [gr-qc]} \BibitemShut
  {NoStop}%
\bibitem [{\citenamefont {Perkins}\ \emph
  {et~al.}(2021{\natexlab{a}})\citenamefont {Perkins}, \citenamefont {Yunes},\
  and\ \citenamefont {Berti}}]{Perkins:2020tra}%
  \BibitemOpen
  \bibfield  {author} {\bibinfo {author} {\bibfnamefont {S.~E.}\ \bibnamefont
  {Perkins}}, \bibinfo {author} {\bibfnamefont {N.}~\bibnamefont {Yunes}},\
  and\ \bibinfo {author} {\bibfnamefont {E.}~\bibnamefont {Berti}},\ }\href
  {https://doi.org/10.1103/PhysRevD.103.044024} {\bibfield  {journal} {\bibinfo
   {journal} {Phys. Rev. D}\ }\textbf {\bibinfo {volume} {103}},\ \bibinfo
  {pages} {044024} (\bibinfo {year} {2021}{\natexlab{a}})},\ \Eprint
  {https://arxiv.org/abs/2010.09010} {arXiv:2010.09010 [gr-qc]} \BibitemShut
  {NoStop}%
\bibitem [{\citenamefont {Barausse}\ \emph {et~al.}(2020)\citenamefont
  {Barausse} \emph {et~al.}}]{Barausse:2020rsu}%
  \BibitemOpen
  \bibfield  {author} {\bibinfo {author} {\bibfnamefont {E.}~\bibnamefont
  {Barausse}} \emph {et~al.},\ }\href
  {https://doi.org/10.1007/s10714-020-02691-1} {\bibfield  {journal} {\bibinfo
  {journal} {Gen. Rel. Grav.}\ }\textbf {\bibinfo {volume} {52}},\ \bibinfo
  {pages} {81} (\bibinfo {year} {2020})},\ \Eprint
  {https://arxiv.org/abs/2001.09793} {arXiv:2001.09793 [gr-qc]} \BibitemShut
  {NoStop}%
\bibitem [{\citenamefont {Gnocchi}\ \emph {et~al.}(2019)\citenamefont
  {Gnocchi}, \citenamefont {Maselli}, \citenamefont {Abdelsalhin},
  \citenamefont {Giacobbo},\ and\ \citenamefont {Mapelli}}]{Gnocchi:2019jzp}%
  \BibitemOpen
  \bibfield  {author} {\bibinfo {author} {\bibfnamefont {G.}~\bibnamefont
  {Gnocchi}}, \bibinfo {author} {\bibfnamefont {A.}~\bibnamefont {Maselli}},
  \bibinfo {author} {\bibfnamefont {T.}~\bibnamefont {Abdelsalhin}}, \bibinfo
  {author} {\bibfnamefont {N.}~\bibnamefont {Giacobbo}},\ and\ \bibinfo
  {author} {\bibfnamefont {M.}~\bibnamefont {Mapelli}},\ }\href
  {https://doi.org/10.1103/PhysRevD.100.064024} {\bibfield  {journal} {\bibinfo
   {journal} {Phys. Rev. D}\ }\textbf {\bibinfo {volume} {100}},\ \bibinfo
  {pages} {064024} (\bibinfo {year} {2019})},\ \Eprint
  {https://arxiv.org/abs/1905.13460} {arXiv:1905.13460 [gr-qc]} \BibitemShut
  {NoStop}%
\bibitem [{\citenamefont {Barack}\ \emph {et~al.}(2019)\citenamefont {Barack}
  \emph {et~al.}}]{Barack:2018yly}%
  \BibitemOpen
  \bibfield  {author} {\bibinfo {author} {\bibfnamefont {L.}~\bibnamefont
  {Barack}} \emph {et~al.},\ }\href {https://doi.org/10.1088/1361-6382/ab0587}
  {\bibfield  {journal} {\bibinfo  {journal} {Class. Quant. Grav.}\ }\textbf
  {\bibinfo {volume} {36}},\ \bibinfo {pages} {143001} (\bibinfo {year}
  {2019})},\ \Eprint {https://arxiv.org/abs/1806.05195} {arXiv:1806.05195
  [gr-qc]} \BibitemShut {NoStop}%
\bibitem [{\citenamefont {Baker}\ \emph {et~al.}(2015)\citenamefont {Baker},
  \citenamefont {Psaltis},\ and\ \citenamefont {Skordis}}]{Baker:2014zba}%
  \BibitemOpen
  \bibfield  {author} {\bibinfo {author} {\bibfnamefont {T.}~\bibnamefont
  {Baker}}, \bibinfo {author} {\bibfnamefont {D.}~\bibnamefont {Psaltis}},\
  and\ \bibinfo {author} {\bibfnamefont {C.}~\bibnamefont {Skordis}},\ }\href
  {https://doi.org/10.1088/0004-637X/802/1/63} {\bibfield  {journal} {\bibinfo
  {journal} {Astrophys. J.}\ }\textbf {\bibinfo {volume} {802}},\ \bibinfo
  {pages} {63} (\bibinfo {year} {2015})},\ \Eprint
  {https://arxiv.org/abs/1412.3455} {arXiv:1412.3455 [astro-ph.CO]}
  \BibitemShut {NoStop}%
\bibitem [{\citenamefont {Koyama}(2016)}]{Koyama:2015vza}%
  \BibitemOpen
  \bibfield  {author} {\bibinfo {author} {\bibfnamefont {K.}~\bibnamefont
  {Koyama}},\ }\href {https://doi.org/10.1088/0034-4885/79/4/046902} {\bibfield
   {journal} {\bibinfo  {journal} {Rept. Prog. Phys.}\ }\textbf {\bibinfo
  {volume} {79}},\ \bibinfo {pages} {046902} (\bibinfo {year} {2016})},\
  \Eprint {https://arxiv.org/abs/1504.04623} {arXiv:1504.04623 [astro-ph.CO]}
  \BibitemShut {NoStop}%
\bibitem [{\citenamefont {Berti}\ \emph {et~al.}(2015)\citenamefont {Berti}
  \emph {et~al.}}]{Berti:2015itd}%
  \BibitemOpen
  \bibfield  {author} {\bibinfo {author} {\bibfnamefont {E.}~\bibnamefont
  {Berti}} \emph {et~al.},\ }\href
  {https://doi.org/10.1088/0264-9381/32/24/243001} {\bibfield  {journal}
  {\bibinfo  {journal} {Class. Quant. Grav.}\ }\textbf {\bibinfo {volume}
  {32}},\ \bibinfo {pages} {243001} (\bibinfo {year} {2015})},\ \Eprint
  {https://arxiv.org/abs/1501.07274} {arXiv:1501.07274 [gr-qc]} \BibitemShut
  {NoStop}%
\bibitem [{\citenamefont {Ferreira}(2019)}]{Ferreira:2019xrr}%
  \BibitemOpen
  \bibfield  {author} {\bibinfo {author} {\bibfnamefont {P.~G.}\ \bibnamefont
  {Ferreira}},\ }\href {https://doi.org/10.1146/annurev-astro-091918-104423}
  {\bibfield  {journal} {\bibinfo  {journal} {Ann. Rev. Astron. Astrophys.}\
  }\textbf {\bibinfo {volume} {57}},\ \bibinfo {pages} {335} (\bibinfo {year}
  {2019})},\ \Eprint {https://arxiv.org/abs/1902.10503} {arXiv:1902.10503
  [astro-ph.CO]} \BibitemShut {NoStop}%
\bibitem [{\citenamefont {Shibata}\ and\ \citenamefont
  {Traykova}(2023)}]{Shibata:2022gec}%
  \BibitemOpen
  \bibfield  {author} {\bibinfo {author} {\bibfnamefont {M.}~\bibnamefont
  {Shibata}}\ and\ \bibinfo {author} {\bibfnamefont {D.}~\bibnamefont
  {Traykova}},\ }\href {https://doi.org/10.1103/PhysRevD.107.044068} {\bibfield
   {journal} {\bibinfo  {journal} {Phys. Rev. D}\ }\textbf {\bibinfo {volume}
  {107}},\ \bibinfo {pages} {044068} (\bibinfo {year} {2023})},\ \Eprint
  {https://arxiv.org/abs/2210.12139} {arXiv:2210.12139 [gr-qc]} \BibitemShut
  {NoStop}%
\bibitem [{\citenamefont {Lara}\ \emph {et~al.}(2023)\citenamefont {Lara},
  \citenamefont {Bezares}, \citenamefont {Crisostomi},\ and\ \citenamefont
  {Barausse}}]{Lara:2022gof}%
  \BibitemOpen
  \bibfield  {author} {\bibinfo {author} {\bibfnamefont {G.}~\bibnamefont
  {Lara}}, \bibinfo {author} {\bibfnamefont {M.}~\bibnamefont {Bezares}},
  \bibinfo {author} {\bibfnamefont {M.}~\bibnamefont {Crisostomi}},\ and\
  \bibinfo {author} {\bibfnamefont {E.}~\bibnamefont {Barausse}},\ }\href
  {https://doi.org/10.1103/PhysRevD.107.044019} {\bibfield  {journal} {\bibinfo
   {journal} {Phys. Rev. D}\ }\textbf {\bibinfo {volume} {107}},\ \bibinfo
  {pages} {044019} (\bibinfo {year} {2023})},\ \Eprint
  {https://arxiv.org/abs/2207.03437} {arXiv:2207.03437 [gr-qc]} \BibitemShut
  {NoStop}%
\bibitem [{\citenamefont {Bezares}\ \emph {et~al.}(2022)\citenamefont
  {Bezares}, \citenamefont {Aguilera-Miret}, \citenamefont {ter Haar},
  \citenamefont {Crisostomi}, \citenamefont {Palenzuela},\ and\ \citenamefont
  {Barausse}}]{Bezares:2021dma}%
  \BibitemOpen
  \bibfield  {author} {\bibinfo {author} {\bibfnamefont {M.}~\bibnamefont
  {Bezares}}, \bibinfo {author} {\bibfnamefont {R.}~\bibnamefont
  {Aguilera-Miret}}, \bibinfo {author} {\bibfnamefont {L.}~\bibnamefont {ter
  Haar}}, \bibinfo {author} {\bibfnamefont {M.}~\bibnamefont {Crisostomi}},
  \bibinfo {author} {\bibfnamefont {C.}~\bibnamefont {Palenzuela}},\ and\
  \bibinfo {author} {\bibfnamefont {E.}~\bibnamefont {Barausse}},\ }\href
  {https://doi.org/10.1103/PhysRevLett.128.091103} {\bibfield  {journal}
  {\bibinfo  {journal} {Phys. Rev. Lett.}\ }\textbf {\bibinfo {volume} {128}},\
  \bibinfo {pages} {091103} (\bibinfo {year} {2022})},\ \Eprint
  {https://arxiv.org/abs/2107.05648} {arXiv:2107.05648 [gr-qc]} \BibitemShut
  {NoStop}%
\bibitem [{\citenamefont {Bezares}\ \emph {et~al.}(2021)\citenamefont
  {Bezares}, \citenamefont {ter Haar}, \citenamefont {Crisostomi},
  \citenamefont {Barausse},\ and\ \citenamefont
  {Palenzuela}}]{Bezares:2021yek}%
  \BibitemOpen
  \bibfield  {author} {\bibinfo {author} {\bibfnamefont {M.}~\bibnamefont
  {Bezares}}, \bibinfo {author} {\bibfnamefont {L.}~\bibnamefont {ter Haar}},
  \bibinfo {author} {\bibfnamefont {M.}~\bibnamefont {Crisostomi}}, \bibinfo
  {author} {\bibfnamefont {E.}~\bibnamefont {Barausse}},\ and\ \bibinfo
  {author} {\bibfnamefont {C.}~\bibnamefont {Palenzuela}},\ }\href
  {https://doi.org/10.1103/PhysRevD.104.044022} {\bibfield  {journal} {\bibinfo
   {journal} {Phys. Rev. D}\ }\textbf {\bibinfo {volume} {104}},\ \bibinfo
  {pages} {044022} (\bibinfo {year} {2021})},\ \Eprint
  {https://arxiv.org/abs/2105.13992} {arXiv:2105.13992 [gr-qc]} \BibitemShut
  {NoStop}%
\bibitem [{\citenamefont {Evstafyeva}\ \emph {et~al.}(2022)\citenamefont
  {Evstafyeva}, \citenamefont {Agathos},\ and\ \citenamefont
  {Ripley}}]{Evstafyeva:2022rve}%
  \BibitemOpen
  \bibfield  {author} {\bibinfo {author} {\bibfnamefont {T.}~\bibnamefont
  {Evstafyeva}}, \bibinfo {author} {\bibfnamefont {M.}~\bibnamefont
  {Agathos}},\ and\ \bibinfo {author} {\bibfnamefont {J.~L.}\ \bibnamefont
  {Ripley}},\ }\href@noop {} {\  (\bibinfo {year} {2022})},\ \Eprint
  {https://arxiv.org/abs/2212.11359} {arXiv:2212.11359 [gr-qc]} \BibitemShut
  {NoStop}%
\bibitem [{\citenamefont {Perkins}\ \emph
  {et~al.}(2021{\natexlab{b}})\citenamefont {Perkins}, \citenamefont {Nair},
  \citenamefont {Silva},\ and\ \citenamefont {Yunes}}]{Perkins:2021mhb}%
  \BibitemOpen
  \bibfield  {author} {\bibinfo {author} {\bibfnamefont {S.~E.}\ \bibnamefont
  {Perkins}}, \bibinfo {author} {\bibfnamefont {R.}~\bibnamefont {Nair}},
  \bibinfo {author} {\bibfnamefont {H.~O.}\ \bibnamefont {Silva}},\ and\
  \bibinfo {author} {\bibfnamefont {N.}~\bibnamefont {Yunes}},\ }\href
  {https://doi.org/10.1103/PhysRevD.104.024060} {\bibfield  {journal} {\bibinfo
   {journal} {Phys. Rev. D}\ }\textbf {\bibinfo {volume} {104}},\ \bibinfo
  {pages} {024060} (\bibinfo {year} {2021}{\natexlab{b}})},\ \Eprint
  {https://arxiv.org/abs/2104.11189} {arXiv:2104.11189 [gr-qc]} \BibitemShut
  {NoStop}%
\bibitem [{\citenamefont {Toubiana}\ \emph {et~al.}(2020)\citenamefont
  {Toubiana}, \citenamefont {Marsat}, \citenamefont {Barausse}, \citenamefont
  {Babak},\ and\ \citenamefont {Baker}}]{Toubiana:2020vtf}%
  \BibitemOpen
  \bibfield  {author} {\bibinfo {author} {\bibfnamefont {A.}~\bibnamefont
  {Toubiana}}, \bibinfo {author} {\bibfnamefont {S.}~\bibnamefont {Marsat}},
  \bibinfo {author} {\bibfnamefont {E.}~\bibnamefont {Barausse}}, \bibinfo
  {author} {\bibfnamefont {S.}~\bibnamefont {Babak}},\ and\ \bibinfo {author}
  {\bibfnamefont {J.}~\bibnamefont {Baker}},\ }\href
  {https://doi.org/10.1103/PhysRevD.101.104038} {\bibfield  {journal} {\bibinfo
   {journal} {Phys. Rev. D}\ }\textbf {\bibinfo {volume} {101}},\ \bibinfo
  {pages} {104038} (\bibinfo {year} {2020})},\ \Eprint
  {https://arxiv.org/abs/2004.03626} {arXiv:2004.03626 [gr-qc]} \BibitemShut
  {NoStop}%
\bibitem [{\citenamefont {Carson}\ \emph {et~al.}(2020)\citenamefont {Carson},
  \citenamefont {Seymour},\ and\ \citenamefont {Yagi}}]{Carson:2019fxr}%
  \BibitemOpen
  \bibfield  {author} {\bibinfo {author} {\bibfnamefont {Z.}~\bibnamefont
  {Carson}}, \bibinfo {author} {\bibfnamefont {B.~C.}\ \bibnamefont
  {Seymour}},\ and\ \bibinfo {author} {\bibfnamefont {K.}~\bibnamefont
  {Yagi}},\ }\href {https://doi.org/10.1088/1361-6382/ab6a1f} {\bibfield
  {journal} {\bibinfo  {journal} {Class. Quant. Grav.}\ }\textbf {\bibinfo
  {volume} {37}},\ \bibinfo {pages} {065008} (\bibinfo {year} {2020})},\
  \Eprint {https://arxiv.org/abs/1907.03897} {arXiv:1907.03897 [gr-qc]}
  \BibitemShut {NoStop}%
\bibitem [{\citenamefont {Yunes}\ \emph {et~al.}(2016)\citenamefont {Yunes},
  \citenamefont {Yagi},\ and\ \citenamefont {Pretorius}}]{Yunes:2016jcc}%
  \BibitemOpen
  \bibfield  {author} {\bibinfo {author} {\bibfnamefont {N.}~\bibnamefont
  {Yunes}}, \bibinfo {author} {\bibfnamefont {K.}~\bibnamefont {Yagi}},\ and\
  \bibinfo {author} {\bibfnamefont {F.}~\bibnamefont {Pretorius}},\ }\href
  {https://doi.org/10.1103/PhysRevD.94.084002} {\bibfield  {journal} {\bibinfo
  {journal} {Phys. Rev. D}\ }\textbf {\bibinfo {volume} {94}},\ \bibinfo
  {pages} {084002} (\bibinfo {year} {2016})},\ \Eprint
  {https://arxiv.org/abs/1603.08955} {arXiv:1603.08955 [gr-qc]} \BibitemShut
  {NoStop}%
\bibitem [{\citenamefont {Maggio}\ \emph {et~al.}(2022)\citenamefont {Maggio},
  \citenamefont {Silva}, \citenamefont {Buonanno},\ and\ \citenamefont
  {Ghosh}}]{Maggio:2022hre}%
  \BibitemOpen
  \bibfield  {author} {\bibinfo {author} {\bibfnamefont {E.}~\bibnamefont
  {Maggio}}, \bibinfo {author} {\bibfnamefont {H.~O.}\ \bibnamefont {Silva}},
  \bibinfo {author} {\bibfnamefont {A.}~\bibnamefont {Buonanno}},\ and\
  \bibinfo {author} {\bibfnamefont {A.}~\bibnamefont {Ghosh}},\ }\href@noop {}
  {\  (\bibinfo {year} {2022})},\ \Eprint {https://arxiv.org/abs/2212.09655}
  {arXiv:2212.09655 [gr-qc]} \BibitemShut {NoStop}%
\bibitem [{\citenamefont {Krishnendu}\ and\ \citenamefont
  {Ohme}(2021)}]{Krishnendu:2021fga}%
  \BibitemOpen
  \bibfield  {author} {\bibinfo {author} {\bibfnamefont {N.~V.}\ \bibnamefont
  {Krishnendu}}\ and\ \bibinfo {author} {\bibfnamefont {F.}~\bibnamefont
  {Ohme}},\ }\href {https://doi.org/10.3390/universe7120497} {\bibfield
  {journal} {\bibinfo  {journal} {Universe}\ }\textbf {\bibinfo {volume} {7}},\
  \bibinfo {pages} {497} (\bibinfo {year} {2021})},\ \Eprint
  {https://arxiv.org/abs/2201.05418} {arXiv:2201.05418 [gr-qc]} \BibitemShut
  {NoStop}%
\bibitem [{\citenamefont {Abbott}\ \emph {et~al.}(2021)\citenamefont {Abbott}
  \emph {et~al.}}]{LIGOScientific:2021sio}%
  \BibitemOpen
  \bibfield  {author} {\bibinfo {author} {\bibfnamefont {R.}~\bibnamefont
  {Abbott}} \emph {et~al.} (\bibinfo {collaboration} {LIGO Scientific, VIRGO,
  KAGRA}),\ }\href@noop {} {\  (\bibinfo {year} {2021})},\ \Eprint
  {https://arxiv.org/abs/2112.06861} {arXiv:2112.06861 [gr-qc]} \BibitemShut
  {NoStop}%
\bibitem [{\citenamefont {Carson}\ and\ \citenamefont
  {Yagi}(2020{\natexlab{a}})}]{Carson:2019kkh}%
  \BibitemOpen
  \bibfield  {author} {\bibinfo {author} {\bibfnamefont {Z.}~\bibnamefont
  {Carson}}\ and\ \bibinfo {author} {\bibfnamefont {K.}~\bibnamefont {Yagi}},\
  }\href {https://doi.org/10.1103/PhysRevD.101.044047} {\bibfield  {journal}
  {\bibinfo  {journal} {Phys. Rev. D}\ }\textbf {\bibinfo {volume} {101}},\
  \bibinfo {pages} {044047} (\bibinfo {year} {2020}{\natexlab{a}})},\ \Eprint
  {https://arxiv.org/abs/1911.05258} {arXiv:1911.05258 [gr-qc]} \BibitemShut
  {NoStop}%
\bibitem [{\citenamefont {Cornish}\ \emph {et~al.}(2011)\citenamefont
  {Cornish}, \citenamefont {Sampson}, \citenamefont {Yunes},\ and\
  \citenamefont {Pretorius}}]{Cornish:2011ys}%
  \BibitemOpen
  \bibfield  {author} {\bibinfo {author} {\bibfnamefont {N.}~\bibnamefont
  {Cornish}}, \bibinfo {author} {\bibfnamefont {L.}~\bibnamefont {Sampson}},
  \bibinfo {author} {\bibfnamefont {N.}~\bibnamefont {Yunes}},\ and\ \bibinfo
  {author} {\bibfnamefont {F.}~\bibnamefont {Pretorius}},\ }\href
  {https://doi.org/10.1103/PhysRevD.84.062003} {\bibfield  {journal} {\bibinfo
  {journal} {Phys. Rev. D}\ }\textbf {\bibinfo {volume} {84}},\ \bibinfo
  {pages} {062003} (\bibinfo {year} {2011})},\ \Eprint
  {https://arxiv.org/abs/1105.2088} {arXiv:1105.2088 [gr-qc]} \BibitemShut
  {NoStop}%
\bibitem [{\citenamefont {Okounkova}\ \emph {et~al.}(2023)\citenamefont
  {Okounkova}, \citenamefont {Isi}, \citenamefont {Chatziioannou},\ and\
  \citenamefont {Farr}}]{Okounkova:2022grv}%
  \BibitemOpen
  \bibfield  {author} {\bibinfo {author} {\bibfnamefont {M.}~\bibnamefont
  {Okounkova}}, \bibinfo {author} {\bibfnamefont {M.}~\bibnamefont {Isi}},
  \bibinfo {author} {\bibfnamefont {K.}~\bibnamefont {Chatziioannou}},\ and\
  \bibinfo {author} {\bibfnamefont {W.~M.}\ \bibnamefont {Farr}},\ }\href
  {https://doi.org/10.1103/PhysRevD.107.024046} {\bibfield  {journal} {\bibinfo
   {journal} {Phys. Rev. D}\ }\textbf {\bibinfo {volume} {107}},\ \bibinfo
  {pages} {024046} (\bibinfo {year} {2023})},\ \Eprint
  {https://arxiv.org/abs/2208.02805} {arXiv:2208.02805 [gr-qc]} \BibitemShut
  {NoStop}%
\bibitem [{\citenamefont {Johnson-McDaniel}\ \emph {et~al.}(2022)\citenamefont
  {Johnson-McDaniel}, \citenamefont {Ghosh}, \citenamefont {Ghonge},
  \citenamefont {Saleem}, \citenamefont {Krishnendu},\ and\ \citenamefont
  {Clark}}]{Johnson-McDaniel:2021yge}%
  \BibitemOpen
  \bibfield  {author} {\bibinfo {author} {\bibfnamefont {N.~K.}\ \bibnamefont
  {Johnson-McDaniel}}, \bibinfo {author} {\bibfnamefont {A.}~\bibnamefont
  {Ghosh}}, \bibinfo {author} {\bibfnamefont {S.}~\bibnamefont {Ghonge}},
  \bibinfo {author} {\bibfnamefont {M.}~\bibnamefont {Saleem}}, \bibinfo
  {author} {\bibfnamefont {N.~V.}\ \bibnamefont {Krishnendu}},\ and\ \bibinfo
  {author} {\bibfnamefont {J.~A.}\ \bibnamefont {Clark}},\ }\href
  {https://doi.org/10.1103/PhysRevD.105.044020} {\bibfield  {journal} {\bibinfo
   {journal} {Phys. Rev. D}\ }\textbf {\bibinfo {volume} {105}},\ \bibinfo
  {pages} {044020} (\bibinfo {year} {2022})},\ \Eprint
  {https://arxiv.org/abs/2109.06988} {arXiv:2109.06988 [gr-qc]} \BibitemShut
  {NoStop}%
\bibitem [{\citenamefont {Shiralilou}\ \emph {et~al.}(2022)\citenamefont
  {Shiralilou}, \citenamefont {Hinderer}, \citenamefont {Nissanke},
  \citenamefont {Ortiz},\ and\ \citenamefont {Witek}}]{Shiralilou:2021mfl}%
  \BibitemOpen
  \bibfield  {author} {\bibinfo {author} {\bibfnamefont {B.}~\bibnamefont
  {Shiralilou}}, \bibinfo {author} {\bibfnamefont {T.}~\bibnamefont
  {Hinderer}}, \bibinfo {author} {\bibfnamefont {S.~M.}\ \bibnamefont
  {Nissanke}}, \bibinfo {author} {\bibfnamefont {N.}~\bibnamefont {Ortiz}},\
  and\ \bibinfo {author} {\bibfnamefont {H.}~\bibnamefont {Witek}},\ }\href
  {https://doi.org/10.1088/1361-6382/ac4196} {\bibfield  {journal} {\bibinfo
  {journal} {Class. Quant. Grav.}\ }\textbf {\bibinfo {volume} {39}},\ \bibinfo
  {pages} {035002} (\bibinfo {year} {2022})},\ \Eprint
  {https://arxiv.org/abs/2105.13972} {arXiv:2105.13972 [gr-qc]} \BibitemShut
  {NoStop}%
\bibitem [{\citenamefont {Carson}\ and\ \citenamefont
  {Yagi}(2020{\natexlab{b}})}]{Carson:2020ter}%
  \BibitemOpen
  \bibfield  {author} {\bibinfo {author} {\bibfnamefont {Z.}~\bibnamefont
  {Carson}}\ and\ \bibinfo {author} {\bibfnamefont {K.}~\bibnamefont {Yagi}},\
  }\href {https://doi.org/10.1103/PhysRevD.101.104030} {\bibfield  {journal}
  {\bibinfo  {journal} {Phys. Rev. D}\ }\textbf {\bibinfo {volume} {101}},\
  \bibinfo {pages} {104030} (\bibinfo {year} {2020}{\natexlab{b}})},\ \Eprint
  {https://arxiv.org/abs/2003.00286} {arXiv:2003.00286 [gr-qc]} \BibitemShut
  {NoStop}%
\bibitem [{\citenamefont {Carson}\ and\ \citenamefont
  {Yagi}(2020{\natexlab{c}})}]{Carson:2020cqb}%
  \BibitemOpen
  \bibfield  {author} {\bibinfo {author} {\bibfnamefont {Z.}~\bibnamefont
  {Carson}}\ and\ \bibinfo {author} {\bibfnamefont {K.}~\bibnamefont {Yagi}},\
  }\href {https://doi.org/10.1088/1361-6382/aba221} {\bibfield  {journal}
  {\bibinfo  {journal} {Class. Quant. Grav.}\ }\textbf {\bibinfo {volume}
  {37}},\ \bibinfo {pages} {215007} (\bibinfo {year} {2020}{\natexlab{c}})},\
  \Eprint {https://arxiv.org/abs/2002.08559} {arXiv:2002.08559 [gr-qc]}
  \BibitemShut {NoStop}%
\bibitem [{\citenamefont {Kuan}\ \emph {et~al.}(2023)\citenamefont {Kuan},
  \citenamefont {Lam}, \citenamefont {Doneva}, \citenamefont {Yazadjiev},
  \citenamefont {Shibata},\ and\ \citenamefont {Kiuchi}}]{Kuan:2023trn}%
  \BibitemOpen
  \bibfield  {author} {\bibinfo {author} {\bibfnamefont {H.-J.}\ \bibnamefont
  {Kuan}}, \bibinfo {author} {\bibfnamefont {A.~T.-L.}\ \bibnamefont {Lam}},
  \bibinfo {author} {\bibfnamefont {D.~D.}\ \bibnamefont {Doneva}}, \bibinfo
  {author} {\bibfnamefont {S.~S.}\ \bibnamefont {Yazadjiev}}, \bibinfo {author}
  {\bibfnamefont {M.}~\bibnamefont {Shibata}},\ and\ \bibinfo {author}
  {\bibfnamefont {K.}~\bibnamefont {Kiuchi}},\ }\href@noop {} {\  (\bibinfo
  {year} {2023})},\ \Eprint {https://arxiv.org/abs/2302.11596}
  {arXiv:2302.11596 [gr-qc]} \BibitemShut {NoStop}%
\bibitem [{\citenamefont {Ma}\ \emph {et~al.}(2023)\citenamefont {Ma},
  \citenamefont {Varma}, \citenamefont {Stein}, \citenamefont {Foucart},
  \citenamefont {Duez}, \citenamefont {Kidder}, \citenamefont {Pfeiffer},\ and\
  \citenamefont {Scheel}}]{Ma:2023sok}%
  \BibitemOpen
  \bibfield  {author} {\bibinfo {author} {\bibfnamefont {S.}~\bibnamefont
  {Ma}}, \bibinfo {author} {\bibfnamefont {V.}~\bibnamefont {Varma}}, \bibinfo
  {author} {\bibfnamefont {L.~C.}\ \bibnamefont {Stein}}, \bibinfo {author}
  {\bibfnamefont {F.}~\bibnamefont {Foucart}}, \bibinfo {author} {\bibfnamefont
  {M.~D.}\ \bibnamefont {Duez}}, \bibinfo {author} {\bibfnamefont {L.~E.}\
  \bibnamefont {Kidder}}, \bibinfo {author} {\bibfnamefont {H.~P.}\
  \bibnamefont {Pfeiffer}},\ and\ \bibinfo {author} {\bibfnamefont {M.~A.}\
  \bibnamefont {Scheel}},\ }\href@noop {} {\  (\bibinfo {year} {2023})},\
  \Eprint {https://arxiv.org/abs/2304.11836} {arXiv:2304.11836 [gr-qc]}
  \BibitemShut {NoStop}%
\bibitem [{\citenamefont {Doneva}\ \emph
  {et~al.}(2022{\natexlab{a}})\citenamefont {Doneva}, \citenamefont
  {Ramazano\u{g}lu}, \citenamefont {Silva}, \citenamefont {Sotiriou},\ and\
  \citenamefont {Yazadjiev}}]{Doneva:2022ewd}%
  \BibitemOpen
  \bibfield  {author} {\bibinfo {author} {\bibfnamefont {D.~D.}\ \bibnamefont
  {Doneva}}, \bibinfo {author} {\bibfnamefont {F.~M.}\ \bibnamefont
  {Ramazano\u{g}lu}}, \bibinfo {author} {\bibfnamefont {H.~O.}\ \bibnamefont
  {Silva}}, \bibinfo {author} {\bibfnamefont {T.~P.}\ \bibnamefont
  {Sotiriou}},\ and\ \bibinfo {author} {\bibfnamefont {S.~S.}\ \bibnamefont
  {Yazadjiev}},\ }\href@noop {} {\  (\bibinfo {year} {2022}{\natexlab{a}})},\
  \Eprint {https://arxiv.org/abs/2211.01766} {arXiv:2211.01766 [gr-qc]}
  \BibitemShut {NoStop}%
\bibitem [{\citenamefont {Lanczos}(1938)}]{Lanczos:1938sf}%
  \BibitemOpen
  \bibfield  {author} {\bibinfo {author} {\bibfnamefont {C.}~\bibnamefont
  {Lanczos}},\ }\href {https://doi.org/10.2307/1968467} {\bibfield  {journal}
  {\bibinfo  {journal} {Annals Math.}\ }\textbf {\bibinfo {volume} {39}},\
  \bibinfo {pages} {842} (\bibinfo {year} {1938})}\BibitemShut {NoStop}%
\bibitem [{\citenamefont {Lovelock}(1972)}]{Lovelock:1972vz}%
  \BibitemOpen
  \bibfield  {author} {\bibinfo {author} {\bibfnamefont {D.}~\bibnamefont
  {Lovelock}},\ }\href {https://doi.org/10.1063/1.1666069} {\bibfield
  {journal} {\bibinfo  {journal} {J. Math. Phys.}\ }\textbf {\bibinfo {volume}
  {13}},\ \bibinfo {pages} {874} (\bibinfo {year} {1972})}\BibitemShut
  {NoStop}%
\bibitem [{\citenamefont {Lovelock}(1971)}]{Lovelock:1971yv}%
  \BibitemOpen
  \bibfield  {author} {\bibinfo {author} {\bibfnamefont {D.}~\bibnamefont
  {Lovelock}},\ }\href {https://doi.org/10.1063/1.1665613} {\bibfield
  {journal} {\bibinfo  {journal} {J. Math. Phys.}\ }\textbf {\bibinfo {volume}
  {12}},\ \bibinfo {pages} {498} (\bibinfo {year} {1971})}\BibitemShut
  {NoStop}%
\bibitem [{\citenamefont {Endlich}\ \emph {et~al.}(2017)\citenamefont
  {Endlich}, \citenamefont {Gorbenko}, \citenamefont {Huang},\ and\
  \citenamefont {Senatore}}]{Endlich:2017tqa}%
  \BibitemOpen
  \bibfield  {author} {\bibinfo {author} {\bibfnamefont {S.}~\bibnamefont
  {Endlich}}, \bibinfo {author} {\bibfnamefont {V.}~\bibnamefont {Gorbenko}},
  \bibinfo {author} {\bibfnamefont {J.}~\bibnamefont {Huang}},\ and\ \bibinfo
  {author} {\bibfnamefont {L.}~\bibnamefont {Senatore}},\ }\href
  {https://doi.org/10.1007/JHEP09(2017)122} {\bibfield  {journal} {\bibinfo
  {journal} {JHEP}\ }\textbf {\bibinfo {volume} {09}},\ \bibinfo {pages}
  {122}},\ \Eprint {https://arxiv.org/abs/1704.01590} {arXiv:1704.01590
  [gr-qc]} \BibitemShut {NoStop}%
\bibitem [{\citenamefont {Cayuso}\ \emph {et~al.}(2017)\citenamefont {Cayuso},
  \citenamefont {Ortiz},\ and\ \citenamefont {Lehner}}]{Cayuso:2017iqc}%
  \BibitemOpen
  \bibfield  {author} {\bibinfo {author} {\bibfnamefont {J.}~\bibnamefont
  {Cayuso}}, \bibinfo {author} {\bibfnamefont {N.}~\bibnamefont {Ortiz}},\ and\
  \bibinfo {author} {\bibfnamefont {L.}~\bibnamefont {Lehner}},\ }\href
  {https://doi.org/10.1103/PhysRevD.96.084043} {\bibfield  {journal} {\bibinfo
  {journal} {Phys. Rev. D}\ }\textbf {\bibinfo {volume} {96}},\ \bibinfo
  {pages} {084043} (\bibinfo {year} {2017})},\ \Eprint
  {https://arxiv.org/abs/1706.07421} {arXiv:1706.07421 [gr-qc]} \BibitemShut
  {NoStop}%
\bibitem [{\citenamefont {Allwright}\ and\ \citenamefont
  {Lehner}(2019)}]{Allwright:2018rut}%
  \BibitemOpen
  \bibfield  {author} {\bibinfo {author} {\bibfnamefont {G.}~\bibnamefont
  {Allwright}}\ and\ \bibinfo {author} {\bibfnamefont {L.}~\bibnamefont
  {Lehner}},\ }\href {https://doi.org/10.1088/1361-6382/ab0ee1} {\bibfield
  {journal} {\bibinfo  {journal} {Class. Quant. Grav.}\ }\textbf {\bibinfo
  {volume} {36}},\ \bibinfo {pages} {084001} (\bibinfo {year} {2019})},\
  \Eprint {https://arxiv.org/abs/1808.07897} {arXiv:1808.07897 [gr-qc]}
  \BibitemShut {NoStop}%
\bibitem [{\citenamefont {Cayuso}\ and\ \citenamefont
  {Lehner}(2020)}]{Cayuso:2020lca}%
  \BibitemOpen
  \bibfield  {author} {\bibinfo {author} {\bibfnamefont {R.}~\bibnamefont
  {Cayuso}}\ and\ \bibinfo {author} {\bibfnamefont {L.}~\bibnamefont
  {Lehner}},\ }\href {https://doi.org/10.1103/PhysRevD.102.084008} {\bibfield
  {journal} {\bibinfo  {journal} {Phys. Rev. D}\ }\textbf {\bibinfo {volume}
  {102}},\ \bibinfo {pages} {084008} (\bibinfo {year} {2020})},\ \Eprint
  {https://arxiv.org/abs/2005.13720} {arXiv:2005.13720 [gr-qc]} \BibitemShut
  {NoStop}%
\bibitem [{\citenamefont {Franchini}\ \emph {et~al.}(2022)\citenamefont
  {Franchini}, \citenamefont {Bezares}, \citenamefont {Barausse},\ and\
  \citenamefont {Lehner}}]{Franchini:2022ukz}%
  \BibitemOpen
  \bibfield  {author} {\bibinfo {author} {\bibfnamefont {N.}~\bibnamefont
  {Franchini}}, \bibinfo {author} {\bibfnamefont {M.}~\bibnamefont {Bezares}},
  \bibinfo {author} {\bibfnamefont {E.}~\bibnamefont {Barausse}},\ and\
  \bibinfo {author} {\bibfnamefont {L.}~\bibnamefont {Lehner}},\ }\href
  {https://doi.org/10.1103/PhysRevD.106.064061} {\bibfield  {journal} {\bibinfo
   {journal} {Phys. Rev. D}\ }\textbf {\bibinfo {volume} {106}},\ \bibinfo
  {pages} {064061} (\bibinfo {year} {2022})},\ \Eprint
  {https://arxiv.org/abs/2206.00014} {arXiv:2206.00014 [gr-qc]} \BibitemShut
  {NoStop}%
\bibitem [{\citenamefont {Cayuso}\ \emph {et~al.}(2023)\citenamefont {Cayuso},
  \citenamefont {Figueras}, \citenamefont {Fran\c{c}a},\ and\ \citenamefont
  {Lehner}}]{Cayuso:2023aht}%
  \BibitemOpen
  \bibfield  {author} {\bibinfo {author} {\bibfnamefont {R.}~\bibnamefont
  {Cayuso}}, \bibinfo {author} {\bibfnamefont {P.}~\bibnamefont {Figueras}},
  \bibinfo {author} {\bibfnamefont {T.}~\bibnamefont {Fran\c{c}a}},\ and\
  \bibinfo {author} {\bibfnamefont {L.}~\bibnamefont {Lehner}},\ }\href@noop {}
  {\  (\bibinfo {year} {2023})},\ \Eprint {https://arxiv.org/abs/2303.07246}
  {arXiv:2303.07246 [gr-qc]} \BibitemShut {NoStop}%
\bibitem [{\citenamefont {Sarbach}\ \emph {et~al.}(2019)\citenamefont
  {Sarbach}, \citenamefont {Barausse},\ and\ \citenamefont
  {Preciado-L\'opez}}]{Sarbach:2019yso}%
  \BibitemOpen
  \bibfield  {author} {\bibinfo {author} {\bibfnamefont {O.}~\bibnamefont
  {Sarbach}}, \bibinfo {author} {\bibfnamefont {E.}~\bibnamefont {Barausse}},\
  and\ \bibinfo {author} {\bibfnamefont {J.~A.}\ \bibnamefont
  {Preciado-L\'opez}},\ }\href {https://doi.org/10.1088/1361-6382/ab2e13}
  {\bibfield  {journal} {\bibinfo  {journal} {Class. Quant. Grav.}\ }\textbf
  {\bibinfo {volume} {36}},\ \bibinfo {pages} {165007} (\bibinfo {year}
  {2019})},\ \Eprint {https://arxiv.org/abs/1902.05130} {arXiv:1902.05130
  [gr-qc]} \BibitemShut {NoStop}%
\bibitem [{\citenamefont {de~Rham}\ \emph {et~al.}(2023)\citenamefont
  {de~Rham}, \citenamefont {Ko\.zuszek}, \citenamefont {Tolley},\ and\
  \citenamefont {Wiseman}}]{deRham:2023ngf}%
  \BibitemOpen
  \bibfield  {author} {\bibinfo {author} {\bibfnamefont {C.}~\bibnamefont
  {de~Rham}}, \bibinfo {author} {\bibfnamefont {J.}~\bibnamefont {Ko\.zuszek}},
  \bibinfo {author} {\bibfnamefont {A.~J.}\ \bibnamefont {Tolley}},\ and\
  \bibinfo {author} {\bibfnamefont {T.}~\bibnamefont {Wiseman}},\ }\href@noop
  {} {\  (\bibinfo {year} {2023})},\ \Eprint {https://arxiv.org/abs/2302.04876}
  {arXiv:2302.04876 [hep-th]} \BibitemShut {NoStop}%
\bibitem [{\citenamefont {Horndeski}(1974)}]{Horndeski:1974wa}%
  \BibitemOpen
  \bibfield  {author} {\bibinfo {author} {\bibfnamefont {G.~W.}\ \bibnamefont
  {Horndeski}},\ }\href {https://doi.org/10.1007/BF01807638} {\bibfield
  {journal} {\bibinfo  {journal} {Int. J. Theor. Phys.}\ }\textbf {\bibinfo
  {volume} {10}},\ \bibinfo {pages} {363} (\bibinfo {year} {1974})}\BibitemShut
  {NoStop}%
\bibitem [{\citenamefont {Richards}\ \emph {et~al.}(2023)\citenamefont
  {Richards}, \citenamefont {Dima},\ and\ \citenamefont
  {Witek}}]{Richards:2023xsr}%
  \BibitemOpen
  \bibfield  {author} {\bibinfo {author} {\bibfnamefont {C.}~\bibnamefont
  {Richards}}, \bibinfo {author} {\bibfnamefont {A.}~\bibnamefont {Dima}},\
  and\ \bibinfo {author} {\bibfnamefont {H.}~\bibnamefont {Witek}},\
  }\href@noop {} {\  (\bibinfo {year} {2023})},\ \Eprint
  {https://arxiv.org/abs/2305.07704} {arXiv:2305.07704 [gr-qc]} \BibitemShut
  {NoStop}%
\bibitem [{\citenamefont {R.}\ \emph {et~al.}(2023)\citenamefont {R.},
  \citenamefont {Most}, \citenamefont {Noronha}, \citenamefont {Witek},\ and\
  \citenamefont {Yunes}}]{R:2022tqa}%
  \BibitemOpen
  \bibfield  {author} {\bibinfo {author} {\bibfnamefont {A.~H.~K.}\
  \bibnamefont {R.}}, \bibinfo {author} {\bibfnamefont {E.~R.}\ \bibnamefont
  {Most}}, \bibinfo {author} {\bibfnamefont {J.}~\bibnamefont {Noronha}},
  \bibinfo {author} {\bibfnamefont {H.}~\bibnamefont {Witek}},\ and\ \bibinfo
  {author} {\bibfnamefont {N.}~\bibnamefont {Yunes}},\ }\href
  {https://doi.org/10.1103/PhysRevD.107.104047} {\bibfield  {journal} {\bibinfo
   {journal} {Phys. Rev. D}\ }\textbf {\bibinfo {volume} {107}},\ \bibinfo
  {pages} {104047} (\bibinfo {year} {2023})},\ \Eprint
  {https://arxiv.org/abs/2212.02039} {arXiv:2212.02039 [gr-qc]} \BibitemShut
  {NoStop}%
\bibitem [{\citenamefont {Elley}\ \emph {et~al.}(2022)\citenamefont {Elley},
  \citenamefont {Silva}, \citenamefont {Witek},\ and\ \citenamefont
  {Yunes}}]{Elley:2022ept}%
  \BibitemOpen
  \bibfield  {author} {\bibinfo {author} {\bibfnamefont {M.}~\bibnamefont
  {Elley}}, \bibinfo {author} {\bibfnamefont {H.~O.}\ \bibnamefont {Silva}},
  \bibinfo {author} {\bibfnamefont {H.}~\bibnamefont {Witek}},\ and\ \bibinfo
  {author} {\bibfnamefont {N.}~\bibnamefont {Yunes}},\ }\href
  {https://doi.org/10.1103/PhysRevD.106.044018} {\bibfield  {journal} {\bibinfo
   {journal} {Phys. Rev. D}\ }\textbf {\bibinfo {volume} {106}},\ \bibinfo
  {pages} {044018} (\bibinfo {year} {2022})},\ \Eprint
  {https://arxiv.org/abs/2205.06240} {arXiv:2205.06240 [gr-qc]} \BibitemShut
  {NoStop}%
\bibitem [{\citenamefont {Doneva}\ \emph
  {et~al.}(2022{\natexlab{b}})\citenamefont {Doneva}, \citenamefont {Va\~n\'o
  Vi\~nuales},\ and\ \citenamefont {Yazadjiev}}]{Doneva:2022byd}%
  \BibitemOpen
  \bibfield  {author} {\bibinfo {author} {\bibfnamefont {D.~D.}\ \bibnamefont
  {Doneva}}, \bibinfo {author} {\bibfnamefont {A.}~\bibnamefont {Va\~n\'o
  Vi\~nuales}},\ and\ \bibinfo {author} {\bibfnamefont {S.~S.}\ \bibnamefont
  {Yazadjiev}},\ }\href {https://doi.org/10.1103/PhysRevD.106.L061502}
  {\bibfield  {journal} {\bibinfo  {journal} {Phys. Rev. D}\ }\textbf {\bibinfo
  {volume} {106}},\ \bibinfo {pages} {L061502} (\bibinfo {year}
  {2022}{\natexlab{b}})},\ \Eprint {https://arxiv.org/abs/2204.05333}
  {arXiv:2204.05333 [gr-qc]} \BibitemShut {NoStop}%
\bibitem [{\citenamefont {Okounkova}(2020)}]{Okounkova:2020rqw}%
  \BibitemOpen
  \bibfield  {author} {\bibinfo {author} {\bibfnamefont {M.}~\bibnamefont
  {Okounkova}},\ }\href {https://doi.org/10.1103/PhysRevD.102.084046}
  {\bibfield  {journal} {\bibinfo  {journal} {Phys. Rev. D}\ }\textbf {\bibinfo
  {volume} {102}},\ \bibinfo {pages} {084046} (\bibinfo {year} {2020})},\
  \Eprint {https://arxiv.org/abs/2001.03571} {arXiv:2001.03571 [gr-qc]}
  \BibitemShut {NoStop}%
\bibitem [{\citenamefont {Silva}\ \emph {et~al.}(2021)\citenamefont {Silva},
  \citenamefont {Witek}, \citenamefont {Elley},\ and\ \citenamefont
  {Yunes}}]{Silva:2020omi}%
  \BibitemOpen
  \bibfield  {author} {\bibinfo {author} {\bibfnamefont {H.~O.}\ \bibnamefont
  {Silva}}, \bibinfo {author} {\bibfnamefont {H.}~\bibnamefont {Witek}},
  \bibinfo {author} {\bibfnamefont {M.}~\bibnamefont {Elley}},\ and\ \bibinfo
  {author} {\bibfnamefont {N.}~\bibnamefont {Yunes}},\ }\href
  {https://doi.org/10.1103/PhysRevLett.127.031101} {\bibfield  {journal}
  {\bibinfo  {journal} {Phys. Rev. Lett.}\ }\textbf {\bibinfo {volume} {127}},\
  \bibinfo {pages} {031101} (\bibinfo {year} {2021})},\ \Eprint
  {https://arxiv.org/abs/2012.10436} {arXiv:2012.10436 [gr-qc]} \BibitemShut
  {NoStop}%
\bibitem [{\citenamefont {Okounkova}\ \emph {et~al.}(2020)\citenamefont
  {Okounkova}, \citenamefont {Stein}, \citenamefont {Moxon}, \citenamefont
  {Scheel},\ and\ \citenamefont {Teukolsky}}]{Okounkova:2019zjf}%
  \BibitemOpen
  \bibfield  {author} {\bibinfo {author} {\bibfnamefont {M.}~\bibnamefont
  {Okounkova}}, \bibinfo {author} {\bibfnamefont {L.~C.}\ \bibnamefont
  {Stein}}, \bibinfo {author} {\bibfnamefont {J.}~\bibnamefont {Moxon}},
  \bibinfo {author} {\bibfnamefont {M.~A.}\ \bibnamefont {Scheel}},\ and\
  \bibinfo {author} {\bibfnamefont {S.~A.}\ \bibnamefont {Teukolsky}},\ }\href
  {https://doi.org/10.1103/PhysRevD.101.104016} {\bibfield  {journal} {\bibinfo
   {journal} {Phys. Rev. D}\ }\textbf {\bibinfo {volume} {101}},\ \bibinfo
  {pages} {104016} (\bibinfo {year} {2020})},\ \Eprint
  {https://arxiv.org/abs/1911.02588} {arXiv:1911.02588 [gr-qc]} \BibitemShut
  {NoStop}%
\bibitem [{\citenamefont {Okounkova}\ \emph {et~al.}(2019)\citenamefont
  {Okounkova}, \citenamefont {Stein}, \citenamefont {Scheel},\ and\
  \citenamefont {Teukolsky}}]{Okounkova:2019dfo}%
  \BibitemOpen
  \bibfield  {author} {\bibinfo {author} {\bibfnamefont {M.}~\bibnamefont
  {Okounkova}}, \bibinfo {author} {\bibfnamefont {L.~C.}\ \bibnamefont
  {Stein}}, \bibinfo {author} {\bibfnamefont {M.~A.}\ \bibnamefont {Scheel}},\
  and\ \bibinfo {author} {\bibfnamefont {S.~A.}\ \bibnamefont {Teukolsky}},\
  }\href {https://doi.org/10.1103/PhysRevD.100.104026} {\bibfield  {journal}
  {\bibinfo  {journal} {Phys. Rev. D}\ }\textbf {\bibinfo {volume} {100}},\
  \bibinfo {pages} {104026} (\bibinfo {year} {2019})},\ \Eprint
  {https://arxiv.org/abs/1906.08789} {arXiv:1906.08789 [gr-qc]} \BibitemShut
  {NoStop}%
\bibitem [{\citenamefont {Witek}\ \emph {et~al.}(2019)\citenamefont {Witek},
  \citenamefont {Gualtieri}, \citenamefont {Pani},\ and\ \citenamefont
  {Sotiriou}}]{Witek:2018dmd}%
  \BibitemOpen
  \bibfield  {author} {\bibinfo {author} {\bibfnamefont {H.}~\bibnamefont
  {Witek}}, \bibinfo {author} {\bibfnamefont {L.}~\bibnamefont {Gualtieri}},
  \bibinfo {author} {\bibfnamefont {P.}~\bibnamefont {Pani}},\ and\ \bibinfo
  {author} {\bibfnamefont {T.~P.}\ \bibnamefont {Sotiriou}},\ }\href
  {https://doi.org/10.1103/PhysRevD.99.064035} {\bibfield  {journal} {\bibinfo
  {journal} {Phys. Rev. D}\ }\textbf {\bibinfo {volume} {99}},\ \bibinfo
  {pages} {064035} (\bibinfo {year} {2019})},\ \Eprint
  {https://arxiv.org/abs/1810.05177} {arXiv:1810.05177 [gr-qc]} \BibitemShut
  {NoStop}%
\bibitem [{\citenamefont {Penrose}(1965)}]{Penrose:1965am}%
  \BibitemOpen
  \bibfield  {author} {\bibinfo {author} {\bibfnamefont {R.}~\bibnamefont
  {Penrose}},\ }\href {https://doi.org/10.1098/rspa.1965.0058} {\bibfield
  {journal} {\bibinfo  {journal} {Proc. Roy. Soc. Lond. A}\ }\textbf {\bibinfo
  {volume} {284}},\ \bibinfo {pages} {159} (\bibinfo {year}
  {1965})}\BibitemShut {NoStop}%
\bibitem [{\citenamefont {Kov\'acs}\ and\ \citenamefont
  {Reall}(2020{\natexlab{a}})}]{Kovacs:2020pns}%
  \BibitemOpen
  \bibfield  {author} {\bibinfo {author} {\bibfnamefont {A.~D.}\ \bibnamefont
  {Kov\'acs}}\ and\ \bibinfo {author} {\bibfnamefont {H.~S.}\ \bibnamefont
  {Reall}},\ }\href {https://doi.org/10.1103/PhysRevLett.124.221101} {\bibfield
   {journal} {\bibinfo  {journal} {Phys. Rev. Lett.}\ }\textbf {\bibinfo
  {volume} {124}},\ \bibinfo {pages} {221101} (\bibinfo {year}
  {2020}{\natexlab{a}})},\ \Eprint {https://arxiv.org/abs/2003.04327}
  {arXiv:2003.04327 [gr-qc]} \BibitemShut {NoStop}%
\bibitem [{\citenamefont {Kov\'acs}\ and\ \citenamefont
  {Reall}(2020{\natexlab{b}})}]{Kovacs:2020ywu}%
  \BibitemOpen
  \bibfield  {author} {\bibinfo {author} {\bibfnamefont {A.~D.}\ \bibnamefont
  {Kov\'acs}}\ and\ \bibinfo {author} {\bibfnamefont {H.~S.}\ \bibnamefont
  {Reall}},\ }\href {https://doi.org/10.1103/PhysRevD.101.124003} {\bibfield
  {journal} {\bibinfo  {journal} {Phys. Rev. D}\ }\textbf {\bibinfo {volume}
  {101}},\ \bibinfo {pages} {124003} (\bibinfo {year} {2020}{\natexlab{b}})},\
  \Eprint {https://arxiv.org/abs/2003.08398} {arXiv:2003.08398 [gr-qc]}
  \BibitemShut {NoStop}%
\bibitem [{\citenamefont {East}\ and\ \citenamefont
  {Ripley}(2021{\natexlab{a}})}]{East:2020hgw}%
  \BibitemOpen
  \bibfield  {author} {\bibinfo {author} {\bibfnamefont {W.~E.}\ \bibnamefont
  {East}}\ and\ \bibinfo {author} {\bibfnamefont {J.~L.}\ \bibnamefont
  {Ripley}},\ }\href {https://doi.org/10.1103/PhysRevD.103.044040} {\bibfield
  {journal} {\bibinfo  {journal} {Phys. Rev. D}\ }\textbf {\bibinfo {volume}
  {103}},\ \bibinfo {pages} {044040} (\bibinfo {year} {2021}{\natexlab{a}})},\
  \Eprint {https://arxiv.org/abs/2011.03547} {arXiv:2011.03547 [gr-qc]}
  \BibitemShut {NoStop}%
\bibitem [{\citenamefont {East}\ and\ \citenamefont
  {Ripley}(2021{\natexlab{b}})}]{East:2021bqk}%
  \BibitemOpen
  \bibfield  {author} {\bibinfo {author} {\bibfnamefont {W.~E.}\ \bibnamefont
  {East}}\ and\ \bibinfo {author} {\bibfnamefont {J.~L.}\ \bibnamefont
  {Ripley}},\ }\href {https://doi.org/10.1103/PhysRevLett.127.101102}
  {\bibfield  {journal} {\bibinfo  {journal} {Phys. Rev. Lett.}\ }\textbf
  {\bibinfo {volume} {127}},\ \bibinfo {pages} {101102} (\bibinfo {year}
  {2021}{\natexlab{b}})},\ \Eprint {https://arxiv.org/abs/2105.08571}
  {arXiv:2105.08571 [gr-qc]} \BibitemShut {NoStop}%
\bibitem [{\citenamefont {East}\ and\ \citenamefont
  {Pretorius}(2022)}]{East:2022rqi}%
  \BibitemOpen
  \bibfield  {author} {\bibinfo {author} {\bibfnamefont {W.~E.}\ \bibnamefont
  {East}}\ and\ \bibinfo {author} {\bibfnamefont {F.}~\bibnamefont
  {Pretorius}},\ }\href {https://doi.org/10.1103/PhysRevD.106.104055}
  {\bibfield  {journal} {\bibinfo  {journal} {Phys. Rev. D}\ }\textbf {\bibinfo
  {volume} {106}},\ \bibinfo {pages} {104055} (\bibinfo {year} {2022})},\
  \Eprint {https://arxiv.org/abs/2208.09488} {arXiv:2208.09488 [gr-qc]}
  \BibitemShut {NoStop}%
\bibitem [{\citenamefont {Corman}\ \emph {et~al.}(2023)\citenamefont {Corman},
  \citenamefont {Ripley},\ and\ \citenamefont {East}}]{Corman:2022xqg}%
  \BibitemOpen
  \bibfield  {author} {\bibinfo {author} {\bibfnamefont {M.}~\bibnamefont
  {Corman}}, \bibinfo {author} {\bibfnamefont {J.~L.}\ \bibnamefont {Ripley}},\
  and\ \bibinfo {author} {\bibfnamefont {W.~E.}\ \bibnamefont {East}},\ }\href
  {https://doi.org/10.1103/PhysRevD.107.024014} {\bibfield  {journal} {\bibinfo
   {journal} {Phys. Rev. D}\ }\textbf {\bibinfo {volume} {107}},\ \bibinfo
  {pages} {024014} (\bibinfo {year} {2023})},\ \Eprint
  {https://arxiv.org/abs/2210.09235} {arXiv:2210.09235 [gr-qc]} \BibitemShut
  {NoStop}%
\bibitem [{\citenamefont {R}\ \emph {et~al.}(2023)\citenamefont {R},
  \citenamefont {Ripley},\ and\ \citenamefont {Yunes}}]{R:2022hlf}%
  \BibitemOpen
  \bibfield  {author} {\bibinfo {author} {\bibfnamefont {A.~H.~K.}\
  \bibnamefont {R}}, \bibinfo {author} {\bibfnamefont {J.~L.}\ \bibnamefont
  {Ripley}},\ and\ \bibinfo {author} {\bibfnamefont {N.}~\bibnamefont
  {Yunes}},\ }\href {https://doi.org/10.1103/PhysRevD.107.044044} {\bibfield
  {journal} {\bibinfo  {journal} {Phys. Rev. D}\ }\textbf {\bibinfo {volume}
  {107}},\ \bibinfo {pages} {044044} (\bibinfo {year} {2023})},\ \Eprint
  {https://arxiv.org/abs/2211.08477} {arXiv:2211.08477 [gr-qc]} \BibitemShut
  {NoStop}%
\bibitem [{\citenamefont {Thaalba}\ \emph {et~al.}(2023)\citenamefont
  {Thaalba}, \citenamefont {Bezares}, \citenamefont {Franchini},\ and\
  \citenamefont {Sotiriou}}]{Thaalba:2023fmq}%
  \BibitemOpen
  \bibfield  {author} {\bibinfo {author} {\bibfnamefont {F.}~\bibnamefont
  {Thaalba}}, \bibinfo {author} {\bibfnamefont {M.}~\bibnamefont {Bezares}},
  \bibinfo {author} {\bibfnamefont {N.}~\bibnamefont {Franchini}},\ and\
  \bibinfo {author} {\bibfnamefont {T.~P.}\ \bibnamefont {Sotiriou}},\
  }\href@noop {} {\  (\bibinfo {year} {2023})},\ \Eprint
  {https://arxiv.org/abs/2306.01695} {arXiv:2306.01695 [gr-qc]} \BibitemShut
  {NoStop}%
\bibitem [{\citenamefont {Lara}\ \emph {et~al.}(2022)\citenamefont {Lara},
  \citenamefont {Bezares},\ and\ \citenamefont {Barausse}}]{Lara:2021piy}%
  \BibitemOpen
  \bibfield  {author} {\bibinfo {author} {\bibfnamefont {G.}~\bibnamefont
  {Lara}}, \bibinfo {author} {\bibfnamefont {M.}~\bibnamefont {Bezares}},\ and\
  \bibinfo {author} {\bibfnamefont {E.}~\bibnamefont {Barausse}},\ }\href
  {https://doi.org/10.1103/PhysRevD.105.064058} {\bibfield  {journal} {\bibinfo
   {journal} {Phys. Rev. D}\ }\textbf {\bibinfo {volume} {105}},\ \bibinfo
  {pages} {064058} (\bibinfo {year} {2022})},\ \Eprint
  {https://arxiv.org/abs/2112.09186} {arXiv:2112.09186 [gr-qc]} \BibitemShut
  {NoStop}%
\bibitem [{\citenamefont {Nakamura}\ \emph {et~al.}(1987)\citenamefont
  {Nakamura}, \citenamefont {Oohara},\ and\ \citenamefont
  {Kojima}}]{Nakamura:1987zz}%
  \BibitemOpen
  \bibfield  {author} {\bibinfo {author} {\bibfnamefont {T.}~\bibnamefont
  {Nakamura}}, \bibinfo {author} {\bibfnamefont {K.}~\bibnamefont {Oohara}},\
  and\ \bibinfo {author} {\bibfnamefont {Y.}~\bibnamefont {Kojima}},\ }\href
  {https://doi.org/10.1143/PTPS.90.1} {\bibfield  {journal} {\bibinfo
  {journal} {Prog. Theor. Phys. Suppl.}\ }\textbf {\bibinfo {volume} {90}},\
  \bibinfo {pages} {1} (\bibinfo {year} {1987})}\BibitemShut {NoStop}%
\bibitem [{\citenamefont {Shibata}\ and\ \citenamefont
  {Nakamura}(1995)}]{Shibata:1995we}%
  \BibitemOpen
  \bibfield  {author} {\bibinfo {author} {\bibfnamefont {M.}~\bibnamefont
  {Shibata}}\ and\ \bibinfo {author} {\bibfnamefont {T.}~\bibnamefont
  {Nakamura}},\ }\href {https://doi.org/10.1103/PhysRevD.52.5428} {\bibfield
  {journal} {\bibinfo  {journal} {Phys. Rev. D}\ }\textbf {\bibinfo {volume}
  {52}},\ \bibinfo {pages} {5428} (\bibinfo {year} {1995})}\BibitemShut
  {NoStop}%
\bibitem [{\citenamefont {Baumgarte}\ and\ \citenamefont
  {Shapiro}(1998)}]{Baumgarte:1998te}%
  \BibitemOpen
  \bibfield  {author} {\bibinfo {author} {\bibfnamefont {T.~W.}\ \bibnamefont
  {Baumgarte}}\ and\ \bibinfo {author} {\bibfnamefont {S.~L.}\ \bibnamefont
  {Shapiro}},\ }\href {https://doi.org/10.1103/PhysRevD.59.024007} {\bibfield
  {journal} {\bibinfo  {journal} {Phys. Rev. D}\ }\textbf {\bibinfo {volume}
  {59}},\ \bibinfo {pages} {024007} (\bibinfo {year} {1998})},\ \Eprint
  {https://arxiv.org/abs/gr-qc/9810065} {arXiv:gr-qc/9810065} \BibitemShut
  {NoStop}%
\bibitem [{\citenamefont {Bona}\ \emph {et~al.}(2003)\citenamefont {Bona},
  \citenamefont {Ledvinka}, \citenamefont {Palenzuela},\ and\ \citenamefont
  {Zacek}}]{Bona:2003fj}%
  \BibitemOpen
  \bibfield  {author} {\bibinfo {author} {\bibfnamefont {C.}~\bibnamefont
  {Bona}}, \bibinfo {author} {\bibfnamefont {T.}~\bibnamefont {Ledvinka}},
  \bibinfo {author} {\bibfnamefont {C.}~\bibnamefont {Palenzuela}},\ and\
  \bibinfo {author} {\bibfnamefont {M.}~\bibnamefont {Zacek}},\ }\href
  {https://doi.org/10.1103/PhysRevD.67.104005} {\bibfield  {journal} {\bibinfo
  {journal} {Phys. Rev. D}\ }\textbf {\bibinfo {volume} {67}},\ \bibinfo
  {pages} {104005} (\bibinfo {year} {2003})},\ \Eprint
  {https://arxiv.org/abs/gr-qc/0302083} {arXiv:gr-qc/0302083} \BibitemShut
  {NoStop}%
\bibitem [{\citenamefont {Bernuzzi}\ and\ \citenamefont
  {Hilditch}(2010)}]{Bernuzzi:2009ex}%
  \BibitemOpen
  \bibfield  {author} {\bibinfo {author} {\bibfnamefont {S.}~\bibnamefont
  {Bernuzzi}}\ and\ \bibinfo {author} {\bibfnamefont {D.}~\bibnamefont
  {Hilditch}},\ }\href {https://doi.org/10.1103/PhysRevD.81.084003} {\bibfield
  {journal} {\bibinfo  {journal} {Phys. Rev. D}\ }\textbf {\bibinfo {volume}
  {81}},\ \bibinfo {pages} {084003} (\bibinfo {year} {2010})},\ \Eprint
  {https://arxiv.org/abs/0912.2920} {arXiv:0912.2920 [gr-qc]} \BibitemShut
  {NoStop}%
\bibitem [{\citenamefont {Alic}\ \emph {et~al.}(2012)\citenamefont {Alic},
  \citenamefont {Bona-Casas}, \citenamefont {Bona}, \citenamefont {Rezzolla},\
  and\ \citenamefont {Palenzuela}}]{Alic:2011gg}%
  \BibitemOpen
  \bibfield  {author} {\bibinfo {author} {\bibfnamefont {D.}~\bibnamefont
  {Alic}}, \bibinfo {author} {\bibfnamefont {C.}~\bibnamefont {Bona-Casas}},
  \bibinfo {author} {\bibfnamefont {C.}~\bibnamefont {Bona}}, \bibinfo {author}
  {\bibfnamefont {L.}~\bibnamefont {Rezzolla}},\ and\ \bibinfo {author}
  {\bibfnamefont {C.}~\bibnamefont {Palenzuela}},\ }\href
  {https://doi.org/10.1103/PhysRevD.85.064040} {\bibfield  {journal} {\bibinfo
  {journal} {Phys. Rev. D}\ }\textbf {\bibinfo {volume} {85}},\ \bibinfo
  {pages} {064040} (\bibinfo {year} {2012})},\ \Eprint
  {https://arxiv.org/abs/1106.2254} {arXiv:1106.2254 [gr-qc]} \BibitemShut
  {NoStop}%
\bibitem [{\citenamefont {Alic}\ \emph {et~al.}(2013)\citenamefont {Alic},
  \citenamefont {Kastaun},\ and\ \citenamefont {Rezzolla}}]{Alic:2013xsa}%
  \BibitemOpen
  \bibfield  {author} {\bibinfo {author} {\bibfnamefont {D.}~\bibnamefont
  {Alic}}, \bibinfo {author} {\bibfnamefont {W.}~\bibnamefont {Kastaun}},\ and\
  \bibinfo {author} {\bibfnamefont {L.}~\bibnamefont {Rezzolla}},\ }\href
  {https://doi.org/10.1103/PhysRevD.88.064049} {\bibfield  {journal} {\bibinfo
  {journal} {Phys. Rev. D}\ }\textbf {\bibinfo {volume} {88}},\ \bibinfo
  {pages} {064049} (\bibinfo {year} {2013})},\ \Eprint
  {https://arxiv.org/abs/1307.7391} {arXiv:1307.7391 [gr-qc]} \BibitemShut
  {NoStop}%
\bibitem [{\citenamefont {Campanelli}\ \emph {et~al.}(2006)\citenamefont
  {Campanelli}, \citenamefont {Lousto}, \citenamefont {Marronetti},\ and\
  \citenamefont {Zlochower}}]{Campanelli:2005dd}%
  \BibitemOpen
  \bibfield  {author} {\bibinfo {author} {\bibfnamefont {M.}~\bibnamefont
  {Campanelli}}, \bibinfo {author} {\bibfnamefont {C.~O.}\ \bibnamefont
  {Lousto}}, \bibinfo {author} {\bibfnamefont {P.}~\bibnamefont {Marronetti}},\
  and\ \bibinfo {author} {\bibfnamefont {Y.}~\bibnamefont {Zlochower}},\ }\href
  {https://doi.org/10.1103/PhysRevLett.96.111101} {\bibfield  {journal}
  {\bibinfo  {journal} {Phys. Rev. Lett.}\ }\textbf {\bibinfo {volume} {96}},\
  \bibinfo {pages} {111101} (\bibinfo {year} {2006})},\ \Eprint
  {https://arxiv.org/abs/gr-qc/0511048} {arXiv:gr-qc/0511048} \BibitemShut
  {NoStop}%
\bibitem [{\citenamefont {Baker}\ \emph {et~al.}(2006)\citenamefont {Baker},
  \citenamefont {Centrella}, \citenamefont {Choi}, \citenamefont {Koppitz},\
  and\ \citenamefont {van Meter}}]{Baker:2005vv}%
  \BibitemOpen
  \bibfield  {author} {\bibinfo {author} {\bibfnamefont {J.~G.}\ \bibnamefont
  {Baker}}, \bibinfo {author} {\bibfnamefont {J.}~\bibnamefont {Centrella}},
  \bibinfo {author} {\bibfnamefont {D.-I.}\ \bibnamefont {Choi}}, \bibinfo
  {author} {\bibfnamefont {M.}~\bibnamefont {Koppitz}},\ and\ \bibinfo {author}
  {\bibfnamefont {J.}~\bibnamefont {van Meter}},\ }\href
  {https://doi.org/10.1103/PhysRevLett.96.111102} {\bibfield  {journal}
  {\bibinfo  {journal} {Phys. Rev. Lett.}\ }\textbf {\bibinfo {volume} {96}},\
  \bibinfo {pages} {111102} (\bibinfo {year} {2006})},\ \Eprint
  {https://arxiv.org/abs/gr-qc/0511103} {arXiv:gr-qc/0511103} \BibitemShut
  {NoStop}%
\bibitem [{\citenamefont {Arest\'e~Sal\'o}\ \emph {et~al.}(2022)\citenamefont
  {Arest\'e~Sal\'o}, \citenamefont {Clough},\ and\ \citenamefont
  {Figueras}}]{AresteSalo:2022hua}%
  \BibitemOpen
  \bibfield  {author} {\bibinfo {author} {\bibfnamefont {L.}~\bibnamefont
  {Arest\'e~Sal\'o}}, \bibinfo {author} {\bibfnamefont {K.}~\bibnamefont
  {Clough}},\ and\ \bibinfo {author} {\bibfnamefont {P.}~\bibnamefont
  {Figueras}},\ }\href {https://doi.org/10.1103/PhysRevLett.129.261104}
  {\bibfield  {journal} {\bibinfo  {journal} {Phys. Rev. Lett.}\ }\textbf
  {\bibinfo {volume} {129}},\ \bibinfo {pages} {261104} (\bibinfo {year}
  {2022})},\ \Eprint {https://arxiv.org/abs/2208.14470} {arXiv:2208.14470
  [gr-qc]} \BibitemShut {NoStop}%
\bibitem [{\citenamefont {Bona}\ \emph {et~al.}(1995)\citenamefont {Bona},
  \citenamefont {Mass\'o}, \citenamefont {Seidel},\ and\ \citenamefont
  {Stela}}]{Bona:1994dr}%
  \BibitemOpen
  \bibfield  {author} {\bibinfo {author} {\bibfnamefont {C.}~\bibnamefont
  {Bona}}, \bibinfo {author} {\bibfnamefont {J.}~\bibnamefont {Mass\'o}},
  \bibinfo {author} {\bibfnamefont {E.}~\bibnamefont {Seidel}},\ and\ \bibinfo
  {author} {\bibfnamefont {J.}~\bibnamefont {Stela}},\ }\href
  {https://doi.org/10.1103/PhysRevLett.75.600} {\bibfield  {journal} {\bibinfo
  {journal} {Phys. Rev. Lett.}\ }\textbf {\bibinfo {volume} {75}},\ \bibinfo
  {pages} {600} (\bibinfo {year} {1995})},\ \Eprint
  {https://arxiv.org/abs/gr-qc/9412071} {arXiv:gr-qc/9412071} \BibitemShut
  {NoStop}%
\bibitem [{\citenamefont {Alcubierre}\ \emph {et~al.}(2003)\citenamefont
  {Alcubierre}, \citenamefont {Br{\"u}gmann}, \citenamefont {Diener},
  \citenamefont {Koppitz}, \citenamefont {Pollney}, \citenamefont {Seidel},\
  and\ \citenamefont {Takahashi}}]{Alcubierre:2002kk}%
  \BibitemOpen
  \bibfield  {author} {\bibinfo {author} {\bibfnamefont {M.}~\bibnamefont
  {Alcubierre}}, \bibinfo {author} {\bibfnamefont {B.}~\bibnamefont
  {Br{\"u}gmann}}, \bibinfo {author} {\bibfnamefont {P.}~\bibnamefont
  {Diener}}, \bibinfo {author} {\bibfnamefont {M.}~\bibnamefont {Koppitz}},
  \bibinfo {author} {\bibfnamefont {D.}~\bibnamefont {Pollney}}, \bibinfo
  {author} {\bibfnamefont {E.}~\bibnamefont {Seidel}},\ and\ \bibinfo {author}
  {\bibfnamefont {R.}~\bibnamefont {Takahashi}},\ }\href
  {https://doi.org/10.1103/PhysRevD.67.084023} {\bibfield  {journal} {\bibinfo
  {journal} {Phys. Rev. D}\ }\textbf {\bibinfo {volume} {67}},\ \bibinfo
  {pages} {084023} (\bibinfo {year} {2003})},\ \Eprint
  {https://arxiv.org/abs/gr-qc/0206072} {arXiv:gr-qc/0206072} \BibitemShut
  {NoStop}%
\bibitem [{\citenamefont {Wald}(1984)}]{Wald:1984rg}%
  \BibitemOpen
  \bibfield  {author} {\bibinfo {author} {\bibfnamefont {R.~M.}\ \bibnamefont
  {Wald}},\ }\href {https://doi.org/10.7208/chicago/9780226870373.001.0001}
  {\emph {\bibinfo {title} {{General Relativity}}}}\ (\bibinfo  {publisher}
  {Chicago Univ. Pr.},\ \bibinfo {address} {Chicago, USA},\ \bibinfo {year}
  {1984})\BibitemShut {NoStop}%
\bibitem [{\citenamefont {Brown}(2011)}]{brown}%
  \BibitemOpen
  \bibfield  {author} {\bibinfo {author} {\bibfnamefont {J.~D.}\ \bibnamefont
  {Brown}},\ }\href {https://doi.org/10.1103/PhysRevD.84.124012} {\bibfield
  {journal} {\bibinfo  {journal} {Phys. Rev. D}\ }\textbf {\bibinfo {volume}
  {84}},\ \bibinfo {pages} {124012} (\bibinfo {year} {2011})},\ \Eprint
  {https://arxiv.org/abs/1109.1707} {arXiv:1109.1707 [gr-qc]} \BibitemShut
  {NoStop}%
\bibitem [{\citenamefont {Figueras}\ \emph {et~al.}(2016)\citenamefont
  {Figueras}, \citenamefont {Kunesch},\ and\ \citenamefont
  {Tunyasuvunakool}}]{Figueras:2015hkb}%
  \BibitemOpen
  \bibfield  {author} {\bibinfo {author} {\bibfnamefont {P.}~\bibnamefont
  {Figueras}}, \bibinfo {author} {\bibfnamefont {M.}~\bibnamefont {Kunesch}},\
  and\ \bibinfo {author} {\bibfnamefont {S.}~\bibnamefont {Tunyasuvunakool}},\
  }\href {https://doi.org/10.1103/PhysRevLett.116.071102} {\bibfield  {journal}
  {\bibinfo  {journal} {Phys. Rev. Lett.}\ }\textbf {\bibinfo {volume} {116}},\
  \bibinfo {pages} {071102} (\bibinfo {year} {2016})},\ \Eprint
  {https://arxiv.org/abs/1512.04532} {arXiv:1512.04532 [hep-th]} \BibitemShut
  {NoStop}%
\bibitem [{\citenamefont {Figueras}\ \emph {et~al.}(2017)\citenamefont
  {Figueras}, \citenamefont {Kunesch}, \citenamefont {Lehner},\ and\
  \citenamefont {Tunyasuvunakool}}]{Figueras:2017zwa}%
  \BibitemOpen
  \bibfield  {author} {\bibinfo {author} {\bibfnamefont {P.}~\bibnamefont
  {Figueras}}, \bibinfo {author} {\bibfnamefont {M.}~\bibnamefont {Kunesch}},
  \bibinfo {author} {\bibfnamefont {L.}~\bibnamefont {Lehner}},\ and\ \bibinfo
  {author} {\bibfnamefont {S.}~\bibnamefont {Tunyasuvunakool}},\ }\href
  {https://doi.org/10.1103/PhysRevLett.118.151103} {\bibfield  {journal}
  {\bibinfo  {journal} {Phys. Rev. Lett.}\ }\textbf {\bibinfo {volume} {118}},\
  \bibinfo {pages} {151103} (\bibinfo {year} {2017})},\ \Eprint
  {https://arxiv.org/abs/1702.01755} {arXiv:1702.01755 [hep-th]} \BibitemShut
  {NoStop}%
\bibitem [{\citenamefont {Gundlach}\ \emph {et~al.}(2005)\citenamefont
  {Gundlach}, \citenamefont {Mart\'in-Garc\'ia}, \citenamefont {Calabrese},\
  and\ \citenamefont {Hinder}}]{Gundlach:2005eh}%
  \BibitemOpen
  \bibfield  {author} {\bibinfo {author} {\bibfnamefont {C.}~\bibnamefont
  {Gundlach}}, \bibinfo {author} {\bibfnamefont {J.~M.}\ \bibnamefont
  {Mart\'in-Garc\'ia}}, \bibinfo {author} {\bibfnamefont {G.}~\bibnamefont
  {Calabrese}},\ and\ \bibinfo {author} {\bibfnamefont {I.}~\bibnamefont
  {Hinder}},\ }\href {https://doi.org/10.1088/0264-9381/22/17/025} {\bibfield
  {journal} {\bibinfo  {journal} {Class. Quant. Grav.}\ }\textbf {\bibinfo
  {volume} {22}},\ \bibinfo {pages} {3767} (\bibinfo {year} {2005})},\ \Eprint
  {https://arxiv.org/abs/gr-qc/0504114} {arXiv:gr-qc/0504114} \BibitemShut
  {NoStop}%
\bibitem [{\citenamefont {Pretorius}(2005)}]{Pretorius:2004jg}%
  \BibitemOpen
  \bibfield  {author} {\bibinfo {author} {\bibfnamefont {F.}~\bibnamefont
  {Pretorius}},\ }\href {https://doi.org/10.1088/0264-9381/22/2/014} {\bibfield
   {journal} {\bibinfo  {journal} {Class. Quant. Grav.}\ }\textbf {\bibinfo
  {volume} {22}},\ \bibinfo {pages} {425} (\bibinfo {year} {2005})},\ \Eprint
  {https://arxiv.org/abs/gr-qc/0407110} {arXiv:gr-qc/0407110} \BibitemShut
  {NoStop}%
\bibitem [{\citenamefont {Torii}\ and\ \citenamefont {Shinkai}(2008)}]{torii}%
  \BibitemOpen
  \bibfield  {author} {\bibinfo {author} {\bibfnamefont {T.}~\bibnamefont
  {Torii}}\ and\ \bibinfo {author} {\bibfnamefont {H.}~\bibnamefont
  {Shinkai}},\ }\href {https://doi.org/10.1103/PhysRevD.78.084037} {\bibfield
  {journal} {\bibinfo  {journal} {Phys. Rev. D}\ }\textbf {\bibinfo {volume}
  {78}},\ \bibinfo {pages} {084037} (\bibinfo {year} {2008})},\ \Eprint
  {https://arxiv.org/abs/0810.1790} {arXiv:0810.1790 [gr-qc]} \BibitemShut
  {NoStop}%
\bibitem [{sup()}]{supp}%
  \BibitemOpen
  \href@noop {} {\bibinfo {title} {{See the ancillary files at}}},\ \bibinfo
  {howpublished} {[\url{URL}]}\BibitemShut {NoStop}%
\bibitem [{\citenamefont {J.~Hinch}(1991)}]{hinch}%
  \BibitemOpen
  \bibfield  {author} {\bibinfo {author} {\bibfnamefont {E.}~\bibnamefont
  {J.~Hinch}},\ }\href@noop {} {\emph {\bibinfo {title} {{Perturbation
  Methods}}}}\ (\bibinfo  {publisher} {{Cambridge University Press}},\ \bibinfo
  {year} {1991})\BibitemShut {NoStop}%
\bibitem [{\citenamefont {Weinberg}(2008)}]{Weinberg:2008hq}%
  \BibitemOpen
  \bibfield  {author} {\bibinfo {author} {\bibfnamefont {S.}~\bibnamefont
  {Weinberg}},\ }\href {https://doi.org/10.1103/PhysRevD.77.123541} {\bibfield
  {journal} {\bibinfo  {journal} {Phys. Rev. D}\ }\textbf {\bibinfo {volume}
  {77}},\ \bibinfo {pages} {123541} (\bibinfo {year} {2008})},\ \Eprint
  {https://arxiv.org/abs/0804.4291} {arXiv:0804.4291 [hep-th]} \BibitemShut
  {NoStop}%
\bibitem [{\citenamefont {Reall}(2021)}]{Reall:2021voz}%
  \BibitemOpen
  \bibfield  {author} {\bibinfo {author} {\bibfnamefont {H.~S.}\ \bibnamefont
  {Reall}},\ }\href {https://doi.org/10.1103/PhysRevD.103.084027} {\bibfield
  {journal} {\bibinfo  {journal} {Phys. Rev. D}\ }\textbf {\bibinfo {volume}
  {103}},\ \bibinfo {pages} {084027} (\bibinfo {year} {2021})},\ \Eprint
  {https://arxiv.org/abs/2101.11623} {arXiv:2101.11623 [gr-qc]} \BibitemShut
  {NoStop}%
\bibitem [{\citenamefont {Clough}\ \emph {et~al.}(2015)\citenamefont {Clough},
  \citenamefont {Figueras}, \citenamefont {Finkel}, \citenamefont {Kunesch},
  \citenamefont {Lim},\ and\ \citenamefont {Tunyasuvunakool}}]{Clough:2015sqa}%
  \BibitemOpen
  \bibfield  {author} {\bibinfo {author} {\bibfnamefont {K.}~\bibnamefont
  {Clough}}, \bibinfo {author} {\bibfnamefont {P.}~\bibnamefont {Figueras}},
  \bibinfo {author} {\bibfnamefont {H.}~\bibnamefont {Finkel}}, \bibinfo
  {author} {\bibfnamefont {M.}~\bibnamefont {Kunesch}}, \bibinfo {author}
  {\bibfnamefont {E.~A.}\ \bibnamefont {Lim}},\ and\ \bibinfo {author}
  {\bibfnamefont {S.}~\bibnamefont {Tunyasuvunakool}},\ }\href
  {https://doi.org/10.1088/0264-9381/32/24/245011} {\bibfield  {journal}
  {\bibinfo  {journal} {Class. Quant. Grav.}\ }\textbf {\bibinfo {volume}
  {32}},\ \bibinfo {pages} {245011} (\bibinfo {year} {2015})},\ \Eprint
  {https://arxiv.org/abs/1503.03436} {arXiv:1503.03436 [gr-qc]} \BibitemShut
  {NoStop}%
\bibitem [{\citenamefont {Andrade}\ \emph {et~al.}(2021)\citenamefont {Andrade}
  \emph {et~al.}}]{Andrade:2021rbd}%
  \BibitemOpen
  \bibfield  {author} {\bibinfo {author} {\bibfnamefont {T.}~\bibnamefont
  {Andrade}} \emph {et~al.},\ }\href {https://doi.org/10.21105/joss.03703}
  {\bibfield  {journal} {\bibinfo  {journal} {J. Open Source Softw.}\ }\textbf
  {\bibinfo {volume} {6}},\ \bibinfo {pages} {3703} (\bibinfo {year} {2021})},\
  \Eprint {https://arxiv.org/abs/2201.03458} {arXiv:2201.03458 [gr-qc]}
  \BibitemShut {NoStop}%
\bibitem [{\citenamefont {Figueras}\ and\ \citenamefont
  {Fran\c{c}a}(2020)}]{Figueras:2020dzx}%
  \BibitemOpen
  \bibfield  {author} {\bibinfo {author} {\bibfnamefont {P.}~\bibnamefont
  {Figueras}}\ and\ \bibinfo {author} {\bibfnamefont {T.}~\bibnamefont
  {Fran\c{c}a}},\ }\href {https://doi.org/10.1088/1361-6382/abb693} {\bibfield
  {journal} {\bibinfo  {journal} {Class. Quant. Grav.}\ }\textbf {\bibinfo
  {volume} {37}},\ \bibinfo {pages} {225009} (\bibinfo {year} {2020})},\
  \Eprint {https://arxiv.org/abs/2006.09414} {arXiv:2006.09414 [gr-qc]}
  \BibitemShut {NoStop}%
\bibitem [{\citenamefont {Figueras}\ and\ \citenamefont
  {Fran\c{c}a}(2022)}]{Figueras:2021abd}%
  \BibitemOpen
  \bibfield  {author} {\bibinfo {author} {\bibfnamefont {P.}~\bibnamefont
  {Figueras}}\ and\ \bibinfo {author} {\bibfnamefont {T.}~\bibnamefont
  {Fran\c{c}a}},\ }\href {https://doi.org/10.1103/PhysRevD.105.124004}
  {\bibfield  {journal} {\bibinfo  {journal} {Phys. Rev. D}\ }\textbf {\bibinfo
  {volume} {105}},\ \bibinfo {pages} {124004} (\bibinfo {year} {2022})},\
  \Eprint {https://arxiv.org/abs/2112.15529} {arXiv:2112.15529 [gr-qc]}
  \BibitemShut {NoStop}%
\bibitem [{\citenamefont {Radia}\ \emph {et~al.}(2022)\citenamefont {Radia},
  \citenamefont {Sperhake}, \citenamefont {Drew}, \citenamefont {Clough},
  \citenamefont {Figueras}, \citenamefont {Lim}, \citenamefont {Ripley},
  \citenamefont {Aurrekoetxea}, \citenamefont {Fran\c{c}a},\ and\ \citenamefont
  {Helfer}}]{Radia:2021smk}%
  \BibitemOpen
  \bibfield  {author} {\bibinfo {author} {\bibfnamefont {M.}~\bibnamefont
  {Radia}}, \bibinfo {author} {\bibfnamefont {U.}~\bibnamefont {Sperhake}},
  \bibinfo {author} {\bibfnamefont {A.}~\bibnamefont {Drew}}, \bibinfo {author}
  {\bibfnamefont {K.}~\bibnamefont {Clough}}, \bibinfo {author} {\bibfnamefont
  {P.}~\bibnamefont {Figueras}}, \bibinfo {author} {\bibfnamefont {E.~A.}\
  \bibnamefont {Lim}}, \bibinfo {author} {\bibfnamefont {J.~L.}\ \bibnamefont
  {Ripley}}, \bibinfo {author} {\bibfnamefont {J.~C.}\ \bibnamefont
  {Aurrekoetxea}}, \bibinfo {author} {\bibfnamefont {T.}~\bibnamefont
  {Fran\c{c}a}},\ and\ \bibinfo {author} {\bibfnamefont {T.}~\bibnamefont
  {Helfer}},\ }\href {https://doi.org/10.1088/1361-6382/ac6fa9} {\bibfield
  {journal} {\bibinfo  {journal} {Class. Quant. Grav.}\ }\textbf {\bibinfo
  {volume} {39}},\ \bibinfo {pages} {135006} (\bibinfo {year} {2022})},\
  \Eprint {https://arxiv.org/abs/2112.10567} {arXiv:2112.10567 [gr-qc]}
  \BibitemShut {NoStop}%
\bibitem [{\citenamefont {Liu}\ \emph {et~al.}(2009)\citenamefont {Liu},
  \citenamefont {Etienne},\ and\ \citenamefont {Shapiro}}]{Liu:2009al}%
  \BibitemOpen
  \bibfield  {author} {\bibinfo {author} {\bibfnamefont {Y.~T.}\ \bibnamefont
  {Liu}}, \bibinfo {author} {\bibfnamefont {Z.~B.}\ \bibnamefont {Etienne}},\
  and\ \bibinfo {author} {\bibfnamefont {S.~L.}\ \bibnamefont {Shapiro}},\
  }\href {https://doi.org/10.1103/PhysRevD.80.121503} {\bibfield  {journal}
  {\bibinfo  {journal} {Phys. Rev. D}\ }\textbf {\bibinfo {volume} {80}},\
  \bibinfo {pages} {121503} (\bibinfo {year} {2009})},\ \Eprint
  {https://arxiv.org/abs/1001.4077} {arXiv:1001.4077 [gr-qc]} \BibitemShut
  {NoStop}%
\bibitem [{\citenamefont {Baumgarte}\ and\ \citenamefont
  {Shapiro}(2010)}]{Baumgarte:2010ndz}%
  \BibitemOpen
  \bibfield  {author} {\bibinfo {author} {\bibfnamefont {T.~W.}\ \bibnamefont
  {Baumgarte}}\ and\ \bibinfo {author} {\bibfnamefont {S.~L.}\ \bibnamefont
  {Shapiro}},\ }\href {https://doi.org/10.1017/CBO9781139193344} {\emph
  {\bibinfo {title} {{Numerical Relativity: Solving Einstein's Equations on the
  Computer}}}}\ (\bibinfo  {publisher} {Cambridge University Press},\ \bibinfo
  {year} {2010})\BibitemShut {NoStop}%
\bibitem [{\citenamefont {Bowen}\ and\ \citenamefont
  {York}(1980)}]{Bowen:1980yu}%
  \BibitemOpen
  \bibfield  {author} {\bibinfo {author} {\bibfnamefont {J.~M.}\ \bibnamefont
  {Bowen}}\ and\ \bibinfo {author} {\bibfnamefont {J.~W.}\ \bibnamefont {York},
  \bibfnamefont {Jr.}},\ }\href {https://doi.org/10.1103/PhysRevD.21.2047}
  {\bibfield  {journal} {\bibinfo  {journal} {Phys. Rev. D}\ }\textbf {\bibinfo
  {volume} {21}},\ \bibinfo {pages} {2047} (\bibinfo {year}
  {1980})}\BibitemShut {NoStop}%
\bibitem [{\citenamefont {Ansorg}\ \emph {et~al.}(2004)\citenamefont {Ansorg},
  \citenamefont {Br{\"u}gmann},\ and\ \citenamefont {Tichy}}]{Ansorg:2004ds}%
  \BibitemOpen
  \bibfield  {author} {\bibinfo {author} {\bibfnamefont {M.}~\bibnamefont
  {Ansorg}}, \bibinfo {author} {\bibfnamefont {B.}~\bibnamefont
  {Br{\"u}gmann}},\ and\ \bibinfo {author} {\bibfnamefont {W.}~\bibnamefont
  {Tichy}},\ }\href {https://doi.org/10.1103/PhysRevD.70.064011} {\bibfield
  {journal} {\bibinfo  {journal} {Phys. Rev. D}\ }\textbf {\bibinfo {volume}
  {70}},\ \bibinfo {pages} {064011} (\bibinfo {year} {2004})},\ \Eprint
  {https://arxiv.org/abs/gr-qc/0404056} {arXiv:gr-qc/0404056} \BibitemShut
  {NoStop}%
\bibitem [{\citenamefont {Aurrekoetxea}\ \emph {et~al.}(2023)\citenamefont
  {Aurrekoetxea}, \citenamefont {Clough},\ and\ \citenamefont
  {Lim}}]{Aurrekoetxea:2022mpw}%
  \BibitemOpen
  \bibfield  {author} {\bibinfo {author} {\bibfnamefont {J.~C.}\ \bibnamefont
  {Aurrekoetxea}}, \bibinfo {author} {\bibfnamefont {K.}~\bibnamefont
  {Clough}},\ and\ \bibinfo {author} {\bibfnamefont {E.~A.}\ \bibnamefont
  {Lim}},\ }\href {https://doi.org/10.1088/1361-6382/acb883} {\bibfield
  {journal} {\bibinfo  {journal} {Class. Quant. Grav.}\ }\textbf {\bibinfo
  {volume} {40}},\ \bibinfo {pages} {075003} (\bibinfo {year} {2023})},\
  \Eprint {https://arxiv.org/abs/2207.03125} {arXiv:2207.03125 [gr-qc]}
  \BibitemShut {NoStop}%
\bibitem [{\citenamefont {Brady}\ \emph {et~al.}(2023)\citenamefont {Brady},
  \citenamefont {Arest\'e~Sal\'o}, \citenamefont {Clough}, \citenamefont
  {Figueras},\ and\ \citenamefont {S}}]{Brady:2023dgu}%
  \BibitemOpen
  \bibfield  {author} {\bibinfo {author} {\bibfnamefont {S.~E.}\ \bibnamefont
  {Brady}}, \bibinfo {author} {\bibfnamefont {L.}~\bibnamefont
  {Arest\'e~Sal\'o}}, \bibinfo {author} {\bibfnamefont {K.}~\bibnamefont
  {Clough}}, \bibinfo {author} {\bibfnamefont {P.}~\bibnamefont {Figueras}},\
  and\ \bibinfo {author} {\bibfnamefont {A.~P.}\ \bibnamefont {S}},\
  }\href@noop {} {\  (\bibinfo {year} {2023})},\ \Eprint
  {https://arxiv.org/abs/2308.16791} {arXiv:2308.16791 [gr-qc]} \BibitemShut
  {NoStop}%
\bibitem [{\citenamefont {Doneva}\ \emph {et~al.}(2023)\citenamefont {Doneva},
  \citenamefont {Arest\'e~Sal\'o}, \citenamefont {Clough}, \citenamefont
  {Figueras},\ and\ \citenamefont {Yazadjiev}}]{Doneva:2023oww}%
  \BibitemOpen
  \bibfield  {author} {\bibinfo {author} {\bibfnamefont {D.~D.}\ \bibnamefont
  {Doneva}}, \bibinfo {author} {\bibfnamefont {L.}~\bibnamefont
  {Arest\'e~Sal\'o}}, \bibinfo {author} {\bibfnamefont {K.}~\bibnamefont
  {Clough}}, \bibinfo {author} {\bibfnamefont {P.}~\bibnamefont {Figueras}},\
  and\ \bibinfo {author} {\bibfnamefont {S.~S.}\ \bibnamefont {Yazadjiev}},\
  }\href@noop {} {\  (\bibinfo {year} {2023})},\ \Eprint
  {https://arxiv.org/abs/2307.06474} {arXiv:2307.06474 [gr-qc]} \BibitemShut
  {NoStop}%
\bibitem [{\citenamefont {Doneva}\ \emph
  {et~al.}(2022{\natexlab{c}})\citenamefont {Doneva}, \citenamefont
  {Collodel},\ and\ \citenamefont {Yazadjiev}}]{Doneva:2022yqu}%
  \BibitemOpen
  \bibfield  {author} {\bibinfo {author} {\bibfnamefont {D.~D.}\ \bibnamefont
  {Doneva}}, \bibinfo {author} {\bibfnamefont {L.~G.}\ \bibnamefont
  {Collodel}},\ and\ \bibinfo {author} {\bibfnamefont {S.~S.}\ \bibnamefont
  {Yazadjiev}},\ }\href {https://doi.org/10.1103/PhysRevD.106.104027}
  {\bibfield  {journal} {\bibinfo  {journal} {Phys. Rev. D}\ }\textbf {\bibinfo
  {volume} {106}},\ \bibinfo {pages} {104027} (\bibinfo {year}
  {2022}{\natexlab{c}})},\ \Eprint {https://arxiv.org/abs/2208.02077}
  {arXiv:2208.02077 [gr-qc]} \BibitemShut {NoStop}%
\bibitem [{\citenamefont {King}\ \emph {et~al.}(2017)\citenamefont {King},
  \citenamefont {Butcher},\ and\ \citenamefont {Zalewski}}]{apocrita}%
  \BibitemOpen
  \bibfield  {author} {\bibinfo {author} {\bibfnamefont {T.}~\bibnamefont
  {King}}, \bibinfo {author} {\bibfnamefont {S.}~\bibnamefont {Butcher}},\ and\
  \bibinfo {author} {\bibfnamefont {L.}~\bibnamefont {Zalewski}},\ }\href
  {https://doi.org/10.5281/zenodo.438045} {\emph {\bibinfo {title} {{Apocrita -
  High Performance Computing Cluster for Queen Mary University of London}}}}
  (\bibinfo {year} {2017})\BibitemShut {NoStop}%
\end{thebibliography}%

\end{document}